%% file: p.tex
\documentclass[a4paper]{article}

\input{p-sty-man}

\begin{document}

\title{The Epstein-Glaser approach to pQFT:\\ Graphs and Hopf algebras}
\author{Alexander Lange\\\small\tt alange@mis.mpg.de\\[2ex]
Max Planck Institute for
Mathematics in the Sciences\\ Inselstr.~22--26, 04275 Leipzig, Germany}
\date{14th July 2004}

\maketitle

\begin{abstract}
The paper aims at investigating perturbative quantum field theory (pQFT)
in the approach of Epstein and Glaser (EG) and, in particular, its formulation in the language of graphs and Hopf algebras (HAs).
Various HAs are encountered, each one associated with a special combination of physical concepts such as normalization, localization, pseudo-unitarity, causal(ity and an associated) regularization, and renormalization.
The algebraic structures, representing the perturbative expansion of the $S$-matrix, are imposed on the operator-valued distributions which are
equipped with appropriate graph indices.
Translation invariance ensures the algebras to be analytically well-defined and graded total symmetry allows to formulate bialgebras.
The algebraic results are given embedded in the physical framework,
which covers the two recent EG versions by Fredenhagen and Scharf
that differ with respect to the concrete recursive implementation of causality.
Besides, the ultraviolet divergences occuring in Feynman's representation are
mathematically reasoned.
As a final result,
the change of the renormalization scheme in the EG framework is modeled via a HA
which can be seen as the EG-analog of Kreimer's HA.

\vspace{3ex}

\normalsize
\noindent{\bf 2003 PACS}: 03.70.+k, 11.10.Gh, 11.15.Bt, 11.55.-m

\end{abstract}

\tableofcontents

\input{p-1}

\input{p-2}

\input{p-3}

\input{p-4}

\input{p-o}

\section*{Acknowledgements}
I would like to thank Prof.~Dirk Kreimer for the invitation to Boston University in fall 2003. I profited a lot from the discussions with him and with Kurush Ebrahimi-Fard.
Moreover, I would like thank the MPI MIS for financial support and especially Prof.~Eberhard Zeidler for his interest in the topic.

\input{p-app}

{\small\bibliographystyle{plain}\bibliography{p-lit-pub}}

\end{document}

%% file: p-sty-man
\usepackage[german,english]{babel}
\usepackage{amscd,amsfonts,amsmath,amssymb}
\usepackage{array}
\usepackage{changebar}
\usepackage{color}
\usepackage{eepic,epic,epsf,epsfig}
\usepackage{enumerate}
\usepackage{fancybox}
\usepackage{float}
\usepackage{latexsym}
\usepackage{picinpar}
\usepackage{stmaryrd}
\usepackage{rotating}
\usepackage{verbatim}
\usepackage{wrapfig}

\input xy
\xyoption{all}

\DeclareSymbolFont{AMSb}{U}{msb}{m}{n}
\DeclareSymbolFontAlphabet{\mathbb}{AMSb}
\usepackage{theorem}

\newtheorem{Thm}{Theorem}[section]
\newtheorem{Cor}[Thm]{Corollary}
{\theorembodyfont{\rmfamily} }
{\theorembodyfont{\rmfamily} \newtheorem{Exa}[Thm]{Example}}
\newtheorem{Lem}[Thm]{Lemma}
\newtheorem{Prn}[Thm]{Proposition}
{\theorembodyfont{\rmfamily} \newtheorem{Rem}[Thm]{Remark}}

\newenvironment{Proof}{{\it Proof.}\rm}{\hspace*{\fill}$_\square$}

\newcommand{\eq}[1]{(\ref{#1})}

\newcommand{\sss}{\scriptscriptstyle}

\newcommand{\sssf}[1]{{\scriptstyle\sf#1}}
\newcommand{\lsf}[1]{{{\text{\raisebox{-1ex}{\LARGE\sf#1}}}}}

%% file: p-1

\section{Introduction}

In perturbative Quantum Field Theory (pQFT) interaction of elementary particles is modeled to describe 
the outcomes of scattering experiments. 
In the approach of Epstein and Glaser (EG) the S(cattering)-matrix is investigated
as a formal power series, 
\begin{align}
\label{S(gtau)=sum...}
S(a\tau)=\sum_{n=0}^\infty\frac{a^n}{n!}T_n(\tau^{\otimes n}),
\quad
\begin{cases}
a&\text{coupling (constant)},\\
\tau&\text{IR regularization (test function)},
\end{cases}
\end{align}
representing an operator-valued distribution acting on the (Fock) space
of free quantum fields,
where in the adiabatic limit, i.e. $\tau\to1$,
the square of the corresponding scalar product,
$|\langle\text{\sf\small final}|S(a\tau)|\text{\sf\small initial}\rangle|^2$,
is supposed to yield the scattering probability.
Based on the interaction term $T_1$, which is assumed to be local,
the higher order terms $T_n$, $n>1$, of the
perturbative expansion can be deduced by `only' applying causality.
In general the series does not converge (in any norm),
but in a few cases it is at least Borel summable,
cf.~\cite{Zi96Q}.
Despite this fact, conceptually, the theory is determined only up to
the so-called renormalization freedom.
However, in this approach, formulated in terms of distributions,
renormalization in the usual sense, as subtraction of counterterms,
does not have to be performed.
This actually reflects the original
motivation of EG
\cite{EG73t},
namely,
to give a proof of locality  (i.e.~renomalizability by local counterterms)
just by proceeding the sketched way which
can be seen as a parallel road to the well-known BPHZ approach
\cite{BP57u,He66p,Zi69c}
(although, referring to the abstract initial setting \eq{S(gtau)=sum...},
coming from the same junction).

The specified title 
of this paper
refers to terminology which so far has not properly been implemented into the considered approach.
Graphs do not appear in the EG framework and its recent versions \cite{Sc01Q,BF99m},
and Hopf algebras (HAs), from the perspective of physics,
constitute a well-developed concept only in the subject of quantum groups,
having decent applications in statistical physics
(see, e.g., Sects.~3.7, 7.6, 8.7 in \cite{KS97Q}).
Here, graphs will serve as indices for the n-point distributions $T_n$,
and the inductive method the latter will be obtained with is modeled by various HAs.

In 1997 Kreimer
\cite{Kr98o} discovered that the counterterms in Zimmermann's forest formula 
\cite{Zi69c},
which underlies renormalization in the Feyman graph approach,
can be interpreted as an antipode condition of a HA.
By then one has obviously overcome the mathematical lack which had been associated with renormalization before.
But the finding of the mathematical structure alone could not fully eliminate the mysticism involved --- the occurance of a combinatorial HA in this context requires further explanation.
The method developed by EG would 
give a closer look at the physics involved.
From a similar point of view
(cf.~\cite{Kr99o}, p.~24),
in 2000 Kreimer
\cite{Kr00c}, p.~9, formulated the research problem of investigating the EG approach under the aspect of HAs; the content of
\cite{GL00c} was not considered as the solution.

A brief history of contributions to the topic of EG-relevant HAs in pQFT reads as follows.
In 2000 Fauser
\cite{Fa00o} describes a HA that combines normal and time ordering.
In the same year, based on the work of Fredenhagen und coworkers
\cite{BF96T,BF97i,DF98a,BF99m}, Pinter
\cite{Pi00t}
discusses the question of uniqueness of EG's n-point distributions
to establish a link to Kreimer's HA.
In 2002 Brouder et al
\cite{BO02q,BS02q}
have taken up the previous two results again;
they revive the work of Rota and Stein
\cite{RS94p,RS94v}
about combinatorial HAs and present a link to quantum groups.
The most recent publications along this line are \cite{BF03l,BF04n}.

In the presented paper, which
uses a very general formulation of pQFT,
Kreimer's HA is reconstructed in the EG approach via a scheme of 
HAs associated with different physical concepts,
such as normalization (referring to the normal product construction),
localization, pseudo-unitarity, causal(ity with an appropriate) regularization,
and eventually renormalization,
cf.~Fig.~\ref{Fig:table}.
\begin{figure}%
{\sf
\setlength{\extrarowheight}{3.5pt}
\begin{tabular}{|c|c|c|c|c|}
\hline
\multicolumn{2}{|c|}{p Q F T}&
\multicolumn{3}{c|}{H o p f / b i - a l g e b r a}\\[.2ex]
\hline
\multicolumn{2}{|c|}{\small
physical  concepts}
&{\small product}&{\small coproduct}&{\small antipode}\\[.5ex]
\hline
\multicolumn{5}{c}{}\\[-1.2ex]
\hline
\multicolumn{2}{|c|}{quantization}
&$\circ$&$\Delta^{\text{\sf N}}$&none\\
\hline
\multicolumn{2}{c|}{}&
\multicolumn{2}{c|}{\raisebox{-.5ex}[.5ex]{\footnotesize S W I C H}}&
\multicolumn{1}{c}{}\\[1ex]
\hline
\multicolumn{2}{|c|}{normalization}
&$\text{\small\sf N}$&&\\
\cline{1-3}
\multicolumn{2}{|c|}{localization}
&$\text{\small\sf L}$&
\raisebox{1.5ex}[-1.5ex]{$\Delta$}&
\raisebox{1.5ex}[-1.5ex]{$(-1)^{|\cdot|}\!\!\sim$}\\
\hline
\multicolumn{2}{c|}{}&
\multicolumn{3}{c|}{\raisebox{-.5ex}[.5ex]{\footnotesize T W I S T\;}}\\[1ex]
\hline
\multicolumn{2}{|r|}{pseudo-untarity}&
$\circ$&&\\
\cline{1-1}\cline{3-3}&&
$\widehat\circ^\exists$&&\raisebox{1.5ex}[-1.5ex]{$\sum$}\\
\cline{2-2}\cline{3-3}\cline{5-5}
\multicolumn{2}{|l|}{\raisebox{1ex}[-1ex]{causal}}&
$\widehat\circ^{\sf F}$&$\Delta$&\\
\cline{3-3}
\multicolumn{2}{|r|}{\raisebox{1.5ex}[-1.5ex]{regularization}}&
$\widehat\circ^{\sf T}$&&{\raisebox{1.5ex}[-1.5ex]{$(-1)^{|\cdot|}\!\!\sim$}}\\
\cline{1-1}\cline{3-3}\cline{5-5}
\multicolumn{1}{|c|}{renormalization}&&
$\underline\circ^\ast$&&$\Sigma$\\
\hline
\end{tabular}
}
\vspace{1ex}
\centering
\caption{Hopf algebras modeling pQFT in the EG approach \label{Fig:table}}
\end{figure}
The over all strategy has been to represent iterated solutions
(e.g.~of the pertubative expansion)
in terms of antipode conditions.
At the beginning of this project,
which has been the topic of the author's thesis
\cite{La03c},
it was not clear what the involved physical instances would be
and neither whether one would be able to realize the aimed correspondence at all.
Due to recent results, the starting point for the construction of the sketched HAs can be fixed by a bialgebra which most naturally models
the quantum character of the considered physical objects. 
The algebra formalizes products of free quantum fields
and the coalgebra normal products of the latter.
By switching the coalgebra into an algebra and adding a shuffling coalgebra which respects graded total symmetry (gtS) one can easily construct a HA which encodes normalization.
Going over to a ring of scalars containing delta distributions,
one can as well include the concept of localization.
Extending this ring by propagators that satisfy the spectral condition
and by translation invariant (tI) distributions, one obtains an algebra
which formalizes the Wick expansion.
The product is again the composition (i.e.~the circle product,
which defines a so-called twist w.r.t.~to the localized normal product).
With the shuffling coproduct and an antipode
(given by the Schmitt formula) which implements pseudo-unitarity,
the resulting (Laplace\footnote{This is the terminology used by Rota and Stein \cite{RS94p,RS94v}. The sketched transfer (i.e.~twist) of a HA into a Laplace HA is referred to as Cliffordization.})
HA models the perturbative expansion.
This structure will have to be extended again
to explicitely implement causality.
In accordance with the well-known ultraviolet problems of pQFT,
causality requires an adequate regularization.
For the three considered approches,
i.e.~the two recent EG versions of Fredenhagen (EGF) and Scharf (EGS) as well as the Feynman approach of Bogoliubov-Shirkov (BS),
which vary with respect to
the concrete product 
and antipode,
this will be realized by localization on causally determined spacetime regions.
In order to finally re-establish the concept of renormalization
in the language of HAs
one has to consider graded commutative versions of the causally regularizing HAs.
Then
one can, in analogy to Kreimer's achievements,
via some kind of deformation,
describe the change of the renormalization scheme in terms of HAs.

The content has been arranged according to 
items just given.
This organizes the paper as follows.
In Sect.~\ref{s:FreeQFT} one introduces free quantum fields, incl.~their algebraic structures with respect to
normal products and localization.
Including interaction,
Sect.~\ref{s:pExp} is about the formal expansion of the $S$-matrix and associated algebras.
In Sect.~\ref{s:cr-pQFT} causality and renormalization get explicitely implemented into pQFT, leading to various HAs. 
More detailed summaries are given at the beginning of each (sub)section.

Some technical remarks: Independent of its mathematical importance,
only propositions which immediately refer to the table are called theorems.
To keep the number of pages in reasonable bounds,
most of the proofs are only sketched (especially when they can be found in
\cite{La03c} or in
some other reference).
There is also 
an appendix with proofs and illustrations that might not really be necessary for understanding the text.

%% file: p-2

\section{Local interaction between free quantized fields\label{s:FreeQFT}}

The basic physical objects in pQFT are free quantized fields.
They are introduced in Sect.~\ref{2.1}, emphasizing physical but also mathematical aspects.
For instance, the free fields can be shown to form a bialgebra
where the normal product serves as a coproduct. 
Two HAs are obtained in Sect.~2.2,
describing normalization and localization.
The localization of free fields is
realized via an appropriate ring of scalars.
In Sect.~2.3 monomial interaction of free fields is modeled.
This requires to apply the adiabatic limit and to implement the concept of classical background fields into the quantum formalism.

\subsection{Free quantized fields\label{2.1}}

The two sections of this first subsection present, one after the other, a physically and a mathematically motivated approach to the subject.
In Sect.~\ref{sssect:setup}
one introduces
(derived anti) particles, (derived) quantum fields, as well as
associated propagators,
and one formulates the causal and the spectral conditions.
In Sect.~\ref{2.1.2}
one constructs
a bialgebra modeling compositions of free quantized fields and normal (co)products of the them.

\subsubsection{The physical set up \label{sssect:setup}}

The inner product of the one particle Hilbert space ${\cal H}_p$,
\begin{align}
\frac{1}{i\hbar}(f,g)^+_p\propto\Delta^+_p(f,g):=(\Delta^+_p\ast g)(f^\ast),
\quad f,g\in{\cal D}_p,
\end{align}
is induced by the fundamental solutions of a wave (i.e.~relativistically
covariant
{\em translation invariant}\, linear
hyperbolic) operator, where $\Delta^+_p$, given via Fourier transformation
$(F\Delta^+_p)(k)=\theta(k^0)(F{\Delta_p})(k)$,
is the positive frequency part
of the difference
$\Delta_p=\Delta_p^R-\Delta_p^A$
between the retarded and advanced Green's functions 
(fulfilling 
${\rm supp}\,\Delta_p^{R|A}(\,\cdot-y)\subseteq
y+\overline{V}_{+|-}$,
$\forall y\in\mathbb{R}^{\rm1+d}$,
where
$\overline{V}_{+|-}$
is 
the closed forward$|$backward lightcone,
i.e.~$\overline{V}_+=\big\{x\in\mathbb{R}^{\rm1+d}\,\big|\,\sqrt{x_1^2+\dots+x_{\rm d}^2}\le x_0\big\}=-\overline{V}_-$).
The appropriate test function space, of Schwartz class say,
is denoted by ${\cal D}_p$.
In addition to the adjoint $f^\ast$ of a test function $f\in{\cal D}_p$, one will also have to apply the transposed of the previous, $\tilde f:=f^{\ast t}$.
Notice, Planck's constant $\hbar$ has already been implemented and the velocity of light $c:=1$.

Following a notation similar\footnote{e.g., covariance is not simulated here}
to \cite{Bo00g},
general indices, e.g.~$\nu$, $\mu$, are supposed to 
determine derived (anti) 
particles, $p=\llbracket\nu\rrbracket$ (and $\tilde p=\llbracket\mu\rrbracket$, resp.),
representing propagators including derivatives in terms 
of the Pauli Jordan function, i.e.
\begin{align}
\Delta_p=\big(\Delta^{\nu\mu}\big)_{\llbracket\nu\rrbracket=p,
\llbracket\mu\rrbracket=\tilde p}
\quad\text{where}\quad
\Delta^{\nu\mu}\overset{\partial^{...}}{\sim} D_{m_\nu}=D^+_{m_\nu}+D^-_{m_\nu}
\end{align}
and $D^\pm_{m_\nu}(x)=\pm\frac{i}{(2\pi)^{\rm1+d}}\int d\vec k\,\frac{e^{\mp
ikx}}{2\omega_\nu(\vec k)}$ where
$\omega_\nu(\vec k)=\sqrt{\vec k^2+\frac{m_\nu^2}{\hbar^2}}$.
Antiparticles $\tilde p=\llbracket\tilde\nu\rrbracket$ are always assumed to
exist, at least formally, defined by
\begin{align}
\Delta^+_{\tilde p}(x)=-s_p\,\Delta^-_p(-x)^t,
\quad\text{where}\quad
s_p:=(-1)^{2\,{\rm spin}(p)}
\end{align}
and where $\Delta^-_p$ symbolizes the negative frequency part.

Quantum fields $\varphi_p$ associated with (anti) particles $p$ are defined as
operators on (the dense subspace of) Fock 
space ${\cal D}^{\rm F}_p=\bigoplus_{n\ge0}S_n^{\sigma_p}{\cal D}_p^{\otimes n}\subset{\cal H}^{\rm F}_p$,
i.e.~in terms
of annihilation 
and creation 
operators,
\begin{align}\label{(phi)=A-(a)+A+(a)}
{\begin{pmatrix}\varphi_p(f)\\ \varphi_{\tilde p}(\tilde{f})\end{pmatrix}}
=A^{-}_p
{\begin{pmatrix} a_p(f)\\ a_{\tilde p}(\tilde{f})\end{pmatrix}}
+A^{+}_p
{\begin{pmatrix}a^\ast_p(f) \\ a^\ast_{\tilde p}(\tilde{f})\end{pmatrix}},
\quad f\in{\cal D}_p,
\end{align}
where for the considered theories
\begin{align}
A^{-}_p=\begin{pmatrix}1&0\\0&\gamma_p\end{pmatrix}
\quad\text{and}\quad
A^{+}_p=\begin{pmatrix}0&1\\1&0\end{pmatrix},
\end{align}
so that by the choice  
of $\gamma_p=-1\in\{\pm1\}$ one can also model
ghost fields.
As usual,
on 
${\cal D}^{\rm F}_p\ni|h\rangle=(h_0,h_1(\cdot),h_2(\cdot,\cdot),\dots)$,
\begin{align}
\label{a_p}
\big(a_{p}(f)|h\rangle\big)_{n}
&=\sqrt{n+1}\,\big(f,h_{n+1}(\,\cdot\,,\dots)\big)^+_{p},\\
\big(a_{p}^\ast(f)|h\rangle\big)_n
&=\sqrt{n}\,S_n^{\sigma_{p}}(f\otimes h_{n-1}),
\end{align}
where (anti) symmetrization characterizes bosons (fermions, resp.),
\begin{align}
S_n^{\sigma_{p}} h_n
=\frac{1}{n!}\sum_{{\sss\pi\in{\cal S}_n}}\sigma_{p}(\pi)\,h_n\circ\pi,
\end{align}
and $\sigma_p(\cdot)$ defines the {\em grading}\,
w.r.t.~the statistics $\sigma_p$,
\begin{align}
\sigma_p(\pi)&=\begin{cases}1\\({\rm sgn}\,\pi)\end{cases}
\text{if}\quad
\sigma_p=\begin{cases}\,\,+1\\(-1,\text{ resp.}).\end{cases}
\end{align}
Derived fields $\varphi_{p'}=\partial^{...}\varphi_p$
are as well,
in order to be treated as free,
considered to be given by
({derived}\, anti) particles $p'\in A'(p)$,
\begin{align}
\Big(\underset{\mu\in
I}{\bigcirc}\partial^\mu\Big)\varphi^\nu(f):=\varphi_\nu\bigg((-1)^{|I|}\Big(\underset{\mu\in
I}{\bigcirc}\partial^\mu\Big)f\bigg),
\quad
I \text{ depends on } (\nu,p').
\end{align}

The graded commutators of free fields have been defined to fulfill 
the {\em causal
property}, i.e.
$\gamma_p=\frac{1}{s_p\sigma_p}\equiv s_p\sigma_p$,
which imposes them,
\begin{align}
\big[\varphi^\nu(f),\varphi^\mu(\tilde g)\big]^\circ_{\sigma_p}&:=
\varphi^\nu(f)\circ\varphi^\mu(\tilde g)-\sigma_p\,\varphi^\mu(\tilde g)\circ\varphi^\nu(f)\\
&\;=i\hbar\,\gamma_p\,\Delta^{\nu\mu}(f,g)\,{\rm
id}_{\cal F}=:i\hbar\,\Delta_\gamma^{\nu\mu}(f,g),
\label{anticomm}
\end{align}
to have a causal support,
i.e.
\begin{align}\label{causalsupp}
\big[\varphi^\nu(f),\varphi^\mu(\tilde g)\big]^\circ_{\sigma_p}=0
\quad\text{if}\quad
{\rm supp}\,f\subseteq\mathbb{R}^{\rm1+d}\setminus\big({\rm supp}\,g+(\overline{V}_+\cup\overline{V}_-)\big),
\end{align}
where $\llbracket\nu\rrbracket=p$ and $\llbracket\mu\rrbracket=\tilde p'$.
Notice, whenever spin and statistics do not coincide the particle is a ghost
\cite{Kr95o}. 
The annihilation and creation part commutators read
(in density notation, with $x,y\in\mathbb{R}^{\rm1+d}$)
\begin{align}\label{+-commrel}
\big[\varphi_{\mp}^\nu(x),\varphi_{\pm}^\mu(y)\big]^\circ_{\sigma_p}
=i\hbar\,\Delta_{\pm\gamma}^{\nu\mu}(x-y),
\end{align}
where the upper case will be applied by Wick's theorem.
The positive frequency part reflects the
{\em spectral condition},
which ensures the propagators' multiplication (as distributions) to be well-defined.
Namely, as $F{\Delta_i^+}\in{\cal S}'(\overline{V}_+)$, i.e.
\begin{align}\label{sc}
{\rm supp}\,F{\Delta_i^+}\subseteq\overline{V}_+,
\quad\forall i\le n,
\end{align}
the so-called Fourier product
\cite{Ob92M} 
\begin{align}
\prod_{i\le n}\Delta^+_i(x):=F^{-1}\Big(F\big({\Delta^+_1(x)}\big)\ast\dots\ast F\big({\Delta^+_n(x)}\big)\Big)
\in{\cal S}'(\mathbb{R}^{\rm1+d}).
\end{align}
Moreover,
cf.~\cite{VD88T}, 
the (associative and commutative) convolution lies in ${\cal S}'(\overline{V}_+)$,
i.e.,
the resulting product satisfies the spectral condition as well.

The relevant counter example is {\em Feynman's (time-ordered) propagator}
\begin{align}
\Delta^{\sf F}:={\sf T}\,\Delta^+.
\end{align}
Applying the causality condition, this
leads to 
the well-known representation,
\begin{align}\label{Delta^F=+}
\Delta^{\sf F\gamma}_p(x)={\sf T}\,\Delta^+_p(x)&:=\theta(x^0)\,\Delta^{+\gamma}_p(x)+\sigma_p\,\theta(-x^0)\,\Delta^{+\gamma}_{\tilde p}(-x)\\
&\;=\theta(x^0)\,\Delta^{+\gamma}_p(x)-\theta(-x^0)\,\Delta^{-\gamma}_p(x),
\end{align}
which shows that the spectral condition does not hold true.

\subsubsection{The initial bialgebra\label{2.1.2}}

Starting with the algebraic considerations for an arbitrarily chosen pQFT, 
let $\rm P$ denote the finite set of particles and ${A}'(p)$ the class of derived (anti) particles associated with $p\in{\rm P}$.
Then the common Fock space is formed by the following 
symmetrized tensor product, 
${\cal H}^{\rm F}_{\rm P}:=\big(S_{|{\rm P}|}\bigotimes_{p\in{\rm P}}\big)\big(
\bigotimes_{q\in A'(p)}{\cal H}^{\rm F}_q\big)$.
Compatible with this definition,
cf.~\eq{anticomm},
the composition of field operators
${\cal F}_{\rm p}:=\{\phi^{i}\equiv\varphi_{{\rm p}_i}(x_i)\,|\,i\in\mathbb{N}\}$
that are 
fixed by a given surjective map
${\rm p}:\mathbb{N}\ni i\mapsto$ ${\rm p}_i\in{\rm P}$
which models the whole scattering scenario,
is commutative only if the associated particles do not belong to the same class of derived (anti) particles,
i.e., $[\varphi_{{\rm p}_1}(x_1),\varphi_{{\rm p}_2}(x_2)]^\circ=0$
if ${\rm p_2}\not\in{A}'({\rm p}_1)$.

The admissible compositions ${\cal F}=\mathbb{C}\langle{\cal F}_{\rm p},+,\circ\rangle/(\dots\circ\phi\circ\dots\circ\phi\circ\cdots)$
of field operators ${\cal F}_{\rm p}$, excluding multiple occurances,
generate an algebra $({\cal F},\mathbb{C},+,\circ,\eta)$
with unity (map) $\eta:=\cdot\;{\rm id}_{{\cal H}^{\rm F}_{\rm P}}$.
According to
the determined brackets' resolution,
i.e.~`$\bigcirc_{i\le n}\phi^i$ with brackets' $=\phi^1\circ(\dots\circ(\phi^{n-1}\circ\phi^n)...)$,
the composition is always assumed to be associative.
In the same spirit, just as the field operators act (linearly) on their domain, which is a (dense) subspace of the Fock space ${\cal H}^{\rm F}_{\rm P}$, distributivity is implemented.
However, repeating the statement from above, the {composition algebra of free fields} is not graded commutative.

In order to model local interaction the composition,
as the following example shows,
is not a good product.
In the limit of coinciding spacetime points the vacuum expectation is not defined,
\begin{align}\label{circ=N+?}
\langle0|\,\varphi^\nu(x)\circ\varphi^\mu(y)\,|0\rangle
=\underbrace{\langle0|\,\varphi^\nu(x)\;\sssf{N}\;\varphi^\mu(y)\,|0\rangle}_{\hspace{2.2em}=0,\;\forall x,y}
+i\hbar
\underbrace{\Delta_{+\gamma}^{\nu\mu}(x-y)}_{\hspace{1em}\longrightarrow_{x\to y}(?)}\equiv0+(?).
\end{align}
The vanishing expectation, i.e.~the first expression on the r.h.s., is induced by the so-called normal product,
\begin{align}
\varphi^\nu(x)\;\sssf{N}\;\varphi^\mu(y)=
\varphi_+^\nu(x)\circ\varphi_-^\mu(y)+\sigma_p\;\varphi_+^\mu(y)\circ\varphi_-^\nu(x).
\end{align}
As already applied in (\ref{+-commrel}, l.h.s.),
the subscript sign $+|-$ denotes the projection
${\rm pr}_\pm:\phi\mapsto\phi_\pm$
onto the creation$|$annihilation part, cf.~\eq{(phi)=A-(a)+A+(a)}.
For more than two factors the normal product is defined by
\begin{align}
{\sf  N}\phi^{\otimes J}:=
\sum_{J_\pm\in{\cal P}_2^0(J)}
\sigma_J(J_\pm)\;\bigcirc\phi^{\otimes J_+}_{+}\circ\bigcirc\phi^{\otimes J_-}_{-},
\end{align}
where $\sigma_J$ denotes the grading sign formulated on tuples $\bigotimes J\equiv(J_1,\dots,J_{n})\in\mathbb{N}^n_{\neq}$ of $n\in\mathbb{N}$ distinct numbers (i.e., $J_i\neq J_j$ if $i\neq j$, $\forall i,j\le n$)
that represent (indices of) particles ${\rm p}_{J_i}$,
where $J_\pm=(J_+,J_-)\in{\cal P}_2^{(0)}(J)$ denotes a 
2-partition of $J$ (resp., incl.~empty parts), and where 
$\phi^{\otimes J}:=\bigotimes_{j\in J}\phi^{j}$.
The grading $\sigma:{\cal S}_n\to\mathbb{Z}_2$, $(J\mapsto K)=\pi\mapsto\sigma(\pi)=\sigma_J(K)$
is a (non-trivial) representation of the permutation group $({\cal S}_n,\circ)$ on $(\mathbb{Z}_2,\cdot\,)$.
In accordance with the definition of
the common Fock space
${\cal H}^{\rm F}_{\rm P}$
it can be given in terms of 
the statistical coefficients $\sigma_p$, i.e.
\begin{align}\label{sigma(pi)=prod}
\sigma(\pi):=\prod_{p\in\rm P}\sigma_p(\pi|_{p})
=\prod_{\substack{p\in\rm P\\{\rm sgn}(\pi|_p)=\sigma_p=-1
}}
\sigma_p(\pi|_{p})
\quad\text{and}\quad\sigma_p(\emptyset):=1,
\end{align}
where
$\pi|_p:
J|_p\mapsto\pi(J|_p)$ denotes a subpermutation of $\pi$ 
which exclusively applies to derived (anti) particles $A'(p)$ w.r.t.~$p$,
defined on the subtuple
$J|_p:=(J_i)_{\substack{(\exists i\le n){\rm p}_{J_i}\in A'(p)}}$.

Important for the (Hopf/bi-)
algebraic scenario going to be presented is the fact that
the normal product
can be denoted with the help of a coproduct,
\begin{align}
\Delta^{\sf N}:=({\rm pr}_+\otimes{\rm pr}_-)\circ\Delta,
\end{align}
which is based on (the gtS version, cf.~right below, of) the shuffling coproduct,
\begin{align}
\Delta\big(\phi^{\bigcirc J}\big)
=\sum_{J_\pm\in{\cal P}_2^0(J)}
\sigma_J(J_\pm)\;\phi^{\bigcirc J_+}\otimes\phi^{\bigcirc J_-}.
\end{align}
\big(Notice, the abbreviation $\phi^{\Box J}:=\bigbox_{j\in J}\phi^j$ is used for any kind of product `$\sssf{\Box}$'.\big)
Immediately one observes that
\begin{align}\label{N=bigcircDelta^N}
{\sf  N} 
=\bigcirc\circ\Delta^{\sf N}
\circ\bigcirc, 
\end{align}
and 
Lemma \ref{DeltaN-coass} in the Appendix
verifies the implicit claim about
$\Delta^{({\sf N})}$ being
(two) coproducts.
Therefore, as one easily verifies,
$\big({\cal F},\mathbb{C},+,\Delta^{(\sf N)}\big)$ forms
a (non) cocommutative
coalgebra
where only $\Delta$ comes with a counit,
\begin{align}
\varepsilon(\phi)=\begin{cases}c&\text{\rm if}\quad\phi=c1,\\0&\text{\rm otherwise}.\end{cases}
\end{align}
Moreover, $\Delta^{(\sf N)}$ as well as $\varepsilon$, can be shown to define homomorphisms w.r.t.~the composition product.
This leads to the 
main result of this subsection.

\begin{Thm}
$\big({\cal F},\mathbb{C},+,\circ,\eta,\Delta^{(\sf N)},\varepsilon\big)$ forms a bialgebra (without counit, resp.).\\[1ex]
\begin{Proof}
All what is left to show is an immediate consequence of Lemma \ref{DeltaN,epsilon-homom}.
\end{Proof}
\end{Thm}

\begin{Rem}
Because of \eq{circ=N+?}, only the structure in parenthesis seems to be relevant for local pQFT.
\end{Rem}

Furthermore, one observes that the latterly introduced objects are equipped with the very quantum theoretic property of {graded total symmetry (gtS)},
which models indistinguishable particles governed by a particular statistics.

\begin{Prn}\label{Prop:gtS}
The normal product $\sf N$, as well as the associated (and, resp., the shuffling) coproduct $\Delta^{({\sf N})}$, 
and the Feynman propagator $\Delta^{\sf F}$ are all 
gtS.\\[1ex]
\begin{Proof}
The gtS of $\Delta^{({\sf N})}$, which holds true because of Lemma \ref{Lem:Delta^(N)gtS}, immediately implies gtS of $\sf N$.
And, applying \eq{Delta^F=+} one gets
\begin{align}\label{gtSofFeynProp}
\Delta^{\sf F\gamma}_p(x)=\sigma_p\,\Delta^{\sf F\gamma}_{\tilde p}(-x)
\end{align}
which, due to the propagator's translation invariance, just expresses gtS.
\end{Proof}
\end{Prn}

\subsection{Normalization and localization\label{2.2}}

In the two Sects., \ref{2.2.1} and \ref{sss:AnotherRing}, HAs modeling the structure underlying the normal and local products are constructed.
With respect to the 
initial bialgebra,
the obtained HAs and the introduced ring of scalars, resp.,
mean a weakening and a specialization of the mathematical structure.

\subsubsection{An appropriate weakening of structure\label{2.2.1}}

As already indicated by \eq{circ=N+?}, the composition algebra of free fields 
does not provide
the adequate 
algebraic structure for modeling local interaction.
Only the normal product has been identified to be suitable
so that, cf.~\eq{N=bigcircDelta^N},
one would prefer to make the latter defining the physically relevant algebra in the considered context of local pQFT.

In fact,
one can drop the `co' and {\em switch}\, the previously constructed
coalgebra into an algebra.
As $\Delta^{\sf N}$ is coassociative the normal product ${\sf N}$ is associative, and therefore,
the normal $\mathbb{C}$-compositions 
${\cal N}:=\{\bigcirc\circ\Delta^{\sf N}(\phi)\,|\,\phi\in{\cal F}\}$ 
constitute
a graded commutative algebra with unity, i.e.~$({\cal N},\mathbb{C},+,\,\sssf{N}\,,\eta)$.
Notice, graded cocommutativity has switched to graded commutativity.
Because of its shared gtS property, 
the right candidate for an additional coalgebra is the one with the shuffling coproduct, i.e.~$({\cal N},\mathbb{C},+,\Delta,\epsilon)$.
Together with the appropriate
antipode,
as the following result states, the structure of the normally ordered free field operators $\cal N$ turnes out to be a Hopf algebra (HA).

\begin{Thm}\label{Thm:N-HA}
$({\cal N},\,\sssf{N}\,,\eta,\Delta,\varepsilon)$ forms a graded [co]commutative bialgebra which moreover, with the antipode\, ${\rm S}:=(-1)^{|\cdot|}\!\!\sim$, i.e.
\begin{align}
{\rm S}\big(\phi^{{\sf N} J}\big)=(-1)^{n}\,\phi^{{\sf N}\widetilde J},
\quad \forall J\in\mathbb{N}^n_{\neq},\;\forall n\in\mathbb{N}
\end{align}
\big(where $\tilde J:=(J_n,\dots,J_1)$\big),
defines a HA.\\[1ex]
\begin{Proof}
The bialgebra part follows from
Lemma \ref{DeltaN,epsilon-homom}.
Notice that graded [co] commutativity is nothing else than gtS, thus realized
for $\sf N$ [and $\Delta^{\sf N}$].
It remains to prove the {\em antipode condition},
i.e.~${\sf N}\circ({\rm S}\otimes{\rm id})\circ\Delta
=\eta\circ\varepsilon={\sf N}\circ({\rm id}\otimes {\rm S})\circ\Delta$.
For the nontrivial case, i.e.~$n>0$,
where $\eta\circ\varepsilon=0$,
one can write the r.h.s. (and similarly the l.h.s.) as a sum over vanishing pairs,
\begin{align*}
{\sf N}&\circ({\rm id}\otimes {\rm S})\circ\Delta\big(\phi^{\bigcirc J}\big)\\
&=\sum_{\substack{J_\pm\in{\cal P}_2^0(J)\\|J_-|\text{ odd}}}\Big(
\sigma_J(J_\pm)\,\phi^{\bigcirc J_+}\,\sssf{N}\,(-1)\,\phi^{\bigcirc\widetilde J_-}+
\underbrace{\sigma_J(K_\pm)\,\phi^{\bigcirc K_+}\,\sssf{N}\,\phi^{\bigcirc\widetilde K_-}}
\Big)\\[-2ex]
&=\hspace{11em}\ldots+\underbrace{\sigma_J(K_\pm)\,\sigma_{J_\pm}(K_\pm)}_{=\,\sigma_J(J_\pm)}\,\phi^{\bigcirc J_+}\,\sssf{N}\,\phi^{\bigcirc\widetilde J_-}\Big)\\[-2ex]
&=0,
\end{align*}
where  $K_+:=J_+'\otimes(j)\otimes J_+''$ and $K_-:=J_-'\otimes J_-''$ provided $J_+=J_+'\otimes J_+''$ and $J_-=J_-'\otimes(j)\otimes J_-''$.
The first brace results from the gtS of $\sf N$ and the second from the
composition laws 
of the grading $\sigma$
(as a group representation).
\end{Proof}
\end{Thm}
Going over 
from the bialgebra of free field operators to the bialgebra of normally ordered field operators, i.e.~${\cal F}\to{\cal N}$,
\begin{align}
\phi^{\bigcirc J}\mapsto\phi^{{\sf N} J},\qquad 
\bigcirc\mapsto{\sf N}\qquad\text{and}\qquad\Delta^{\sf N}\mapsto\Delta
\end{align}
(as well as $\eta\mapsto\eta$),
referred to as {\em normalization}\, here, this defines a functor
that obviously weakens the algebraic structure,
{switching} from a non-(graded)[co] commutative to a graded [co]commutative bialgebra.

\subsubsection{Another ring of scalars \label{sss:AnotherRing}}

In order to model local interaction one has to describe 
different free particles localized at the same spacetime point. In the chosen mathematical context of distributions that can be realized by applying Dirac's delta distribution.
At the current stage of developing the (Hopf) algebraic formalism
the latter can be implemented by changing to another ring of scalars.

With tuples $J\in\mathbb{N}^n$ of $n\in\mathbb{N}$
numbers (representing particles)
one associates the following products of delta distributions,
\begin{align}
\delta^{\Pi J}\equiv \Pi\delta^{\otimes J}:=
\begin{cases}
\prod_{i<n}\delta(x_{J_{i+1}}-x_{J_i})&\text{if}\quad J\in\mathbb{N}^n_{\neq},\\
0&\text{otherwise}.
\end{cases}
\end{align}
Generated by these distributions and its derivatives,
one obtains an algebra of localizing 
of scalars, i.e.
\begin{align}
\mathbb{L}:=\mathbb{C}\oplus\mathbb{C}\big\langle\big\{\partial^\beta\delta^{K} \,\big|\,\beta\in\mathbb{N}^2, 
\;K\in\mathbb{N}^{2}_{\neq}
\big\},+,\cdot\big\rangle\big/\big(\delta^K\cdot\partial^\beta\delta^K,\;\delta^K-\delta^{\tilde K}\big),
\end{align}
having excluded multiple occurances of $\delta^{\Pi J}$,
even w.r.t.~permuations of its indices,
and added the unity.
The localized quantum field operators
$\phi^{{\sf L} J}:=\delta^{\Pi J}\phi^{{\sf N} J}$
are introduced 
as a $\mathbb{C}$-submodule,
\begin{align}
{\cal L}:=\bigoplus_{J\in\mathbb{N}^n_{\neq},\,n\ge0}\mathbb{C}\big\{\phi^{{\sf L} J}\,\big|\,
\phi^j\in{\cal F}_{\rm p},\,\forall j\in J\big\},
\end{align}
of $\mathbb{L}\otimes{\cal N}=:{\cal N}_{\mathbb{L}}$
equipped with the (associative and distributive) product
${\sf L}:\phi^{{\sf L} J}\otimes\phi^{{\sf L} K}\mapsto\phi^{{\sf L}(J\otimes K)}$ which 
varies from the one that is naturally
given on ${\cal N}_{\mathbb{L}}$,
i.e.~$\Pi\otimes{\sf N}\circ({\rm id}\otimes\tau\otimes{\rm id})=:{\sf N}$
(where $\tau:J\otimes K\mapsto K\otimes J$ denotes the flip operation).
The latter two 
spaces carry as well the over all announced stucture.

\begin{Thm}
$({\cal N}_\mathbb{L},\mathbb{L},+,\sssf{N},\eta,\Delta,\varepsilon,{\rm S})$
and its substructure\, $({\cal L},\mathbb{C},+,\sssf{L},\eta,\Delta,\varepsilon,{\rm S})$
form graded [co]commutative HAs over the ring $\mathbb{L}$ and, resp., the field $\mathbb{C}$.\\[1ex]
\begin{Proof}
Except of the 
modifications,
this is again
the content of
Theorem \ref{Thm:N-HA}.
\end{Proof}
\end{Thm}

\subsection{Interaction\label{2.3}}

This last subsection contains necessary supplements.
In Sect.~\ref{2.3.1}, which covers the more mathematical ones,
monomial couplings and the adiabatic limit are explained.
Sect.~\ref{sssect:PhysSuppl} deals with the more physical supplements,
introducing background fields and defining the interaction Hamiltonian.

\subsubsection{Monomial couplings and the adiabatic limit\label{2.3.1}}

The Hamiltonian modeling
local interaction will be formulated by the adiabatic limit of coupled localized field operators. In the considered physical approach the coupling ${\sf\Sigma}_c$
is 
given by a monomial in 
$\cal L$.
Let $M\in\mathbb{N}^m_{\neq}$ symbolize a tuple of $m\in\mathbb{N}$
particles.
Then
\begin{align}
{\sf \Sigma}_b\big(\phi^{{\sf L}M}\big)=\sum_{\overline\mu\in M}\sum_{|\alpha|\le\omega_b}b^{\alpha,\overline\mu}\big(\partial^{\alpha}\delta^{\Pi\overline\mu}\big)\phi^{{\sf N}\overline\mu}
\end{align}
formalizes a coupling monomial,
$b$ labels the 
associated coupling coefficients
$b^{\alpha,\overline\mu}\in\mathbb{C}$
where $\alpha$ serves as a suitable multi-index,
and $\omega_b\ge0$ denotes the order (of singularity).
Spelled out explicitely,
${\sf \Sigma}_{b}\phi^{{\sf L}M}\equiv{\sf \Sigma}_{b}\varphi^{{\sf L}M}(x)$
formally depends on $m$ spacetime variables $x\equiv(x_1,\dots,x_m)\in\mathbb{R}^{({\rm1+d})\times m}$, even though
only one of them is independent
(because of the localization which uses delta distributions).

The adiabatic limit replaces those $m$ variables by a single one,
i.e. 
\begin{align}
{\sf A}_{x_i}{\sf\Sigma}_b\phi^{{\sf L}M}&:=
\lim_{\tau\to1}\big\langle{\sf \Sigma}_{b}\varphi^{{\sf L}M}(x),\tau(x_1,\dots,x_{i-1},x_{i+1},\dots,x_{m})\big\rangle_{m-1}\\
&\;=\big\langle{\sf \Sigma}_{b}\varphi^{{\sf L}M}(\dots,x_i,\dots),1(x_1,\dots,x_{i-1},x_{i+1},\dots,x_{m})
\big\rangle_{m-1},
\end{align}
where, as well as in general, $\langle t(x),\tau(x)\rangle_{m}
:=\int_{\mathbb{R}^{({\rm1+d})\times m}}t(x)\,\tau(x)\,dx_1\cdots dx_m$
denotes the distributions' dual bracket
and $1(x)$ symbolizes an arbitrary test function in ${\cal D}_{{\rm p}_M}^{\otimes}:=\bigotimes_{i\le m}{\cal D}_{{\rm p}_{M_i}}$ with value `1' in a neighborhood of the origin $x=0\in\mathbb{R}^{({\rm1+d})\times m}$.
Of course, one may introduce another variable $x_\bullet\in\mathbb{R}^{\rm1+d}$.
Then
\begin{align}
{\sf A}_{x_\bullet}{\sf \Sigma}_{b}\phi^{{\sf L}M}
=\delta(x_\bullet)\ast{\sf A}_{x_1}{\sf \Sigma}_{b}\phi^{{\sf L}M}
=\big\langle\delta(x_\bullet-x_1)\,{\sf \Sigma}_{b}\varphi^{{\sf L}M}(x),1^{\Pi M}(x)\big\rangle_{m},
\end{align}
which yields the prototype of a so-called interaction Hamiltonian.

In the algebraic context of normalization and localization
the composition of monomial couplings and adiabatic limits is quite straightforward.
For any $i\le n$, let
$\overline M_i\in\mathbb{N}^{m_i}_{\neq}$
denote an $m_i\in\mathbb{N}$ tuple of particles
and $b_i$ the associated coupling with order $\overline\omega_{b_i}$. Furthermore, form tuples of those latter objects and let $y\in\mathbb{R}^{({\rm1+d})\times n}$.
Then
\begin{align}
\underset{i\le n}{\lsf{N}}\;
{\sf A}_{y_i}{\sf\Sigma}_{b_i}\phi^{{\sf L}\overline M_i}
=
{\sf A}^{\otimes\overline M}_{y}\,{\sf\Sigma}_{\Pi b}\,\phi^{{\sf N}\circ{\sf L}\overline M},
\end{align}
where ${\sf A}^{\otimes\overline M}_y={\sf A}^{\overline M_1}_{y_1}\cdots{\sf A}^{\overline M_n}_{y_n}$ specifies the application of the adiabatic limit
(in the obvious way) and
\begin{align}
{\sf\Sigma}_{\Pi b}\,\phi^{{\sf N}\circ{\sf L}\overline M}=
\sum_{\widehat\mu\in \overline M}\sum_{|\overline\alpha|\le\overline\omega_b}\prod_{i\le n}b_i^{\overline\alpha_i,\widehat\mu_i}\big(\partial^{\Pi\overline\alpha}\delta^{\Pi^2\widehat\mu}\big)
\phi^{{\sf N}^{\circ2}\widehat\mu}.
\end{align}

\subsubsection{Physical supplements \label{sssect:PhysSuppl}}

To make the coupling monomial 
a physical magnitude 
it might be supplemented by a further coupling constant, 
and to get a reasonable physical theory some of those prototype Hamiltonians
with vanishing order of singularity might be additively combined.

Moreover, the concrete physical model, fixed by the interaction Hamiltonian,
may contain classical background fields
which are given by scalar (e.g.~real or Grassmann) valued test functions $\tau_p\in{\cal D}_p$,
supposed to smear the operator valued distributions.
Those are 
non-quantized 
fields that,
in order to be equipped with the required indices, are as well
labeled by particles $p\in{\rm C}\equiv{\rm P}\setminus{\rm Q}$,
referred to as virtual ones.
The particles considered so far are renamed ${\rm Q}\subseteq{\rm P}$.
Notice that virtual particles $p\neq(\emptyset)\in{\rm C}$, i.e.~if not just representing a non-indexed real valued test function $\tau_{(\emptyset)}$,
may also have to follow a non-trivial statistics.

Let the disjoint union
${\cal I}={\cal C}\;\dot\cup\;{\cal Q}\subset{\cal P}(\mathbb{N})$
denote a system of subsets of basic indices,
i.e.~$\{c_j\}\in{\cal C}$ and ${\cal Q}_j\in{\cal Q}$
where $j\le m\in\mathbb{N}$,
supposed specify virtual particles
$p_j={\rm p}_{c_j}\in{\rm C}$,
that label the associated
couplings $a_{p_j}$, $b_{p_j}$,
as well as tuples of particles
$Q_j=({\rm p}_i)_{i\in{\cal Q}_j}\subset{\rm Q}$
and their coordinates,
$x_{c_j}$ 
and $x_{Q_j}=(x_i)_{i\in {\cal Q}_j}$.
Then the smeared abstract interaction Hamiltonian reads
\begin{align}
H_{\cal I}\big(\tau^{\oplus{\cal C}}\big)=\sum_{j\le m}a_{p_j}\,\Big\langle{\sf A}_{x_{c_j}}\big({\sf\Sigma}_{b_{p_j}}\,\varphi^{{\sf L}Q_j}(x_{Q_j})\big),\tau_{c_j}(x_{c_j})\Big\rangle_1.
\end{align}

\begin{Exa}\label{Exa:QED}
Quantum electrodynamics (QED) in an external electromagnetic field
is, using the conventional notions (e.g.,
$\overline\psi,\psi$ for the spinorial positron and electron,
$A^{\rm cl},A$ for the classical and quantized electromagnetic field,
all 
equipped with localized spacetime indices, and colons for the normal product), described 
by the interaction Hamiltonian,
$\frac{e}{i\hbar}:\overline\psi(x)\gamma^\mu\psi(x):\big(A_\mu(x)+A^{\rm cl}_\mu(x)\big)$.
Therefore, one has to identify $a=\frac{e}{i\hbar}$, $c=\gamma$, and
$\varphi\in\{\overline\psi,\psi,A\}$.
The two sorts of test functions are real and Grassmann valued, i.e.
$\tau\in\{\tau_{(\emptyset)},A^{\rm cl}\}$.
\end{Exa}

%% file: p-3

\section{Perturbative expansion\label{s:pExp}}

There is a HA underlying the 
expansion of the $S$-matrix.
Its antipode condition represents the property of pseudo-unitarity.
In order to achieve this result at the end of 
Sect.~\ref{3.2},
one has to construct algebras of operator-valued distributions
that model the Wick expansion.
The main physical ingredients
that ensure a well-defined analytical behavior
(relevant for the scalar parts of the n-point distributions)
are translation invariance 
and the spectral condition. 
Rather unusual for the EG approach,
in addition to vertex tuples, graphs
are chosen as indices for n-point distributions.
They are introduced in Sect.~\ref{3.1}.

\subsection{The concept of graphs\label{3.1}}

Graphs, vertices, and their constituents in the context of pQFT are 
defined in Sect.~\ref{3.1.1}.
Moreover, a correspondence between
configuration space variables and vertex tuples of graphs
is pointed out.
Algebraic properties of graphs and their symmetric versions are introduced
in Sect.~\ref{3.1.2}.

\subsubsection{Vertices and graphs\label{3.1.1}}

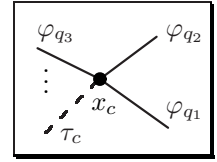
\begin{wrapfigure}{r}{11em}
\setlength{\shadowsize}{1pt}\setlength{\unitlength}{1ex}
\centering\shadowbox{
\begin{picture}(15,12)\thicklines
\put(6,6){\dashline{1}(-.2,-.2)(-4.7,-4.7)}\put(3,0.5){\makebox(1,1){$\tau_{c}$}}
\put(6,6){\line(7,-5){6}}\put(13,2.7){\makebox(1,1){$\varphi_{q_1}$}}
\put(6,6){\line(4,3){5}}\put(13,9.5){\makebox(1,1){$\varphi_{q_2}$}}
\put(6,6){\line(-10,5){5.5}}\put(1.7,9.5){\makebox(1,1){$\varphi_{q_3}$}}
\put(6,6){\circle*{1}}\put(5.9,3.1){\makebox(1,1){$x_c$}}
\put(0.8,6){\makebox(1,1){$\vdots$}}
\end{picture}}
\caption{\label{Fig:Vertex}A vertex.}
\end{wrapfigure}

The elements of the index set ${\cal I}$ in Sect.~\ref{sssect:PhysSuppl},
and correspondingly the
prototype interaction Hamiltonians,
will be interpreted as vertices, cf.~Fig.~\ref{Fig:Vertex}. 

Starting from an initial set U of vertices, the perturbative expansion 
will lead to objects that can be interpreted as graphs ${\rm G_U}={\cal G}({\rm U})$ formed by those vertices. Furthermore, renormalization will
allow to reinterprete graphs as vertices,
i.e.~$\Omega:{\rm G_U}\to{\rm W}$,
which produces new vertices in $\rm W\setminus U$. 
With all these vertices one will as well form graphs in $\rm G_W={\cal G}({\rm W})$
that can again, by forgetting about the inner graph structure,
be shrunken to vertices via $\Omega:{\rm G_W}\to{\rm W}$.
\begin{figure}[h]\newcommand{\ouset}[1]{\overset{}{\underset{}{#1}}}
\begin{align*}
^{\xymatrix{{\rm G}_{\rm W}\;\ar@(ur,ul)[rrr]_{\Omega}&&&\;
\ouset{\rm W}\;\ar@{->}[lll]^{\cal G}\ar@(r,r)[d]^{\txt{laws of\\physics}}
\\
{\rm G}_{\rm V}\;
\ar@(r,dl)[urrr]^{\Omega}&&&\;\ouset{\rm V}\;\ar@{->}[lll]^{\cal G}\ar@{_{(}->}[u]
\\
{\rm G}_{\rm U}\;
\ar@(r,d)[uurrr]_{\Omega}&&&\;\ouset{\rm U}\;\ar@{->}[lll]_{\cal G}\ar@{_{(}->}[u]
}}
\end{align*}
\vspace{-6ex}
\caption{
Generation of graphs and (of new) vertices.
\label{Fig:NewVertices}}
\vspace{2ex}
\end{figure}
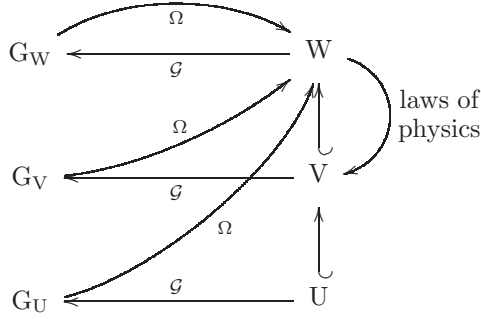
This rough description of the induction, cf.~Fig.~\ref{Fig:NewVertices},
has to be supplemented by the remark that physical requirements,
e.g.~symmetries,
restrict the number of physically relevant vertices V.

Formally,
the pairs $v\equiv(P,b)\in{\rm Q}^{n}\times\mathbb{C}^{|{\cal P}_{n}({\omega_b})|\times\Pi|P|}$
of particle tuples $P={\rm P}(v)$ and coupling coefficients $b={\rm b}(v)$,
where ${\cal P}_{n}({\omega_b})=\big\{\alpha\in\mathbb{N}^n\,\big|\,\sum_{i\le n}\alpha_i\equiv|\alpha|\le\omega_b\big\}$ and $n\equiv|P|$,
are supposed to
define vertices $v\in\{\rm U,V,W\}$.
Graphs are defined as
pairs $\Gamma\equiv(\overline v,l)$ of a vertex tuple $\overline v={\rm V}(\Gamma)$ and a $|\overline v|\times|\overline v|$ matrix
$l={\rm l}(\Gamma)$
of (sets of 
internal)
lines
$l_{rs}\subseteq\big\{(i,j)\in\mathbb{N}_{|\overline p|}\times\mathbb{N}_{|\overline q|}\,\big|\,\gamma_{\overline p_i\overline q_j}\neq0$, $\overline p\equiv\overline v_r$, $\overline q\equiv\overline v_s\big\}$,
forming a collection of index pairs that represent two matching particles,
where so-called tadpoles are excluded, i.e. $l_{rr}=\emptyset$.

Keeping the notations,
the iterated tuple ${\rm I}(\Gamma)\equiv({I}_r)_{r\le|\overline v|}$ of particles that support the internal lines of a graph $\Gamma$, referred to as representing the internal half-lines, can be reobtained by 
${I}_r=\big({\rm P}_i(\overline v_r)\big)_{i\in{\rm l}_r}$ where ${\rm l}_r=\{{\rm pr}_1\lambda\,|\,\lambda\in l_{rs}$, $s\le|\overline v|\}$.
Thus,
the tuple ${\rm E}(\Gamma)\equiv({E}_r)_{r\le|\overline v|}$ 
representing the external half-lines is given
by
${E}_r=\big({\rm P}_i(\overline v_r)\big)_{i\in\overline{\rm l}_r}$
where $\overline{\rm l}_r=\mathbb{N}_{|{\rm P}(\overline v_r)|}\setminus{\rm l}_r$.

By introducing graph related cardinalities Euler's 
formula is stated, i.e.
\begin{align}\label{eulerformula}
\begin{matrix}
|\Gamma|=\#\Gamma-|l|+|\overline v|,\\\\
\text{\hspace{6em}where}
\end{matrix}\hspace{1.2em}
\begin{cases}
|\Gamma|\\
\#\Gamma\\
|l|\\
|\overline v|
\end{cases}\hspace{-.3em}
\begin{matrix}\text{denotes the}\\
\text{number of\hspace{.45em}}
\end{matrix}\quad
\begin{cases}
\text{connected components},\\
\text{independent loops},\\
\text{(internal) lines},\\
\text{vertices}.
\end{cases}
\end{align}
The number of lines can be obtained by $|l|=\sum_{r<s\le|\overline v|}|l_{rs}|$.

The map ${\rm p}$ which identifies particles ${\rm p}_i$ via basic indices $i\in I\subset\mathbb{N}$,
and thus provides 
a correspondence with spacetime variables $x_i\in\mathbb{R}^{\rm1+d}$,
can consistently be defined to as well
identify vertex tuples 
and associated classes of graphs.
In this context one will 
have to consider a bigger system
${\cal J}:={\cal C}\;\dot\cup\;{\cal Q}\subseteq{\cal P}(\mathbb{N})$
of subsets of 
indices than ${\cal I}$ in Sect.~\ref{sssect:PhysSuppl},
i.e. $c_J:=\dot\bigcup_{j\in J}\{c_j\}$ and $Q_J:=\dot\bigcup_{j\in J}Q_j$
satisfying $c_J\cap c_K=\emptyset$ and $Q_J\cap Q_K=\emptyset$,
for $J,K\in{\cal P}(\mathbb{N})$.

\begin{Lem}
There are a system ${\cal J}$ and a surjective map\,
${\rm p}:\mathbb{N}\to{\rm P}$,
\begin{align*}
n\mapsto
\begin{cases}
{\rm p}^{\rm C}_n\in{\rm C}\\
{\rm p}^{\rm Q}_n\in{\rm Q}
\end{cases}
\text{if}\quad n\in
\begin{cases}
\mathbb{N}_{\rm C},\\
\mathbb{N}_{\rm Q},
\end{cases}
\text{where}\quad
\mathbb{N}_{\rm C}\cap\mathbb{N}_{\rm Q}=\emptyset,
\quad\text{such that}
\end{align*}
$(\forall\overline v\in\bigcup_{n\ge0}W^n)$
$(\exists J\in\mathbb{N}^n_{\neq})$
$(\forall j\in J)$
${\rm b}(\overline v_j)=b_{{\rm p}^{\rm C}({c_j})}$
$\wedge$
${\rm P}(\overline v_j)=\big({\rm p}^{\rm Q}(i)\big)_{i\in Q_j}$.\\[1ex]
\begin{Proof} 
Decompose the operation of forming graphs,
i.e.~${\cal G}=\bigcup_{n\le0}{\cal G}_n$,
so that ${\cal G}_{n}:{\cal P}({\rm W})\to{\cal P}({\rm G}_{\rm W}),\;{\rm W}_{k}\mapsto{\cal G}_n({{\rm W}_{k}})$ yields graphs $\Gamma$ with $n=|{\rm V}(\Gamma)|$ vertices, ${\rm V}(\Gamma)\in{\rm W}_{k}^{\times n}$.
Then the existence of ${\cal J}$ and p is guaranteed by the 
inductive construction of
${\rm W}=\bigcup_{k\ge1}{\rm W}_k$,
and thus $\rm G_W=\bigcup_{n\ge1}{\cal G}_n({\rm W}_n)$,
from finite sets ${\rm W}_{k+1}=\{\Omega(\Gamma)\in{\rm W}\,|\,\Gamma\in{\cal G}_k({\rm W}_k)\}$
starting with ${\rm W}_1={\rm U}$.
\end{Proof}
\end{Lem}
Notice that vertices and virtual particles 
immediately correspond via ${\rm p}^{\rm C}$.
In Sect.~\ref{s:cr-pQFT} one will 
have to indicate 
spacetime variables $x_i$ associated with
vertices $\overline v_j=\Gamma_j\sim{\rm p}^{\rm C}_i$ of tuples $\overline v$ and graphs $\Gamma\equiv(\overline v,{l})$. This will be denoted by
\begin{align}
x_\bullet:\overline v_j\mapsto x_{v_j}:=x_i\quad\text{and}\quad
\Gamma_j\mapsto x_{\Gamma_j}:=x_i,
\quad\text{resp}.
\end{align}

\subsubsection{Algebras of graphs\label{3.1.2}}

For the algebraic considerations, graphs are always assumed to form a
complex vector space
which respects the $\mathbb{N}$-grading corresponding to the number of vertices,
i.e. $\big({\rm G}_{\rm V},\mathbb{C},+\big)$
where
${\rm G}_{\rm V}=\bigoplus_{n\ge0}\mathbb{C}{\cal G}_n({\rm V})$.

Let $\Gamma^i\equiv(\overline v^i,l^i)$, 
$i\le2$, denote two graphs.
There are $|\overline v^1|\times|\overline v^2|$ matrices
$l^{12}\in\mathbb{M}_{|\overline v^1|\times|\overline v^2|}\big({\cal P}(\mathbb{N}^{\times2})\big)$
that yield a new graph from the previous ones, i.e.
\begin{align}
\Gamma^1\sqcup_{l^{12}}\Gamma^2:=
\Bigg(\overline v^1\otimes\overline v^2,
\begin{pmatrix}
l^1&l^{12}\\\tilde l^{12}& l^2
\end{pmatrix}
\Bigg), 
\end{align}
so that the required symmetry, 
$l_{rs}=\tilde l_{sr}:=\{({\rm pr}_2\lambda,{\rm pr}_1\lambda)\,|\,\lambda\in l_{sr}\}$,
is realized.
The admissible ones, i.e. $l^{12}\in{\rm l}(\Gamma^1,\Gamma^2)$,
are called concatenation matrices.
Cf. Example \ref{Exa:Graphs}.
In contrast to the matrices of lines,
$|l^{12}|=\sum_{\substack{r\le|\overline v^1|,s\le|\overline v^2|}}|l^{12}_{rs}|$.

This composition leads to the {Wick product} of graphs which is defined by
$\sqcup_\Sigma=\sum_{l^{12}\in{\rm l}(\cdot,\cdot)}\circ\,\sqcup_{l^{12}}$.
Moreover, by linear extension one [as well] considers gtS versions of graphs,
$\sqcap\Gamma:=\frac{1}{n!}\sum_{\pi\in{\cal S}_n}\sigma(\pi)\,\pi(\Gamma)\in
\bigoplus_{n\ge0}S_n^\sigma\,\mathbb{C}{\cal G}_n({\rm V})=:{\rm G}_{\rm V}^{\sigma}$,
where the permutation $\pi(\Gamma):=\big(\pi(\overline v),\pi(l)\big)$
acts on the vertex tuple and on the matrix of lines,
i.e.~$\pi(\overline v):=(\overline v_{\pi(r)})_{r\le n}$
and $\pi(l):=(l_{\pi(r)\pi(s)})_{r,s\le n}$,
and the grading is reduced to the graph's vertex tuple,
i.e.~$\sigma_\Gamma(\pi(\Gamma)):=\sigma_{{\rm V}(\Gamma)}({\rm V}\circ\pi(\Gamma))$,
discussed right below in Sect.~\ref{sss:ansatz}.

\begin{Lem}
$\big({\rm G}_{\rm V}^{[\sigma]},\mathbb{C},+,\sqcup_\Sigma,\emptyset\big)$ forms an algebra [in both cases]
where the empty graph serves as unity.\\[1ex]
\begin{Proof} Cf.~Sect.~3.2.2 in \cite{La03c}, pp.~75ff.
\end{Proof}
\end{Lem}
Associativity can be illustrated introducing the inverse concept of concatenation, i.e.~separation into subgraphs that, w.r.t.~the reference graph, are completely determined by their vertices.
Let 
$(\Gamma^1,\dots,\Gamma^k)\in{\cal P}_k^{(0)}(\Gamma)$
denote a $k$-partition
(resp., with possibly empty subgraphs)
of the graph $\Gamma\in{\rm G}_{\rm V}^{[\sigma]}$
iff there are a permutation\footnote{which depending on the context, e.g.~when showing associativity of $\sqcup_\Sigma$, might be fixed by $\pi={\rm id}$}
$\pi\in{\cal S}_k$
and
appropriate concatenation matrices, 
$l^1,\dots,l^{k-1}$,
such that
$\pi(\Gamma)=\Gamma^1\sqcup_{l^1}\big(\dots\sqcup_{l^{k-1}}\Gamma^k)...\big)$
[appropriately extended].

Let $\Gamma\in{\rm G}_{\rm V}^{[\sigma]}$.
Then ${\rm G}_\Gamma^{[\sigma]}:={\cal P}(\Gamma)
=\{\Gamma'\,|\,(\Gamma',...)\in{\cal P}_2^0(\Gamma)\}
\subset{\rm G}_{\rm V}^{[\sigma]}$ denotes the powerset
which contains all subgraphs of $\Gamma$.
Graphs in ${\rm G}_{\rm V}^{[\sigma]}$
can be composed as $\Gamma$-restricted graphs,
i.e.~via $\sqcup_\Gamma=(\sigma^0_\Gamma\circ\sqcup_\Sigma)\,\sqcup_\Sigma$,
where
\begin{align}\label{1_G}
\sigma^0_\Gamma:{\rm G}_{\rm V}^{[\sigma]}\to\{-1,0,1\},
\quad\Gamma'\mapsto\begin{cases}\sigma_{\Gamma}(\Gamma')&\text{\rm if}\quad \Gamma'\in{\rm G}_\Gamma^{[\sigma]},\\0&\text{\rm otherwise},\end{cases}
\end{align}
and $\sigma_{\Gamma}(\Gamma')=\sigma_{\Gamma'}(\Gamma)\equiv\sigma(\pi)$,
regarding $\pi:\Gamma'\mapsto\Gamma$ as a permutation in ${\cal S}_{|{\rm V}(\Gamma')|}$.

\begin{Prn}
$\big({\rm G}_{\rm V}^{[\sigma]},\mathbb{C},+,\sqcup_\Gamma,\emptyset\big)$ forms an algebra,
and $\big({\rm G}_\Gamma^{[\sigma]},\dots\big)$ identifies its largest zero divisor free subalgebra $\big({\rm G}_{\Gamma'}^\sigma,\dots\big)$, $\Gamma'\in{\rm G}_{\rm V}^{[\sigma]}$.\\[1ex]
\begin{Proof}
Associativity of $\sqcup_\Sigma$ implies associativity of $\sqcup_\Gamma$.
The other statements are obvious.
\end{Proof}
\end{Prn}

\subsection{The expansion of the $S$-matrix\label{3.2}}

In the EG approach,
the $S$-matrix
is assumed to be expanded
in terms of operator-valued gtS n-point distributions indexed by vertex tuples.
This ansatz and its 
property of being an infrared regularization
are content of Sect.~\ref{sss:ansatz}.
In contrast to the usual pQFT approach, EG have not mixed Wick and time ordering.
Therefore, in Sect~\ref{3.2.2}, one forms an algebra which models Wick ordering without an explicite implementation of causality.
Translation invariance (tI) of the n-point distributions' scalar parts
and the spectral condition for the propagators are
the properties that ensure the algebraic constructions underlying 
the Wick expansion
to be built on well-defined analytical grounds.
In contrast to the 
EG approach,
in Sect.~\ref{sss:GaVD},
one introduces
n-point distributions indexed by graphs.
The usual ones, indexed by vertex tuples, are reproduced from the latter. 
Algebras of both these distributions, 
each one equipped with the shuffling coproduct and an antipode which represents
the Schmitt formula,
form HAs that represent 
the perturbative expansion and
especially,
as it is shown in Sect.~\ref{3.2.4},
the property of pseudo-unitarity.

\subsubsection{Infrared regularization, the ansatz, and gtS
\label{sss:ansatz}}

In the considered approach the $S$-matrix is supposed to depend on
coupling constants $a_p$ and
normalized test functions
$\tau_p\in{\cal D}_p$ 
(i.e.~$\tau_{p}(0)=1$)
that, representing external classical fields $p\in {\rm C}$,
yield an infrared regularization.
To obtain concrete physical quantities the {adiabatic limit}
has to be taken,
i.e.
\begin{align}
S\big((a\tau)^{\oplus}_C\big)=\lim_{\epsilon\to0}S\big((a\tau_\epsilon)^{\oplus}_C\big),
\quad
\forall\tau:\tau\circ\epsilon\equiv\tau_\epsilon\overset{L^\infty(K)}{\longrightarrow}_{\epsilon\to0}1,\,\forall K\Subset\mathbb{R}^{\rm1+d}.
\end{align}
However, this 
task will not be regarded in the paper.
Notice,
as the vertices ${\rm V}$ correspond to the set ${\cal C}$ in Sect.~\ref{sssect:PhysSuppl},
which yields the virtual particles ${\rm C}=\{{\rm p}_i\,|\,i\in{\cal C}\}$,
they will be used as indices as well.

The expansion of the $S$-matrix is given
w.r.t.~a renormalization scheme ${\cal R}$
(only discussed in Sect.~\ref{ss:Ren})
and
a (somehow related) 
set ${\rm V}$ of vertices.
The ansatz,
chosen in accordance with the original ones 
of Bogoliubov and Shirkov (BS)
\cite{BS55p}, EG, and Scharf \cite{Sc95F},
reflects the Dyson series (cf.~Example \ref{Exa:Dyson}),
i.e.
\begin{align}\label{BSEGSansatz}
S^{\cal R}_{\rm V}\big((a\tau)^{\oplus{\rm V}}\big)
=\sum_{n\ge0}\frac{1}{n!} 
\sum_{\substack{\overline v\in{\rm V}^n}}a^{\Pi\overline v}\,
\Big\langle T^{\cal R}_{\overline v}(y),\tau^{\sqcap\overline v}(y)\Big\rangle_n,
\end{align}
where
$T_{\overline v}\in{C}\big(S^\sigma_{|\overline v|}{\cal D}^{\otimes}_{\overline v},{\cal H}^{\rm F}_{\rm P}\big)$
are operator-valued $n$-point distributions,
i.e.~strongly continuous mappings,
indexed by vertices.
The interaction Hamiltonian prototypes can be reidentified as
$H_{p}=i\hbar\, a_p T_{(p)}$.

The vertex distributions are assumed to be gtS,
i.e.~$T_{\overline v}=\sigma(\pi)T_{\pi(\overline v)}\circ\pi$,
$\forall\pi\in S_{n}$.
Here, $\sigma$ symbolizes the grading on vertex tuples,
given again by \eq{sigma(pi)=prod},
but
$\pi|_p:\overline v|_p\mapsto\pi(\overline v|_p)$
denotes a subpermutation of $\pi$
defined on
$\overline v|_p=(\overline v_k)_{k'\text{\rm odd}}$ where
$k'=\sum_{q\in{A'}(p)}1_{q=v_i\wedge v=\overline v_k}$
(cf.~\cite{La03c}, p.~45).
This is again a representation of ${\cal S}_n$ on $\mathbb{Z}_2$,
and one also writes $\sigma_{U}(V):=\sigma(U\mapsto V)$, for 
vertex tuples $U,V\in\mathbb{N}^n_{\neq}$.

\begin{Lem}
The two definitions of $\sigma$ are consistent,
i.e.~$\sigma_{U}(V)=\sigma_{[U]}\big([V]\big)$
where, here, $[\cdot]$ symbolizes the formal transfer from a vertex tuple into a particle tuple by simply omiting brackets and couplings.
\\[1ex]
\begin{Proof}
Cf.~Lemma 2.1.7 in \cite{La03c}, p.~45.
\end{Proof}
\end{Lem}

Of course, the test functions
associated with vertex tuples are gtS as well,
$\tau^{\sqcap\overline v}\equiv\tau^{\sqcap\overline c}:=S^\sigma_{n}\tau^{\otimes\overline c}\in S^\sigma_{n}{\cal D}^{\otimes}_{\overline c}
\equiv S^\sigma_{n}{\cal D}^{\otimes}_{\overline v}$,
where $\overline c\in{\rm C}^{n}_{\neq}$
denotes the corresponding tuple of virtual particles.
Again, the definitions are consistent.

\begin{Lem}
$\sigma_{\overline v}\big(\pi(\overline v)\big)=
\sigma_{\overline c}\big(\pi(\overline c)\big)$,
$\forall\pi\in{\cal S}_n$.\\[1ex]
\begin{Proof} Applying the permutation $\pi$ to both sides of the dual bracket
in \eq{BSEGSansatz},
this yields $\sigma_{\overline v}\big(\pi(\overline v)\big)\,
\sigma_{\overline c}\big(\pi(\overline c)\big)=1$.
Hence, the statement is due to the fact that $\sigma(\pi)\in\{\pm1\}$.
\end{Proof}
\end{Lem}

\subsubsection{Algebras of n-point distributions\label{3.2.2}}

Founded on Wick's theorem, referring to the expansion rule of normal products
w.r.t.~the composition product
(which can also be considered as twist or so-called Cliffordization),
one is going to 
construct an algebra of graph distributions 
over a carefully chosen ring of 
translation invariant (tI)
scalar
distributions.
The tI 
distributions
$t^J\in{\cal S}'(\mathbb{R}^{({\rm1+d})\times n})$,
$J\in\mathbb{N}^n_{\neq}$, i.e.
\begin{align}
t^J\equiv t_J(\tau)=t_J\big(\tau(\cdot-z^{\times n})\big),
\quad\forall\tau\in{\cal S}(\mathbb{R}^{({\rm1+d})\times n}),
\;\forall z\in\mathbb{R}^{\rm1+d},
\end{align}
are supposed to form an $\mathbb{N}$-graded algebra w.r.t.~the direct product, i.e.
\begin{align}
\mathbb{T}:=\mathbb{C}\oplus\mathbb{C}\big\langle\big\{t^{K} 
\,\big|\,K\in\mathbb{N}^n_{\neq},\,n\in\mathbb{N}
\big\},+,\cdot\,\big\rangle\big/\big(t^{(\dots,k,\dots)}\cdot t^{(\dots,k,\dots)}\big),
\end{align}
covering the (algebra 
of) localizing scalars $\mathbb{L}$ and
the 
positive frequency part propagators $\Delta^J_+\equiv\Delta^{+\gamma}_{{\rm p}_{J_1}}\big(x_{J_1}-x_{J_2}\big)$
with derivatives, i.e.
\begin{align}
\mathbb{P}:=\mathbb{C}\oplus\mathbb{C}\big\langle\big\{\partial^\beta\Delta_+^{K} \,\big|\,\beta\in\mathbb{N}^2, 
\;K\in\mathbb{N}^{2}_{\neq},\,\gamma_K\neq0
\big\},+,\cdot\,\big\rangle\big/\big(\Delta_+^{\widetilde K}\cdot\partial^\beta\Delta_+^{K}\big),
\end{align}
that, w.r.t.~the Fourier product and in accordance with the spectal condition
(cf.~end of Sect.~\ref{sssect:setup}), form an algebra 
as well.
Precisely, 
$\mathbb{L}$ and $\mathbb{P}$ are ideals in $\mathbb{T}$.
Combining the two previously defined algebras, 
one gets another commutative algebra, 
\begin{align}
\mathbb{T_P}:=\mathbb{T}\otimes\mathbb{P}\big/\big(t\otimes p-p\otimes t,\,
\cdots\hspace{-.5em}\underbrace{t^{(\dots,K_{1|2},\dots,K_{2|1},\dots)}}_{\neq t^{(...,K_{1|2},...)}t^{(...,K_{2|1},...)}
}
\otimes\,\cdots\Delta_+^{K}
\big)
\end{align}
which, on the level of distributions, encodes the Fourier and the direct product, and additonally the well-defined product of distributions with continous functions.
Including localization, the related commutative algebra 
$(\mathbb{T_{PL}},\mathbb{C},+,\cdot\,,1)$
is given by
\begin{align}
\mathbb{T_{PL}}:=\mathbb{T_P}\otimes\mathbb{L}\big/\big(t\otimes l-l\otimes t,\;
\cdots\underbrace{t^{(\dots,K_{1|2},\dots,K_{2|1},\dots)}}_{\text{\rm irreducible in $\mathbb{T_P}$}}\otimes\cdots\delta^K-\cdots\otimes\cdots\delta^K
\big).
\end{align}
Again, the substructures are ideals, i.e.~$\mathbb{T_{PL}}\subset\mathbb{T_{P}}\subset\mathbb{T}$.
Similar to Sect.~\ref{sss:AnotherRing}, those algebras will mainly
be applied as rings.

Turning to the operator side,
let ${\cal T}_{\rm V}:=\bigoplus_{n\ge0}\bigcup\,\{{\cal T}_\Gamma\,|\,\Gamma\in{\rm G}_{\rm V},\,|{\rm V}(\Gamma)|=n\}$
be the $\mathbb{N}$-graded space
of graph-indexed 
$n$-point distributions
\begin{align}
{\cal T}_\Gamma=\Big\{
\lsf{$\sf\Sigma$}_{{\rm b}(\Omega\Gamma)}\,
t_{{\rm I}(\Gamma)}(\bullet)\;
{\sf A}^{\otimes{\rm V}(\Gamma)}_{\bullet}\phi^{{\sf N}\circ{\sf L}{\rm E}(\Gamma)}\in{C}\big(S^\sigma_{n}{\cal D}^{\otimes}_{{\rm V}(\Gamma)},{\cal H}^{\rm F}_{\rm P}\big) \,\Big|\,t_{{\rm I}(\Gamma)}\in\mathbb{T}\Big\},
\end{align}
where,
in terms of the couplings at the vertices of ${\rm V}(\Gamma)$,
the coupling at the vertex $\Omega\Gamma$ reads
\begin{align}
{\rm b}(\Omega\Gamma)=\prod_{\substack{
r<s\le n}}
\,
\prod_{\substack{(i,j)\in{\rm l}_{rs}(\Gamma)}}
\,
\sum_{\substack{\mu\\\llbracket\mu\rrbracket\in{P}_{ri}}}
\,
\sum_{\substack{\nu\\\llbracket\nu\rrbracket\in{P}_{sj}}}
\;
\prod_{k\le n}
b_{{\rm V}_k(\Gamma)}^{\alpha_k,P_k}.
\end{align}
Then ${\cal T}_{\rm V}$ is obviously a submodule of $\mathbb{T_{PL}}\otimes{\cal N}$, 
and it is algebraically closed w.r.t.~the composition product.

\begin{Prn}
$\big({\cal T}_{\rm V},\mathbb{T_{PL}},+,\circ,\eta\big)$ 
forms an algebra.\\[1ex]
\begin{Proof}
This is the content of Prop.~3.3.2 in \cite{La03c}, p.~85,
which in the presented approach serves as the alternative to
EG's Theorem 0 in \cite{EG73t}, p.~229.
In order to illustrate the crucial points,
let $T^i\in{\cal T}_{\Gamma^i}$ with scalars $t^i\equiv t_{{\rm I}(\Gamma^i)}\in\mathbb{T}$,
for $\Gamma^i\in{\rm G}_{\rm V}$ and $i\le2$,
and use the abbreviation
$N(\bullet):={\sf A}^{\otimes{\rm V}(\bullet)}_{x^\otimes_{{\rm V}(\bullet)}}\phi^{{\sf N}\circ{\sf L}{\rm E}(\bullet)}$.
Then
\begin{align}\label{T1oT2=sumInTw}
T^{1}\circ\,T^{2}
=\sum_{l\in{\rm l}{(\Gamma^1,\Gamma^2)}}
(i\hbar)^{|l|}\,
\lsf{$\sf\Sigma$}_{{\rm b}(\Omega(\Gamma^1\sqcup_l\Gamma^2))}\,
t^{l}\,
N(\Gamma^1\sqcup_l\Gamma^2)\in{\cal T}_{\rm V},
\end{align}
having an well-defined associated scalar distribution which is again tI, i.e.
\begin{align}
t^{l}(y^1\otimes y^2)
=t^{1}(y^1)\,t^{2}(y^2)\,
\amalg^{l}_{(\Gamma^1,\Gamma^2)}(y^1,y^2)\in\mathbb{T_P}.
\end{align}
The most r.h.s.~expression abbreviates a certain product
of positive frequency propagators
(that, cf.~below, 
are also referred to as Wick contractions\footnote{
The symbol `$\amalg$' is chosen in accorance with BS' notation of a Wick contraction,
\begin{align*}
{\sf N}
(\ldots\otimes\;\underset{\xymatrix@1{\ar@{^|-^|}[rr]&&}}{\phi^i\;\otimes\;\cdots\;\otimes\;\phi^j}\;\otimes\dots)
=[\phi_{-}^{i},\phi_{+}^{j}]^\otimes_{\sigma}\;
{\sf N}(\ldots\otimes\dots\otimes\dots).
\end{align*}
}),
\begin{align}
\amalg^l_{(\Gamma^1,\Gamma^2)}(y^1,y^2)
:=\prod_{\substack{r\le|P^1|,\,s\le|P^2|}}\;
\prod_{\substack{(i,j)\in l_{rs}}}\Delta_{+\gamma}^{P^1_{ri}P^2_{sj}}(y^1_r-y^2_s)
\in\mathbb{P},
\end{align}
being well defined due to the spectral condition \eq{sc}.
Besides tI, for the analytical part,
the proof of the algebraic 
part rests upon
\end{Proof}
\end{Prn}

\begin{Lem} {\rm (Wick's theorem).} Keeping the notations, then 
\begin{align}\label{NoN=N}
N(\Gamma^1)\circ N(\Gamma^2)
=\sum_{l\in{\rm l}(\Gamma^1,\Gamma^2)}(i\hbar)^{|l|}\amalg^l_{(\Gamma^1,\Gamma^2)}
N(\Gamma^1\sqcup_l\Gamma^2).
\end{align}
\begin{Proof}
On the level of free
field operators
the expansion reads
\begin{align}\label{NN=N}
\phi^{{\sf N} I}\otimes\phi^{{\sf N} J}=
\sum_{\substack{\Lambda\subseteq\mathbb{N}_{|I|}\times\mathbb{N}_{|J|}}}
\bigg(\prod_{\lambda\in\Lambda}
\Big[\phi^{I_{\lambda_1}}_-,\phi^{J_{\lambda_2}}_+\Big]^\otimes_{\sigma}\bigg)
\;\phi^{{\sf N}(I\otimes_\Lambda J)},
\end{align}
provided
$[\cdot,\cdot]_\sigma^\otimes\propto{\rm id}_{{\cal H}^{\rm F}_{\rm P}}$,
where $I\otimes_\Lambda J=(I_i)_{i\in\mathbb{N}_{|I|}\setminus\{\lambda_1|\lambda\in\Lambda\}}\otimes(J_i)_{i\in\mathbb{N}_{|J|}\setminus\{\lambda_2|\lambda\in\Lambda\}}$,
for tuples $I,J$.
This is proven by successive insertion of
{Wick contractions}, 
\begin{align*}
\phi^{i}_{-}\otimes\phi^{j}_{+}=\big[\phi^{i}_{-},\phi^{j}_{+}\big]^\otimes_{\sigma}+\sigma_{(i,j)}\big((j,i)\big)\,\phi^{j}_{+}\otimes\phi^{i}_{-},
\end{align*}
moving the $\phi^{j}_+$'s  to the left and the $\phi^{i}_-$'s to the right. 
Due to \eq{+-commrel}, the bracket in \eq{NN=N} can be replaced by propagators.
In accordance with the definition of $\mathbb{T_{PL}}$, the localizing
delta distribution's 
are supposed to 
dominate (i.e.~erase) the propagators,
which reduces \eq{NoN=N} to \eq{NN=N}.
\end{Proof}
\end{Lem}

\begin{Rem}
The associativity of the algebra, i.e. $N(\Gamma^1)\circ\big(N(\Gamma^2)\circ N(\Gamma^3)\big)=\big(N(\Gamma^1)\circ N(\Gamma^2)\big)\circ N(\Gamma^3)$,
is equivalent to the 2-cocycle condition for $\amalg$,
i.e. $\amalg_{(\Gamma^1,\Gamma^2\sqcup_\Gamma\Gamma^3)}\amalg_{(\Gamma^2,\Gamma^3)}=\amalg_{(\Gamma^1,\Gamma^2)}\amalg_{(\Gamma^1\sqcup_\Gamma\Gamma^2,\Gamma^3)}$,
$\forall\Gamma^i\in{\rm G}_\Gamma$, $i\le3$,
which is obviously fulfilled.
In the context of \cite{RS94p,RS94v} and \cite{BF03l},
$\amalg_{(\cdot,\cdot)}$ is called a Laplace pair.
There, the relation \eq{NoN=N} is regarded as a so-called
Cliffordization or twist, resp.
\end{Rem}

\subsubsection{Graph and vertex distributions\label{sss:GaVD}}

In this subsection the physically relevant solution of the pertubation expansion
is assumed to be given --- formally, without having introduced the essential concepts yet.
This will only be done in Sect.~\ref{s:cr-pQFT}.

Graph distributions ${\cal R}^{\sf T}_{\rm G_V}:=\bigoplus_{n\ge0}\mathbb{C}\{{T}_\Gamma^{\cal R}\,|\,\Gamma\in{\rm G}_{\rm V},\,|{\rm V}(\Gamma)|=n\}$,
supposed to solve the perturbative expansion by respecting causality,
are introduced via a
vector space homomorphism,
\begin{align}\label{Tbullet:}
T_\bullet^{\cal R}:{\rm G}_{\rm V}\to{\cal T}_\Gamma,\quad
\Gamma\mapsto{T}_\Gamma^{\cal R}
:=(i\hbar)^{\#\Gamma-|\Gamma|}\,
\lsf{$\sf\Sigma$}_{{\rm b}(\Omega\Gamma)}\,
t^{\cal R}_{{\rm I}(\Gamma)}\,
N(\Gamma),
\end{align}
fixing the associated scalar valued
n-point distributions
$t^{\cal R}_{{\rm I}(\Gamma)}\in S^\sigma_{n}{\cal D}^{\otimes}_{{\rm V}(\Gamma)}\subset\mathbb{T}$
which are supposed to be obtained 
w.r.t.~a renormalization scheme ${\cal R}$,
and via a gtS requirement (which can only be motivated in Sect.~\ref{s:cr-pQFT}, cf.~Remark \ref{Rem:gtS}),
\begin{align}\label{TgtS}
{T}_{\pi(\Gamma)}^{\cal R}=\sigma(\pi)\,{T}_{\Gamma}^{\cal R},
\quad\forall\pi\in{\cal S}_{|{\rm V}(\Gamma)|}.
\end{align}

\begin{Rem}
Due to gtS, one may alternatively have introduced 
$T^{\cal R}_\bullet$ as a vector space isomorphism, ${\rm G}_{\rm V}^\sigma\to{\cal T}_\Gamma$.
\end{Rem}

\begin{Lem}\label{Lem:independence} 
Let ${\Gamma^1}, {\Gamma^2}$ be independent,
i.e. $\nexists c\in\mathbb{C}$ s.t.
$\sqcap\Gamma^1=c$ $\sqcap\Gamma^2$.
Then
$c_1T_{\Gamma^1}+c_2T_{\Gamma^2}=0$, $\forall c_1,c_2\in\mathbb{C}$,
implies that
$T_{\Gamma^1}^{\cal R}=T_{\Gamma^2}^{\cal R}=0$.\\[1ex]
\begin{Proof}
This is an immediate consequence of the previous remark.
\end{Proof}
\end{Lem}
Defining the homomorphism $T^{\cal R}_\bullet$ only for graphs
$\rm G_\Gamma\subset G_V$,
this leads to the space ${\cal R}^{\sf T}_{\rm G_\Gamma}$
of $\Gamma$-restricted graph distributions with a finite grading.
But, not only on this vector space,
the product 
is 
given by 
\begin{align}
\circ^{\sf T}_{\Gamma}:
T_{\Gamma^1}^{\cal R}\otimes T_{\Gamma^2}^{\cal R}\mapsto
T_{\Gamma^1\sqcup_\Gamma\Gamma^2}^{\cal R}.
\end{align}

\begin{Lem}
Let 
$\Gamma,\Gamma'\in{\rm G}_{\rm V}$ and $\ast\in\{{\rm V},\Gamma'\}$. Then
$\big({\cal R}^{\sf T}_{{\rm G}_\ast},\mathbb{C},+,\circ^{\sf T}_{\Gamma},\eta\big)$ forms an algebra
(which is free of zero divisors only
if $\ast\subseteq\Gamma$).\\[1ex]
\begin{Proof}
Obviously, ${\rm T}^{\cal R}_{\bullet}:{\rm G}_\ast\to{\cal R}^{\sf T}_{{\rm G}_\ast}$ defines an algebra homomorphism.
\big(Cf.~\eq{1_G}, if $\Gamma^1\not\subseteq\Gamma$ or $\Gamma^2\not\subseteq\Gamma$
then $T_{\Gamma^1}\circ^{\sf T}_\Gamma T_{\Gamma^2}\equiv T_{\Gamma^1\sqcup_\Gamma\Gamma^2}=T_0\equiv0$.\big)
\end{Proof}
\end{Lem}

Consider the following
restrictions of the composition product, obtained by redefining the r.h.s.~of \eq{T1oT2=sumInTw},
\begin{align}
T^1\circ T^2
&=:\sum_{l\in{\rm l}(\Gamma^1,\Gamma^2)}T^1\circ_l T^2
=:\sum_{\substack{\Upsilon\in{\rm G}_{\rm V},\,(\Gamma^1,\Gamma^2)\in{\cal P}_2^0(\Upsilon)}}T^1\circ_{\Upsilon} T^2,
\end{align}
where
$T^1\circ_l T^2=T^1\circ_{\Upsilon}T^2$
iff $\exists\pi\in{\cal S}_{|\Upsilon|}$ s.t.~$l=(\mathbb{I}\quad\emptyset)\cdot{\rm l}(\pi\circ\Upsilon)\cdot(\emptyset\quad\mathbb{I})^t$.
For any fixed
graph $\Gamma\in{\rm G}_{\rm V}$,
this induces
a {$\Gamma$-restricted product} on the tI distributions
${\cal T}_{\rm V}$,
i.e.
$\circ_\Gamma:{\cal T}_{\Gamma^1}\otimes{\cal T}_{\Gamma^2}\to{\cal T}_{\Gamma^1\sqcup_\Gamma\Gamma^2}$,
\begin{align}
T^1\circ_\Gamma T^2
:=\begin{cases}\sigma_{\Upsilon}(\Gamma)\;T^1\circ_{\Upsilon} T^2&\text{if}\quad
\exists!\,\Upsilon\in{\cal P}(\Gamma)
\text{ s.t. }
(\Gamma^1,\Gamma^2)\in{\cal P}_2^0(\Upsilon),\\
0&\text{otherwise}.
\end{cases}
\end{align}
${\cal R}_{{\rm G}_\Gamma}:=\bigoplus_{n\ge0}
\mathbb{C}\big\{\bigcirc_{\Gamma}\,\bigotimes_{i\le r}T^{\cal R}_{\Gamma^i}\,\big|\,\Gamma^i\in{\rm G}_\Gamma,\,\sum_{i\le r}|{\rm V}(\Gamma^i)|=n\big\}$
denotes
the corresponding space of distributions
defining a $\mathbb{C}$-submodule of ${\cal T}_{\rm V}$
and hence a $\mathbb{C}$-algebra, i.e.
$({\cal R}_{{\rm G}_\Gamma},\mathbb{C},+,\circ_{\Gamma},\eta)$.

\begin{Rem}
The consistency of the choice of the scalar $(i\hbar)^{\#\Gamma-|\Gamma|}$
in \eq{Tbullet:}
can easily be checked
with the help of Euler's formula \eq{eulerformula},
cf.~Corollary 3.3.8 in \cite{La03c}, p.~90.
\end{Rem}

In this context
the vertex distributions used in the ansatz, cf.~\eq{BSEGSansatz}, reappear
as a sum over graph distributions, i.e.
\begin{align}\label{sumT=T_v}
T^{\cal R}_{\overline v}=
\sum_{\substack{\Gamma\in{\rm G}_{\rm V},{\rm V}(\Gamma)=\overline v}}T^{\cal R}_\Gamma,
\quad
\forall\overline v\in{\rm V},
\end{align}
so that the space 
${\cal R}^{\sf T}_{\rm V}:=\bigoplus_{n\ge0}\mathbb{C}\{{T}_{\overline v}^{\cal R}\,|\,\overline v\in{\rm V},\,|\overline v|=n\}$
of vertex distributions
can be introduced restricting the 
homomorphism
$T^{\cal R}_\bullet$
to be defined on
the vector space
of vertex tuples, i.e.~$\bigoplus_{n\ge0}\mathbb{C}\big\{\overline v\,\big|\,|\overline v|=n\big\}\equiv{\rm V}
\to{\cal T}_{\rm V}$,
$\overline v\mapsto T^{\cal R}_{\overline v}$.

\begin{Lem}
Let $\circ^{\sf T}:T_{\overline v^1}^{\cal R}\otimes T_{\overline v^2}^{\cal R}\mapsto T_{\overline v^1\otimes\overline v^2}^{\cal R}$. Then
$\big({\cal R}^{\sf T}_{\rm V},\mathbb{C},+,\circ^{\sf T},\eta\big)$
forms an algebra.\\[1ex]
\begin{Proof}
Obviously, $T^{\cal R}_{\bullet}:{\rm V}\to{\cal T}_{\rm V}$
defines an algebra homomorphism.
\end{Proof}
\end{Lem}
Furthermore,
${\cal R}_{{\rm V}}:=\bigoplus_{n\ge0}
\mathbb{C}\big\{\bigcirc_{i\le r}T^{\cal R}_{\overline v^i}\,\big|\,\overline v^i\in{\rm V},\,\sum_{i\le r}|\overline v^i|=n\big\}$
denotes
the 
space of vertex distributions
which defines a $\mathbb{C}$-submodule of ${\cal T}_{\rm V}$
and hence a $\mathbb{C}$-algebra, i.e.
$({\cal R}_{\rm V},\mathbb{C},+,\circ,\eta)$.

\subsubsection{HAs underlying pseudo-unitarity\label{3.2.4}}

The perturbative expansion is now going to be modeled by a HA.
In accordance with the power series ansatz \eq{BSEGSansatz}, let
\begin{align}
\sum_{n\ge0}\frac{1}{n!}\sum_{\overline v\in {\rm V}^{n}}
a^{\Pi\overline v}\,
\big\langle
\widetilde{T}_{\overline v}^{\cal R}(y),\tau^{\sqcap\overline v}(y)
\big\rangle_n:=S^{\cal R}_{\rm V}\big((a\tau)^{\oplus{\rm V}}\big)^{-1},
\end{align} 
where the {inverse} n-point distributions
$\widetilde T_{\overline v}\in{C}\big(S^\sigma_{|\overline v|}{\cal D}^{\otimes}_{\overline v},{\cal H}^{\rm F}_{\rm P}\big)$,
supposed to satisfy gtS as well,
are denoted by a tilde.
Formally,
the inverse of the $S$-matrix is characterized by
{relation of pseudo-unitarity}, i.e.
\begin{align}
S^{\cal R}_{\rm V}(\xi)^{-1}\circ S^{\cal R}_{\rm V}(\xi)={\rm id}_{{\cal H}^{\rm F}_{\rm P}}=S^{\cal R}_{\rm V}(\xi)\circ S^{\cal R}_{\rm V}(\xi)^{-1},
\quad
\forall\xi\in{\cal D}^\oplus_{\rm V}\big(\mathbb{R}^{1+{\rm d}}\big).
\end{align}

\begin{Rem}
The probabilistic interpretation of quantum theory is ensured by requiring
{unitarity}.
But, for gauge theories with their unphysical states,
unitarity can only be {pseudo} (i.e.~true on a subspace of
of the Fock space
${\cal H}^{\rm F}_{\rm P}$), i.e.
$S(\xi)^{-1}=S(\xi)^{K}$,
where $K$ is a pseudo-adjoint (i.e.~adjoint for a subdomain).
That has been investigated extensively by Scharf and collaborators in the nineties,
cf.~\cite{DH95C,Kr95a,Gr00q}.
\end{Rem}
In terms of n-point
distributions 
the relation of pseudo-unitarity reads as follows.

\begin{Lem}
Let $T_Y,T_Z\in{\cal R}_{\rm V}$, where $Y,Z\in{\rm V}$,
and let $\emptyset\neq X\in{\rm V}$. Then
\begin{align}\label{EGSpreAC}
\sum_{(Y,Z)\in{\cal P}_2^0(X)}\sigma_X(Y\otimes Z)\;\widetilde T_Y\circ T_Z
=0=
\sum_{(Y,Z)\in{\cal P}_2^0(X)}\sigma_X(Y\otimes Z)\;T_Y\circ\widetilde T_Z.
\end{align}
Furthermore, 
$\widetilde T_Y,\widetilde T_Z\in{\cal R}_{{\rm V}}$ and therefore
the inverse distributions are
gtS.\\[1ex]
\begin{Proof}
To obtain \eq{EGSpreAC},
perform the
power 
expansion
(cf.~Lemma 3.3.15 in \cite{La03c}, p.~94).
The other claim is a side result.
\end{Proof}
\end{Lem}
This statement can be rewritten
assuming the tilde to be linear,
i.e.~assigning the tilde to graph distributions
in the following straightforward way,
\begin{align}\label{sumTildeT=TildeT_v}
\sum_{\Gamma\in{\rm G}_{\rm V},{\rm V}(\Gamma)=\overline v}\widetilde T_\Gamma^{\cal R}:=\widetilde T_{\overline v}^{\cal R}.
\end{align}

\begin{Cor}
Let $\emptyset\neq\Gamma'\in{\rm G}_\Gamma$ and $\Gamma\in{\rm G}_{\rm V}$. Then
\begin{align}
\sum_{(\Gamma^1,\Gamma^2)\in{\cal P}_2^0(\Upsilon)}\hspace{-1em}
\sigma_\Gamma(\Gamma^1\sqcup_\Gamma\Gamma^2)~\widetilde T_{\Gamma^1}^{\cal R}\circ_\Gamma T_{\Gamma^2}^{\cal R}
=0=\sum_{(\Gamma^1,\Gamma^2)\in{\cal P}_2^0(\Upsilon)}\hspace{-1em}
\sigma_\Gamma(\Gamma^1\sqcup_\Gamma\Gamma^2)~T_{\Gamma^1}^{\cal R}\circ_\Gamma\widetilde T_{\Gamma^2}^{\cal R},
\end{align}
\begin{Proof}
Insert the expansions \eq{sumT=T_v} and \eq{sumTildeT=TildeT_v} into \eq{EGSpreAC}
and apply 
Lemma \ref{Lem:independence}
(cf.~Corollary 3.3.17 in \cite{La03c}, p.~97).
\end{Proof}
\end{Cor}
The previous two equations,
if interpreted as antipode conditions,
motivate to consider
${\cal R}_{\rm V}$ and ${\cal R}_{{\rm G}_\Gamma}$, resp.,
as HAs. 

\begin{Thm}
$({\cal R}_\ast,\mathbb{C},+,\circ_{(\Gamma)},\eta,\Delta,\varepsilon,{\rm S}_{(\circ_\Gamma)})$, 
where $\ast={\rm V}$ (resp., $={\rm G}_\Gamma$, for any $\Gamma\in{\rm G}_{\rm V}$),
forms a HA.
The tilde serves as antipode, i.e.
${\rm S}_{(\circ_\Gamma)}T:=\widetilde T$,
which, e.g.~in ${\cal R}_{{\rm G}_\Gamma}$, is explicitely given by 
\begin{align}\label{S=sum}
{\rm S}_{\circ_\Gamma}T_{\Gamma'}^{\cal R}=\sum_{r=1}^{|{\rm V}(\Gamma')|}(-1)^r
\sum_{(\Gamma^1,\dots,\Gamma^r)\in{\cal P}_r(\Gamma')}
\sigma_\Gamma(\Gamma^1\sqcup_\Gamma\cdots\sqcup_\Gamma\Gamma^r)~T_{\Gamma^1}^{\cal R}\circ_\Gamma\dots\circ_\Gamma T_{\Gamma^r}^{\cal R}.
\end{align}
\begin{Proof}
It only remains to show the representation of the antipode. This will done by induction over the number of vertices.
Assume \eq{S=sum} is fulfilled for $|{\rm V}(\Gamma')|=n$,
which is true for $n=1$. Then, starting with the antipode condition, 
one reproduces \eq{S=sum},
\begin{align*}
{\rm S}_{\circ_\Gamma}T_{\Upsilon}^{\cal R}
&=-T_{\Upsilon}^{\cal R}-\sum_{(\Gamma',\Gamma'')\in{\cal P}_2(\Upsilon)}
\sigma_\Gamma(\Gamma'\sqcup_\Gamma\Gamma'')~{\rm S}_{\circ_\Gamma}T_{\Gamma'}^{\cal R}\circ_\Gamma T_{\Gamma''}^{\cal R}
\\
&=-T_{\Upsilon}^{\cal R}
-\sum_{r=1}^{n}(-1)^r
\sum_{(\Gamma',\Gamma'')\in{\cal P}_2(\Upsilon)}
\sum_{(\Gamma^1,\dots,\Gamma^r)\in{\cal P}_r(\Gamma')}\\ 
&\hspace{6em}
\sigma_\Gamma(\Gamma'\sqcup_\Gamma\Gamma'')\,
\sigma_\Gamma(\Gamma^1\sqcup_\Gamma\cdots\sqcup_\Gamma\Gamma^r)~T_{\Gamma^1}^{\cal R}\circ_\Gamma\dots\circ_\Gamma T_{\Gamma^r}^{\cal R}\circ_\Gamma T_{\Gamma''}^{\cal R}\\
&=-\sum_{r=0}^{n}(-1)^r
\sum_{(\Gamma^1,\dots,\Gamma^r,\Gamma'')\in{\cal P}_{r+1}(\Upsilon)}\sigma_\Gamma(\Gamma^1\sqcup_\Gamma\cdots\sqcup_\Gamma\Gamma^r\sqcup_\Gamma\Gamma'')\dots\\
&=\sum_{r=1}^{n+1}(-1)^r\sum_{(\Gamma^1,\dots,\Gamma^r,\Gamma'')\in{\cal P}_{r}(\Upsilon)}\dots,
\end{align*}
for the graph $\Upsilon$ with $|{\rm V}(\Upsilon)|=n+1$
and hence for any such graph.
\end{Proof}
\end{Thm}

\begin{Rem}
Rota and Stein \cite{RS94p,RS94v} have obtained this kind of antipode for their {\em plethystic HAs}
and refer to the general version of \eq{S=sum} as {\em Schmitt formula}.
\end{Rem}

%% file: p-4

\section{Renormalized causal pQFT \label{s:cr-pQFT}}

The main concepts that determine the $S$-matrix of a considered pQFT,
causality and renormalization,
will enter the
(algebraic) formalism
in this section.
The presentation
is given in terms of graph distributions,
in contrast to vertex distributions used in the original papers.
In Sect.~\ref{s:caus}, which is
devoted to causality,
the required
regularization of
distributions
is always performed as a localization
on an appropriate causally determined spacetime region in configuration space,
excluding so-called diagonals.
In Sect.~\ref{ss:Ren} renormalization is treated.
As the EG approach is a recursive method, it is free of subdivergences.
A change of the renormalization scheme, however, requires all the structure that is contained in Kreimer's HA of renormalization.

\subsection{Causal regularization \label{s:caus}}

The basic notions of (non) (weak) causally determined spacetime regions
(resp.) are introduced
in  Sect.~\ref{sss:caus}.
Furthermore, causality of the $S$-matix is characterized in terms of
n-point distributions.
In Sect.~\ref{sss:2EG}, the two recent EG versions
by Fredenhagen und Scharf
(which vary with respect to the concrete
recursive implementation of causality)
are presented in the framework of causal regularization,
and in Sect.~\ref{sss:RegHAs},
they are reformulated in terms of HAs.
Sect.~\ref{sss:BS-Feynman} deals with the conventional, the Feynman approach to pQFT which, because of mixing Wick with time ordering,
yields an explicite representation of graph distributions.
However, the required coarse localization
motivates the occurrence of subdivergences.

\subsubsection{Causality\label{sss:caus}}

On sets of spacetime points, two versions of causality will be introduced.
Let $X,Y\subseteq\mathbb{R}^{\rm1+d}$.
The relation of {causal ordering}
$\scriptstyle\lesssim$ is given by
\begin{align}
X\cap(Y+\overline{V}_+)=\emptyset\Leftrightarrow:X\lesssim Y\Leftrightarrow Y\gtrsim X,
\end{align} 
and the relation of {weak causal ordering}
$\scriptstyle\lesssim^\exists$ (i.e., $\supseteq$ $\scriptstyle\lesssim$)
is given by
\begin{align}
\exists(x,y)\in X\times Y~\text{s.t.}~\{x\}\lesssim\{y\}\Leftrightarrow:X\lesssim^\exists Y\Leftrightarrow Y\gtrsim^\exists X.
\end{align}
These two relations are {not reflexive},
i.e.~$\exists X\not\lesssim^{(\exists)} X$.
Therefore,
coinciding spacetime points cannot be compared with respect to causality.
This fact, however, makes renormalization look acceptable from a physics point of view. 

Accordingly, one distinguishes two sorts of configuration space regions.
For the one sort, the {causal cones}, the spacetime points are causally related, whereas for the other, the {diagonals}, which somehow model coinciding spacetime points, they are not.
Let $P\in{\cal P}_r(\Gamma)$, where $\Gamma\in{\rm G}_{\rm V}$ and $r\le n\equiv|{\rm V}(\Gamma)|$.
Then
\begin{align}
{\rm cone}_P^{(\exists)}:=\big\{y\in\mathbb{R}^{({\rm1+d})\times n}\,\big|\,
\{y^\otimes_{P_1}\}\gtrsim^{(\exists)}\dots\gtrsim^{(\exists)}\{y^\otimes_{P_r}\}\big\}
\end{align}
define (two) tI open cones on the configuration space
$\mathbb{R}^{({\rm1+d})\times n}:=\times_{i\le n}\mathbb{R}^{\rm1+d}$.
Furthermore, the sets
of $r$ coinciding 
points,
\begin{align}
{\rm diag}_P:=\big\{y\in\mathbb{R}^{({\rm1+d})\times n}\,\big|\,
y_i=y_j,\,\forall i,j\in P_s,\,\forall s\le r
\big\},
\end{align}
are called the {(sub)diagonals} w.r.t.~$P$ (if $P\neq(\Gamma)$, resp.).
Subtracting 
\begin{align}
{\rm diag}_r^n
:=\bigcup_{s<r}\bigcup_{P\in{\cal P}_s(I)}{\rm diag}_P
\end{align}
from the configuration space,
one obtains 
the {causally determined region} which contains at
least $r$ {non-coinciding
spacetime points}, i.e.
\begin{align}
\mathbb{R}^{({\rm1+d})\times n}_{\,r\sss\neq}
:=\mathbb{R}^{({\rm1+d})\times n}\,\setminus{\rm diag}_r^n.
\end{align}
One uses the abbreviations, 
$\mathbb{R}^{({\rm1+d})\times n}_{\,\sss\times}:=\mathbb{R}^{({\rm1+d})\times n}_{\,2\sss\neq}$
and $\mathbb{R}^{({\rm1+d})\times n}_{\,\sss\neq}:=\mathbb{R}^{({\rm1+d})\times n}_{\,n\sss\neq}$.

\begin{Lem}\label{Lem:Cones}
Let $\Gamma\in{\rm G}_{\rm V}$ and $r\le n\equiv|{\rm V}(\Gamma)|$. Then
\begin{align}\label{cone0}
\bigcup_{P\in{\cal P}_r(\Gamma)}{\rm cone}_P
=\mathbb{R}^{({\rm1+d})\times n}_{r\neq}.
\end{align}
Let $\Gamma,\Gamma_1,\Gamma''\in{\rm G}_{\rm V}$
where $(\Gamma_1,\Gamma'')\in{\cal P}_2(\Gamma)$ and $|{\rm V}(\Gamma_1)|=1$. Then
\begin{align}\label{cone1}
{\rm cone_{(\Gamma_1,\Gamma'')}^{\exists}}
\,\cup\,
{\rm cone_{(\Gamma'',\Gamma_1)}^{\exists}}
=
\mathbb{R}^{({\rm1+d})\times n}_\times
\end{align}
where the two cones are represented by
\begin{align}\label{cone2}
{\rm cone}_{\sss\bigg\{
\begin{matrix}
\scriptstyle(\Gamma_1,\Gamma'')\\
\scriptstyle(\Gamma'',\Gamma_1)
\end{matrix}}^{\exists}
=\bigcup_{(\Gamma_Q,\Gamma_P)\in{\cal P}_2(\Gamma),
\,\Gamma_1\subseteq\sss\bigg\{
\begin{matrix}
\scriptstyle\Gamma_Q\\
\scriptstyle\Gamma_P
\end{matrix}}
{\rm cone}
_{(\Gamma_Q,\Gamma_P)}.
\end{align}
\begin{Proof}
These relations follow immediately from the definitions above.
\end{Proof}
\end{Lem}

In the EG approach, the causality assumption of the $S$-matrix is crucial.
Causality holds true, i.e.~the $S$-matrix is causal, iff
\big($\forall\xi^i\in{\cal D}^\oplus_{\rm V}(\mathbb{R}^{\rm1+d})$, $i\le2$\big)
\begin{align}
{\rm supp}\,\xi^1\gtrsim{\rm supp}\,\xi^2
\Rightarrow
S_{\rm V}^{\cal R}(\xi^1+\xi^2)=S_{\rm V}^{\cal R}(\xi^1)\circ S_{\rm V}^{\cal R}(\xi^2).
\end{align}
There is an equivalent definition of causality in terms of graph distributions.
\begin{Prn} The $S$-matrix is causal iff
${\rm supp}\,\xi^1\gtrsim{\rm supp}\,\xi^2$ implies
\begin{align}
T_{\Gamma'}^{\cal R}(\xi^1\sqcap\xi^2)=\sigma_{\Gamma'}(\Gamma^1\sqcup_\Gamma\Gamma^2)~T_{\Gamma^1}^{\cal R}(\xi^1)\circ_\Gamma T_{\Gamma^2}^{\cal R}(\xi^2),
\end{align}
$\forall(\Gamma^1,\Gamma^2)\in{\cal P}_2(\Gamma')$, $\forall\Gamma'\subseteq\Gamma\in{\rm G}_{\rm V}$,
and 
$\forall\xi^i\in{\cal D}^\otimes_{{\rm V}(\Gamma^i)}$,
$i\le2$.\\[1ex]
\begin{Proof}
This is due to
power expansion
(cf.~Lemma 4.1.4 in \cite{La03c}, p.~103).
\end{Proof}
\end{Prn}
From now on, one assumes that the $S$-matrix is causal.

\subsubsection{Regularization of two EG versions\label{sss:2EG}}

EG's method can be understood as an induction over subgraphs
where the implementation of causality 
leads to localized
n-point
distributions (which are gtS and have tI scalar parts)
of the form
$T_{\Gamma'}^{r\sss\neq}
:=T_{\Gamma'}^{\cal R}\big|_{\mathbb{R}^{({\rm1+d})\times n}_{\,r\neq}}\in{\cal R}^{\sf T}_{{\rm G}_\Gamma}\big(\mathbb{R}^{({\rm1+d})\times n}_{\,r\sss\neq}\big)$,
$r\le n$.
For this subsection,
which introduces the EG versions by Fredenhagen und Scharf,
let $\Gamma\in{\rm G}_{\rm V}$ and $\Gamma'\in{\cal P}(\Gamma)$
where $n\equiv|{\rm V}(\Gamma')|$.

\paragraph{EG \`a la Fredenhagen (EGF)}
For each graph $\Gamma$ and 
each $r\le n$ 
one chooses a partition of unity 
$(\chi_P)_{P\in{\cal P}_r(\Gamma)}$
of class $C^\infty$ subordinate to the {finite open covering}
$({\rm cone}_P)_{P\in{\cal P}_r(\Gamma)}$
of the {metric space}
$\mathbb{R}^{({\rm1+d})\times n}_{\,r\sss\neq}$,
i.e.
\begin{align}
\sum_{P\in{\cal P}_r(\Gamma)}
&\chi_P(x)=1,
\qquad\forall x\in\mathbb{R}^{({\rm1+d})\times n}_{\,r\sss\neq},\\
\text{where}\qquad
&0\le\chi_P\in C^\infty
\big(\mathbb{R}^{({\rm1+d})\times n}_{\,r\sss\neq}\big)\\
\text{and}\qquad&\text{\rm supp }\chi_P
\subset {\rm cone}_P.
\end{align}
It is well-known that partitions of {this kind} exist (e.g., cf.~\cite{CD82A}, p.~214).

\begin{Prn}\label{Prn:EGF}
The graph distributions are (uniquely) determined by
\begin{align}\label{Tx-EGF}
T_{\Gamma'}^\chi:=\sum_{(\Gamma^1,\dots,\Gamma^r)\in{\cal P}_r(\Gamma')}
\hspace{-1em}
\chi_{(\Gamma^1,\dots,\Gamma^r)}~\sigma_{\Gamma}(\Gamma^1\sqcup_\Gamma\dots\sqcup_\Gamma\Gamma^r)~T_{\Gamma^1}^{\cal R}\circ_\Gamma\dots\circ_\Gamma T_{\Gamma^r}^{\cal R}
\end{align}
(independently of $\chi$),
i.e.~$T_{\Gamma'}^{r\sss\neq}=T_{\Gamma'}^\chi\big|_{\mathbb{R}^{({\rm1+d})\times n}_{\,r\neq}}\in{\cal R}^{\sf T}_{{\rm G}_\Gamma}\big(\mathbb{R}^{({\rm1+d})\times n}_{\,r\sss\neq}\big)$.
\\[1ex]
\begin{Proof} By applying \eq{cone0} one can straightforwardly generalize \cite{BF99m},
where only the case of $r=2$ is considered
(cf.~Prop.~4.2.6 in \cite{La03c}, p.~110).
\end{Proof}
\end{Prn}

\paragraph{EG \`a la Scharf (EGS)}
The genuine approach is based on two expressions,
\begin{align}
\text{the}\quad
\begin{cases}\text{retarded}\\\text{advanced}\end{cases} 
\text{distribution}
\quad
\begin{cases}
R_{(\Gamma_1,\Gamma'')}:=T_{\Gamma'}+R'_{(\Gamma_1,\Gamma'')},\\
A_{(\Gamma_1,\Gamma'')}:=T_{\Gamma'}+A'_{(\Gamma_1,\Gamma'')},
\end{cases}
\end{align}
where, applying Scharf's notation,
\begin{align}
R'_{(\Gamma_1,\Gamma'')}
&:=\sum_{\substack{
(\Gamma^1,\Gamma^2)\in{\cal P}_2(\Gamma'),
\,\Gamma_1\subseteq\Gamma^1}}
\sigma_{\Gamma}(\Gamma^1\sqcup\Gamma^2)\;
T_{\Gamma^1}^{\cal R}\circ_\Gamma\widetilde T_{\Gamma^2}^{\cal R},\\
A'_{(\Gamma_1,\Gamma'')}
&:=\sum_{\substack{(\Gamma^1,\Gamma^2)\in{\cal P}_2(\Gamma'),
\,\Gamma_1\subseteq\Gamma^2}}
\sigma_{\Gamma}(\Gamma^1\sqcup\Gamma^2)\;
\widetilde T_{\Gamma^1}^{\cal R}\circ_\Gamma T_{\Gamma^2}^{\cal R}.
\end{align}
These expressions vanish when restricted to appropriate cones.

\begin{Lem}\label{Lem:EGS}
Let $\Gamma,\Gamma_1,\Gamma''\in{\rm G}_{\rm V}$
where $(\Gamma_1,\Gamma'')\in{\cal P}_2(\Gamma)$
and $|{\rm V}(\Gamma_1)|=1$.
Then 
\begin{align}
A_{(\Gamma_1,\Gamma'')}
\big|_{\rm cone_{(\Gamma'',\Gamma_1)}^{\exists}}
=0=
R_{(\Gamma_1,\Gamma'')}
\big|_{\rm cone_{(\Gamma_1,\Gamma'')}^{\exists}}.
\end{align}
\begin{Proof}
One will have to apply the property \eq{cone2} of the cones of weak causality,
cf.~Lemma 4.2.12 in \cite{La03c}, pp.~113 f.
\end{Proof}
\end{Lem}
Again, but formally varying from the named approach, one considers a partition of unity.
For each triple of graphs 
$\Gamma,\Gamma_1,\Gamma''\in{\rm G}_{\rm V}$
where
$(\Gamma_1,\Gamma'')\in{\cal P}_2(\Gamma)$
and $|{\rm V}(\Gamma_1)|=1$ one chooses 
a partition of unity
$(\chi^\exists_s)_{s\le2}$ of class $C^\infty$
subordinate to the finite open covering
${\rm cone}^{\exists}:=$ 
$\big({\rm cone}_{(\Gamma_1,\Gamma'')}^{\exists},\,
{\rm cone}_{(\Gamma'',\Gamma_1)}^{\exists}\big)$
of the metric space $\mathbb{R}^{({\rm1+d})\times n}_\times$,
i.e.
\begin{align}
&\chi_{1}^\exists(x)+\chi_{2}^\exists(x)
=1,
\qquad\forall x\in\mathbb{R}^{({\rm1+d})\times n}_\times,\\
\text{where}\qquad
&0\le\chi^\exists_s\in C^\infty
\big(\mathbb{R}^{({\rm1+d})\times n}_\times\big)\\
\text{and}\qquad&\text{\rm supp }\chi^\exists_s
\subset {\rm cone}_s^{\exists},
\qquad
s\le2.
\end{align}

\begin{Exa}
One may use the partitions of unity $\chi$
introduced for the  EGF approach
to define such a 
$\chi^\exists$,
i.e.
$\chi_{s}^\exists:=\sum_{\substack{P\in{\cal P}_2(\Gamma'),\,
\Gamma_1\subseteq P_s}}\chi_P$,
$s\le2$.
\end{Exa}

\begin{Prn}\label{Prn:EGS}
The graph distributions are (uniquely)
determined 
by
\begin{align}\label{Tx-EGS}
T_{\Gamma'}^\exists:=-\chi_1^\exists R'_{(\Gamma_1,\Gamma'')}-\chi_2^\exists A'_{(\Gamma_1,\Gamma'')}
\end{align}
(independently of $\chi^\exists$),
i.e.~$T_{\Gamma'}^{\times}=T_{\Gamma'}^\exists\big|_{\mathbb{R}^{({\rm1+d})\times n}_{\,\times}}\in{\cal R}_{{\rm G}_\Gamma}^{\sf T}\big(\mathbb{R}^{({\rm1+d})\times n}_{\,\times}\big)$.
\\[1ex]
\begin{Proof}
This is a consequence of \eq{cone2} and Lemma \ref{Lem:EGS} (cf.~Prop.~4.2.15, p.~116).
\end{Proof}
\end{Prn}

\begin{Rem}
In this EG version the n-point distribution is actually obtained from
a so-called
splitting of
the 
{causal distribution}
$D:=R'-A'=R-A$.
Here, the mapping
$D\mapsto\chi_2^\exists D$ defines the regularized splitting,
i.e.
\begin{align}
T_{\Gamma'}^\times=\chi_2^\exists D_{(\Gamma_1,\Gamma'')}\big|_{\mathbb{R}^{({\rm1+d})\times n}_{\,\times}}- R'_{(\Gamma_1,\Gamma'')}\big|_{\mathbb{R}^{({\rm1+d})\times n}_{\,\times}}.
\end{align}
\end{Rem}

\subsubsection{Regularizing HAs \label{sss:RegHAs}}

As the causal regularization
is understood to be a localization of 
distributions to subregions of 
non-coinciding spacetime points of configuration spaces,
it can be modeled algebraically
via 
$A^0$-modules
$\overline{A}=A^0\,\overline1\oplus\bigoplus_{n\ge0}\overline A^n$
that are extensions of
$\mathbb{N}$-graded (configuration space depending) 
modules ${A}=\bigoplus_{n\ge0}{A}^n$,
where
$\overline A^n=\bigoplus_{r\le n}{A}^n
\big(\mathbb{R}^{({\rm1+d})\times n}_{r\sss\neq}\big)$
and $\overline1$ denotes the formally defined unity.
Moreover, the concrete structures will be algebras.
The instances 
for the latter modules (resp., algebras) are
the (algebra of) n-point distributions ${\cal R}_{{\rm G}_\Gamma}$
as well as their tensor product structure,
$C^\infty\otimes{\cal R}_{{\rm G}_\Gamma}$,
with $C^\infty$ functions (multiplied pointwise).

The concrete regularization for the EGF (resp., EGS) version
is given as a subalgebra
of
$\overline{C^\infty\otimes{\cal R}}_{{\rm G}_\Gamma}$
which is isomorphic to $\overline{\cal R}_{{\rm G}_\Gamma}$ as an algebra,
i.e. $\big(\widehat{\cal R}^{(\exists)}_{{\rm G}_\Gamma},
\mathbb{C},+,\widehat\circ_\Gamma^{(\exists)},\overline1\big)$,
where
\begin{align}
\overline{\cal R}_{{\rm G}_\Gamma}
\,\overset{\iota}{\cong}\,
\widehat{\cal R}^{(\exists)}_{{\rm G}_\Gamma}
\,
\overset{\vartheta^{(\exists)}}{\hookrightarrow}
\,
\overline{C^\infty\otimes{\cal R}}_{{\rm G}_\Gamma}
\end{align}
and, for at least $r\in\mathbb{N}$ non-coinciding spacetime points,
the embedding algebra homomorphism 
\begin{align}
\vartheta^{(\exists)}:
\widehat\bigcirc_\Gamma^{(\exists)}\iota\Big(T^\otimes_{P}
\big|_{\mathbb{R}^{({\rm1+d})\times n}_{r\sss\neq}}\Big)
=:
\widehat\bigcirc_\Gamma^{(\exists)}\widehat T^\otimes_{P}
\mapsto\vartheta^{(\exists)}_{P}\otimes
\bigcirc_\Gamma T^\otimes_{P}
\big|_{\mathbb{R}^{({\rm1+d})\times n}_{r\sss\neq}}
\end{align}
is realized via functions
$\vartheta^{(\exists)}_{P}\in
C^\infty\big(\mathbb{R}^{({\rm1+d})\times n}_{r\sss\neq}\big)$
indexed with $r$-partitions $P\in{\cal P}_r^0(\Gamma')$
of graphs $\Gamma'\in{\rm G}_\Gamma$
where $r\le n\equiv|{\rm V}(\Gamma')|$.
Those causally ordering functions (coF)
will be used to model 
the considered regularizing algebra.
One of their general properties 
is due to the defining
isomorphism,
i.e.~$\vartheta^{(\exists)}_{(\Gamma')}=1$,
$\forall\Gamma'\in{\rm G}_\Gamma$.
The {co-structure} is introduced 
by the gtS shuffling coproduct
where, e.g., 
$\Delta(\overline1)=\overline1\otimes\overline1$
and
$\varepsilon(\overline1)=1$.
Moreover,
the (two) algebras above define bialgebras,
i.e.~$\big(\widehat{\cal R}^{(\exists)}_{{\rm G}_\Gamma},\mathbb{C},+,\widehat\circ_\Gamma^{(\exists)},\eta,\Delta,\varepsilon\big)$.

Notice, one will always use the following abbreviation,
$T_{\Gamma'}=\widehat T_{(\Gamma')}\in\widehat{\cal R}^{(\exists)}_{{\rm G}_\Gamma}$.

\paragraph{Modeling EGF}
The coFs are given in terms of the partition of unity, i.e.
\begin{align}
\vartheta_P:=\frac{1}{2}\chi_{\breve P},
\quad\forall P\in{\cal P}^0_r(\Gamma'),
\,\forall\Gamma'\in{\rm G}_\Gamma,
\,\forall r\le 
|{\rm V}(\Gamma')|,
\end{align}
where $\breve P\in{\cal P}_s(\Gamma')$, for $s\le r$, denotes the tuple $P$ reduced by empty graph entries.
This motivates a further general property of $\vartheta$, i.e.
\begin{align}
\vartheta_{P}=\vartheta_{Q}
\quad\text{if}\quad
\breve P=\breve Q~\text{ and }~|P|=|Q|.
\end{align}

\begin{Thm}$
\big(\widehat{\cal R}_{{\rm G}_\Gamma},\mathbb{C},+,\widehat\circ_\Gamma,\eta,\Delta,\varepsilon,{\rm S}_{\widehat\circ_\Gamma}\big)$
forms a HA where
\begin{align}
{\rm S}_{\widehat\circ_\Gamma}T_{\Gamma'}
=\begin{cases}
T_\emptyset&\text{\rm if}\quad\Gamma'=\emptyset,\\
-T_{\Gamma'}&\text{\rm otherwise}.
\end{cases}
\end{align}
\begin{Proof}
It remains to check the antipode condition.
This can be done by
rewriting EGF's induction (Prop.~\ref{Prn:EGF}),
i.e.
$0=\frac{1}{2}T_{\Gamma'}^{\times}-T_{\Gamma'}^{\chi/2}$.
Let $L:=\mathbb{R}^{({\rm1+d})\times n}_{\times}$, then
\begin{align*}
0&=\Big(1-\frac{1}{2}\Big)\chi_{(\Gamma')}T_{\Gamma'}\big|_L
-\frac{1}{2}\sum_{(\Gamma^1,\Gamma^2)\in{\cal P}_2(\Gamma')}
\chi_{(\Gamma^1,\Gamma^2)}\;\sigma_{\Gamma'}(\Gamma^1\sqcup_\Gamma\Gamma^2)\;T_{\Gamma^1}\circ_\Gamma T_{\Gamma^2}\Big|_L\\
&=
\vartheta_{(\Gamma',\emptyset)}\otimes T_{\Gamma'}\big|_L
-\vartheta_{(\emptyset,\Gamma')}\otimes T_{\Gamma'}\big|_L
-\sum_{(\Gamma^1,\Gamma^2)\in{\cal P}_2(\Gamma')}
\hspace{-1em}
\vartheta_{(\Gamma^1,\Gamma^2)}\otimes
\sigma_{\Gamma'}(...)\;\dots
\Big|_L\\
&=T_{\Gamma'}\;\widehat\circ_\Gamma\;T_{\emptyset}
+T_{\emptyset}\;\widehat\circ_\Gamma\,(-T_{\Gamma'})+
\!\!\sum_{(\Gamma^1,\Gamma^2)\in{\cal P}_2(\Gamma')}
\!\sigma_{\Gamma'}(\Gamma^1\sqcup_\Gamma\Gamma^2)
\,T_{\Gamma^1}\;\widehat\circ_\Gamma\,(-T_{\Gamma^2})\\
&=\widehat\bigcirc_\Gamma\circ({\rm id}\otimes{\rm S}_{\widehat\circ_\Gamma})\circ\Delta T_{\Gamma'}.
\end{align*}
The other side of the antipode condition follows similarly.
\end{Proof}
\end{Thm}

\paragraph{Modeling EGS}
There are two extra properties of $\vartheta^\exists$, i.e.
\begin{align}
\vartheta^\exists_{P}=
\begin{cases}
\vartheta^\exists_{(\dots\sqcup_\Gamma P_{i-1},P_i,P_{i+1}\sqcup_\Gamma\cdots)}
&\text{if}\quad\exists\Gamma_1\subseteq P_i
\text{ s.t. }
y_1=x_{\Gamma_1},
\\
\vartheta^\exists_{\sqcup_\Gamma P}
&\text{otherwise},
\end{cases}
\end{align}
and
\begin{align}
\vartheta^\exists_P=
\begin{cases}
\vartheta^\exists_{(\Gamma_1,\Gamma'')}\\
\vartheta^\exists_{(\Gamma'',\Gamma_1)}
\end{cases}
\text{if}\quad\exists\Gamma_1\subseteq 
\begin{cases}
P_1\\P_{r}
\end{cases}
\text{s.t. }
y_1=x_{\Gamma_1},
\end{align}
so that the coFs can again be given in terms of the partition of unity, i.e.
\begin{align*}
\vartheta_P^\exists:=
\begin{cases}
\chi_s^\exists
&\text{if}\quad
\exists\Gamma_1\subseteq P_s
\text{ s.t. }
y_1=x_{\Gamma_1}
\text{ and }
P\in{\cal P}_2(\Gamma'),\\
0&\text{otherwise},
\end{cases}
\end{align*}
$\forall P\in{\cal P}^0_r(\Gamma')$,
$\Gamma'\in{\rm G}_\Gamma$,
$r\le|{\rm V}(\Gamma')|$.

\begin{Thm}
$\big(\widehat{\cal R}_{{\rm G}_\Gamma}^{\exists},\mathbb{C},+,\widehat\circ_\Gamma^{\exists},\eta,\Delta,\varepsilon,{\rm S}_{\widehat\circ_\Gamma^{\exists}}\big)$
forms a HA where
\begin{align}
{\rm S}_{\widehat\circ^\exists_\Gamma}T_{\Gamma'}
=\begin{cases}
0&\text{\rm if}\quad
\exists\Gamma_1\subseteq\Gamma'\text{ \rm s.t. } y_1=x_{\Gamma_1},\\
{\rm S}_{\circ_\Gamma} T_{\Gamma'}&\text{\rm otherwise}.
\end{cases}
\end{align}
\begin{Proof}
Again, it remains to check the antipode condition.
By rewriting the $R$-part of EGS' induction (cf. Lemma~\ref{Lem:EGS}),
i.e.
$0=\chi_1^\exists\,R_{(\Gamma_1,\Gamma'')}\big|_L=\chi_1^\exists\,T_{\Gamma'}^\exists+\chi_1^\exists\,R'_{(\Gamma_1,\Gamma'')}\big|_L$,
where $L:=\mathbb{R}^{({\rm1+d})\times n}_{\times}$,
one obtains
\begin{align*}
0&=\chi_1^\exists\, T_{\Gamma'}\big|_L
+\chi_1^\exists
\sum_{(\Gamma^1,\Gamma^2)\in{\cal P}_2(\Gamma'),\,\Gamma_1\subseteq\Gamma^1}
\sigma_{\Gamma}(\Gamma^1\sqcup_\Gamma\Gamma^2)\;
T_{\Gamma^1}\circ_\Gamma{\rm S}_{\circ_\Gamma}T_{\Gamma^2}\Big|_L\\
&=\vartheta_{(\Gamma',\emptyset)}^\exists\otimes T_{\Gamma'}\big|_L
+\!\!\sum_{(\Gamma^1,\Gamma^2)\in{\cal P}_2(\Gamma'),\,\Gamma_1\subseteq\Gamma^1}
\!\!\vartheta_{(\Gamma^1,\Gamma^2)}^\exists\otimes
\sigma_{\Gamma'}(\Gamma^1\sqcup_\Gamma\Gamma^2)\;
T_{\Gamma^1}\circ_\Gamma{\rm S}_{\circ_\Gamma} T_{\Gamma^2}\Big|_L\\
&=\sum_{(\Gamma^1,\Gamma^2)\in{\cal P}_2^0(\Gamma'),\,\Gamma_1\subseteq\Gamma^1}
\vartheta_{(\Gamma^1,\Gamma^2)}^\exists\otimes
\sigma_{\Gamma'}(\Gamma^1\sqcup_\Gamma\Gamma^2)\;
T_{\Gamma^1}\circ_\Gamma{\rm S}_{\circ_\Gamma} T_{\Gamma^2}\Big|_L
\\&=\sum_{(\Gamma^1,\Gamma^2)\in{\cal P}_2^0(\Gamma')}
\sigma_{\Gamma'}(\Gamma^1\sqcup_\Gamma\Gamma^2)\;
T_{\Gamma^1}\;\widehat\circ^\exists_\Gamma\;{\rm S}_{\widehat\circ^\exists_\Gamma} T_{\Gamma^2}
\\&\equiv 
\widehat\bigcirc^\exists_\Gamma\circ\big({\rm id}\otimes{\rm S}_{\widehat\circ^\exists_\Gamma}\big)\circ\Delta T_{\Gamma'}.
\end{align*}
Similarly, 
the other side of the antipode condition follows from the $A$-part.
\end{Proof}
\end{Thm}

\subsubsection{BS' Feynman approach\label{sss:BS-Feynman}}

In the conventional approach to pQFT Feynman propagators are used.
This leads to an explicite representation of graph distributions
in momentum space which, without (iterated) renormalization, suffers from
so-called ultraviolet
(sub) divergences.
The distributions,
when going over to configuration space in this subsection,
are shown to be determined only on 
$L:=\mathbb{R}^{({\rm1+d})\times n}_{\,\sss\neq}$.
The aspect of causality has already been investigated by BS \cite{BS55p},
but only the work of BPHZ \cite{BP57u,He66p,Zi69c}
could verify
the correctness of the common method of curing (sub)divergences
via renormalization (over so-called forests of subgraphs, nowadays
encoded by the antipode condition of Kreimer's HA \cite{Kr98o}, resp.).

\begin{Lem}\label{Lem:T-Wick}
Let $\Gamma'=\Gamma^1\sqcup_l\Gamma^2$ be a graph
and $P'\equiv P^1\otimes P^2=({\rm P}\circ{\rm V}_i(\Gamma'))_{i\le n}$,
$n\equiv|{\rm V}(\Gamma')|$,
the associated tuples of particles.
Localized on $L$,
time-ordered (cf.~{\rm\eq{T-ordering}}) Wick contractions can be
represented by Feynman contractions, i.e.\footnote{
The symbol `$\Pi$' is chosen in accorance with BS' notation of a Feynman contraction,
\begin{align*}
{\sf N}
(\ldots\otimes\;\overset{\xymatrix@1{\ar@{_|-_|}[rr]&&}}{\phi^i\;\otimes\;\cdots\;\otimes\;\phi^j}\;\otimes\dots)
={\sf T}[\phi_{-}^{i},\phi_{+}^{j}]^\otimes_{\sigma}\;
{\sf N}(\ldots\otimes\dots\otimes\dots).
\end{align*}
}
\begin{align}\label{Tamalg=Pi}
({\sf T}\,\amalg)^l_{(\Gamma^1,\Gamma^2)}\big|_L=
\Pi^l_{(\Gamma^1,\Gamma^2)}\big|_L
\quad\text{where}\quad
\Pi^l_{(\Gamma^1,\Gamma^2)}
:=\prod_{\substack{(i,j)\in l_{rs}\\
r\le|P^1|,\,s\le|P^2|}}
\Delta_{{\sf F}\gamma}^{P^1_{ri}P^2_{sj}}.
\end{align}
\begin{Proof}
As chronological cycles $C\subset M$ can be excluded by localization on $L$ (cf.~third line), one confirms
\begin{align*}
\Pi^l_{(\Gamma^1,\Gamma^2)}\big|_L
&=\prod_{\substack{(i,j)\in l_{rs}\\r\le|P^1|,\,s\le|P^2|}}
\Big(
\theta_{(r,s)}\,\Delta_{+\gamma}^{P^1_{ri}P^2_{sj}}
+\sigma\,\theta_{(s,r)}\,\Delta_{+\gamma}^{P^2_{sj}P^1_{ri}}
\Big)
\,\bigg|_L
\\
&=\sum_{\substack{M\subseteq
M^{12}\cup M^{21}
\\(u,v)\in M\Leftrightarrow(v,u)\not\in M}}\,
\prod_{\substack{(i,j)\in\overline l_{uv}\\(u,v)\in M}}
\theta_{(u,v)}\;
\sigma_{{\rm I}(\Gamma')}(P'_{u},P'_{v})\;
\Delta_{+\gamma}^{P'_{ui}P'_{vj}}\,\bigg|_L
\\
&=\sum_{\substack{M\subseteq
M^{12}\cup M^{21}
\\\not\exists C\subset M\,\text{cycle}}}\,
\prod_{\substack{(u,v)\in M}}
\theta_{(u,v)}\;
\sigma_{{\rm I}(\Gamma')}(\Gamma'_u\sqcup_\emptyset\Gamma'_v)\;
\amalg_{(\Gamma'_u,\Gamma'_v)}^{\overline l_{uv}}\bigg|_L
\\
&=\sum_{\substack{\pi\in{\cal S}_n}}
\theta_{\pi(\Gamma')}\;\sigma_{{\rm I}(\Gamma^1\sqcup_l\Gamma^2)}\big(\pi(\Gamma')\big)
\;
\prod_{i<j\le n}
\;
\amalg_{(\pi\circ\Gamma'_i,\pi\circ\Gamma'_j)}^{{\pi\circ\overline l}_{ij}}
\,\bigg|_L\\
&=({\sf T}\,\amalg)^{l}_{(\Gamma^1,\Gamma^2)}\big|_L,
\end{align*}
where the (omitted) variables in $\Delta_{+\gamma}^{\ast\ast}(y_{\ast\ast})$ are related as follows,
\begin{align*}
y_{uv}:=y'_u-y'_v=\begin{cases}
y_r-y_s=:y_{rs}
\\y_s-y_r=:y_{sr}\end{cases}
\text{if}\quad
(u,v)=\begin{cases}(r,s+|\nu^1|),
\\(s+|\nu^1|,r).
\end{cases}
\end{align*}
The sets
$\begin{cases}M^{12}\\M^{21}\end{cases}$
\hspace{-1.0em} collect the corresponding 
pairs $(u,v)\in\mathbb{N}_n\times\mathbb{N}_n$
of indices,
and the bar over the concatenation matrix $l\in{\rm l}(\Gamma^1,\Gamma^2)$
denotes its conversion into a matrix of lines,
$\overline{l}:=\begin{pmatrix}\emptyset&l\\
\tilde l
&\emptyset\end{pmatrix}$.
\end{Proof}
\end{Lem}

\begin{Rem}
Restricted to $x_{uv}\in
\mathbb{R}^{1\times n}_{\sss\neq}\times\mathbb{R}^{{\rm d}\times n}\subset L$,
the products of step functions vanish whenever their variables form a chronological cycle $C$, i.e.,
\begin{align*}
\prod_{(u,v)\in C}\theta_{(u,v)}(x_{uv})
\equiv\prod_{(u,v)\in C}\theta(x^0_{uv})
=0
\quad\text{if}\quad
\sum_{(u,v)\in C}x^0_{uv}=0.
\end{align*}
The localization on 
$L$ is sufficient, 
namely
${\rm supp}\,\Delta^{\ast\ast}_{+\gamma}\subseteq\overline V_+\cup\overline V_-=:\overline V$,
which yields ${\rm supp}\,\theta_{(\ast,\ast)}\Delta^{\ast\ast}_{+\gamma}\subseteq L\cap\overline V\subset\mathbb{R}^{1\times n}_{\sss\neq}\times\mathbb{R}^{{\rm d}\times n}$.
\end{Rem}
Without localization,
a simple consideration
about the 
role of cycles,
which correspond to so-called loops in Feyman diagrams,
can motivate the phenomenon of renormalization freedom.

\begin{Exa}\label{Exa:RoleOfLoops}
Let $(r,s,t)$ be 
a cycle, i.e.
$\theta(x_{rs}^0)\,\theta(x_{st}^0)\,\theta(x_{tr}^0)
\big|_{\mathbb{R}^{1\times3}_{\sss\neq}\times\mathbb{R}^{{\rm d}\times3}}=0$,
and apply tI. Then, on the subspace 
generated by $x_r=x_t$, 
\begin{align*}
\theta(x^0)\,\theta(-x^0)\big|_{(\mathbb{R}\setminus\{0\})\times\mathbb{R}^{\rm d}}=0
\quad\text{where}\quad
x:=x_{rs}=-x_{st}.
\end{align*}
The restriction cannot be dropped,
as one realizes by applying the l.h.s.~to certain distributions $t$.
E.g., let $t=\delta$ be the Dirac distribution,
and let the product be defined via a regularization of $\theta$,
e.g. $\theta_n(x^0):=\frac{1}{2}\big(1+\tanh(nx^0)\big)$.
Then, induced by this particular regularization,
\begin{align*}
\theta(x^0)\,\theta(-x^0)\,t(x):=
\lim_{n\to\infty}\theta_n(x^0)\,\theta_n(-x^0)\,t(x)
=\frac{1}{4}\,\delta(x^0),
\end{align*}
which obviously spoils associativity (i.e.~the algebraic structure),
\begin{align*}
\frac{1}{4}\,\delta(x)=\theta(x^0)\,\theta(-x^0)\,\delta(x)\neq
\big(\theta(x^0)\,\theta(-x^0)\big)\,\delta(x)
=0\,\delta(x).
\end{align*}
This is consistent with the  no-go theorem of Schwartz \cite{Sc54s}.
\end{Exa}

\begin{Prn}\label{Prop:ttP}
Localized on $L$, 
graph distributions 
can be represented by 
time-orded products
$T_{\Gamma'}^{\neq}=T_{\Gamma^1}^{\cal R}\circ^{\sf T}_\Gamma\dots\circ^{\sf T}_\Gamma T_{\Gamma^r}^{\cal R}\big|_{L}\in{\cal R}^{\sf T}_{{\rm G}_\Gamma}(L)$,
where
\begin{align}
T_{\Gamma'}^{\neq}
=(i\hbar)^{\#\Gamma'-|\Gamma'|}\lsf{$\sf\Sigma$}_{{\rm b}(\Omega\Gamma')}\,
t_{{\rm I}(\Gamma^1)}^{\cal R}t_{{\rm I}(\Gamma^2)}^{\cal R}
\Pi_{(\Gamma^1,\Gamma^2)}^l\,
N(\Gamma')\Big|_L,
\end{align}
for $r:=2$ and $\Gamma'\equiv\Gamma^1\sqcup_l\Gamma^2$.\\[1ex]
\begin{Proof}
The scalar part rests upon Lemma \ref{Lem:T-Wick} and the structural part upon
\end{Proof}
\end{Prn}
\begin{Lem} {\rm (Wick's theorem of time-ordering).}
Keeping the notations,
\begin{align}
N(\Gamma^1)\circ^{\sf T}_{\Gamma}N(\Gamma^2)\big|_L
=\sum_{\substack{l\in{\rm l}(\Gamma^1,\Gamma^2)\\\Gamma^1\sqcup_l\Gamma^2=\Gamma^1\sqcup_\Gamma\Gamma^2}}\hspace{-.5em}
(i\hbar)^{|l|}\,
\Pi^l_{(\Gamma^1,\Gamma^2)}
N(\Gamma^1\sqcup_l\Gamma^2)\Big|_L.
\end{align}
\begin{Proof}
This follows from 
\eq{NoN=N}
and \eq{Tamalg=Pi} (cf.~Lemma 4.2.25 in \cite{La03c}, p.~122).
\end{Proof}
\end{Lem}
\begin{Rem}
Associativity of $\circ^{\sf T}_\Gamma$ corresponds with $\Pi_{(\cdot,\cdot)}$ being a 2-cocyle.
In addition to $\amalg$, also $\Pi$ can be considered as a Laplace pair.  
\end{Rem}

Explicitely, the graph distributions
\eq{Tbullet:}
can be characterized 
by
\begin{Prn} {\rm (Feynman diagrams).}
Localized on $L$, 
graph distributions are (uniquely) determined by their
scalar parts,
\begin{align}\label{FD}
t^{\cal R}_{\rm I(\Gamma')}\big|_L=\Pi^\Gamma_{\Gamma'}\big|_L
\quad\text{where}\quad
\Pi^\Gamma_{\Gamma'}=
\prod_{\substack{r<s\le n}}\;
\prod_{\substack{\Gamma'_r\sqcup_\Gamma\Gamma'_s\in{\cal P}(\Gamma')}}\Pi^{{\rm l}(\Gamma')_{rs}}_{(\Gamma'_r,\Gamma'_s)}
\end{align}
(independent of the renormalization scheme ${\cal R}$).\\[1ex]
\begin{Proof}
This is a consequence of Prop.~\ref{Prop:ttP}.
The calculation of \eq{FD} is  similar to the one in the proof of Lemma \ref{Lem:T-Wick} (cf.~(4.139) in \cite{La03c}, p.~127).
\end{Proof}
\end{Prn}

\begin{Rem}\label{Rem:gtS}
Immediately, \eq{gtSofFeynProp}, \eq{Tamalg=Pi}, and \eq{FD} imply gtS of $t_{{\rm I}(\Gamma')}\big|_L$. 
\end{Rem}

For completion,
one will as well present an algebraic formulation of the BS regularization
in the sense of Sect.~\ref{sss:RegHAs}.
The the corresponding coFs read
\begin{align}\vartheta^{\sf T}:
\widehat\bigcirc_\Gamma^{\sf T}\widehat T^{\otimes}_{P}
\mapsto\vartheta^{\sf T}_{P}\otimes
\bigcirc^{\sf T}_\Gamma T^{\otimes}_{P}
\big|_L 
\quad\text{where}\quad
\vartheta^{\sf T}_P:=1\big|_L, 
\end{align}
$\forall P\in{\cal P}^0_r(\Gamma'),
\,\forall\Gamma'\in{\rm G}_\Gamma,
\,\forall r\le 
|{\rm V}(\Gamma')|$.

\begin{Thm}
$\big(\widehat{\cal R}^{\sf T}_{{\rm G}_\Gamma},\mathbb{C},+,\widehat\circ^{\sf T}_\Gamma,\eta,\Delta,\varepsilon,{\rm S}_{\widehat\circ^{\sf T}_\Gamma}\big)$ forms a HA where
\begin{align}
{\rm S}_{\widehat\circ^{\sf T}_\Gamma}T_{\Gamma'}=(-1)^{|{\rm V}(\Gamma')|}T_{\widetilde\Gamma'}.
\end{align}
\begin{Proof}
It remains to check the antipode condition.
Due to 
the claimed form of the antipode,
the calculation is analogous to the one that proves Theorem \ref{Thm:N-HA}. Formally, one would only have to substitute
${\sf N}:={\sf T}$ and $\phi^{\bigcirc J}:=T_{\Gamma'}$.
\end{Proof}
\end{Thm}

\subsection{Renormalization\label{ss:Ren}}

Using EGF's method \cite{BF99m} of extending localized distributions to the
(smallest) diagonal, as explained in Sect.~\ref{sss:extension},
the concept of renormalization can be
implemented for both EG versions.
Sect.~\ref{sss:PhysMean} briefly reminds of the physical meaning of renormalization.
Finally, by weakening the structure of causally regularizing HAs
one is able to construct EG counterparts of Kreimer's HA
in Sect.~\ref{sss:Kreimer},
modeling the change of the renormalization scheme.
For the whole subsection,
let $\ast\in\{{\sf F},\exists,{\sf T}\}$ symbolize the 
reference to one of the three causal regularizations
where, to 
include EGF,
$\widehat{\cal R}^{\sf F}_{{\rm G}_\Gamma}:=\widehat{\cal R}^{}_{{\rm G}_\Gamma}$ and $\widehat\circ_\Gamma^{\sf F}:=\widehat\circ_\Gamma$.

\subsubsection{Implementation via an extension of distributions\label{sss:extension}}

Renormalization can be understood as a 
method to (re)obtain elements of $\widehat{\cal R}_{{\rm G}_\Gamma}^\ast$
which are not determined by causality.
Here it is introduced, 
without reference to the concrete
regularization, via (two) projective maps,\footnote{i.e., $\widehat{\rm ren}\Big|_{\widehat{\cal R}_{{\rm G}_\Gamma}^{\ast n}(\mathbb{R}^{({\rm1+d})\times n})}={\rm id}$}
\begin{align}\label{renProj}
\widehat{\rm ren}\equiv
\;\widehat{}\;\circ{\rm ren}
:\widehat{\cal R}_{{\rm G}_\Gamma}^{\ast n}\big(\mathbb{R}^{({\rm1+d})\times n}_{r\sss\neq}\big)\to{\cal R}_{{\rm G}_\Gamma}^{\ast n}\hookrightarrow
\widehat{\cal R}_{{\rm G}_\Gamma}^{\ast n}.
\end{align}
In particular,
when performing the (causally regularizing) induction step,
this map (as indicated below)
implements the renormalization scheme ${\cal R}$
depending on the underlying graph $\Gamma'$,
\begin{align}
{\rm ren}^{\cal R}_{\Gamma'}:T_{\Gamma'}^{r\sss\neq}\mapsto T_{\Gamma'}^{\cal R},\quad\forall r\le n.
\end{align}

For both EG versions, renormalization is 
performed as an
extension of (the) graph distributions
\big(in \eq{Tx-EGF} and \eq{Tx-EGS}, where $r:=2$\big)
to the 
diagonal ${\rm diag}_2^n$,
determined by the scalar parts of the operator-valued distributions,
i.e.
\begin{align}
T_{\Gamma'}^{\cal R}:={\rm ren}^{\cal R}_{\Gamma'}\,T^\times_{\Gamma'}\equiv(i\hbar)^{\#\Gamma'-|\Gamma'|}\,
\lsf{$\sf\Sigma$}_{{\rm b}(\Omega\Gamma')}\,
{\rm ext}^{\varrho,k}_{\Gamma'}
\big(t^\times_{{\rm I}(\Gamma)}\big)
\,N(\Gamma').
\end{align}
Reflecting the involved {renormalization freedom (i.e.~ambiguity)},
the extension map 
${\rm ext}^{\varrho,k}_{\Gamma'}:{\cal D}_{{\rm I}(\Gamma')}^{\prime\otimes}\big(\mathbb{R}^{({\rm1+d})\times n}_{\times}\big)\to{\cal D}_{{\rm I}(\Gamma')}^{\prime\otimes}$
is indexed by an auxiliary function $\varrho\in{\cal S}(\mathbb{R}^{({\rm1+d})\times n})$
and a set (abbr.~by $k$) of 
complex constants $k^{\alpha,{\rm I}(\Gamma')}_{\,\Gamma'}$
where $|\alpha|\le\omega_{\Gamma'}$.
Depending on the associated graph $\Gamma'$,
the instances for $\varrho$ and $k$
can up to a few restrictions be chosen 
arbitrarily.
The concrete choice of $(\varrho_{\Gamma'},k_{\Gamma'})$
fixes a particular renormalization scheme, i.e.
${\cal R}\simeq\big\{(\varrho_{\Gamma'},k_{\Gamma'})\,\big|\,\Gamma'\text{ divergent}\big\}$.
According to the assumed property of {tI},
the relevant dimension reduces to
${\rm dim=(1+d)}(n-1)$,
and hence, the diagonal
${\rm diag}_2^n\subset\mathbb{R}^{({\rm1+d})\times n}$
goes over 
into the origin
$\{0\}\subset\mathbb{R}^{\rm dim}$.
For the BS approach, 
renormalization can only be given formally, i.e.
$T_{\Gamma'}^{\cal R}
:={\rm ren}^{\cal R}\big(T_{\Gamma^1}^{\cal R}\;\widehat\circ^{\sf T}_\Gamma\;T_{\Gamma^2}^{\cal R}\big)
\equiv
T_{\Gamma^1}^{\cal R}\circ^{\sf T}_\Gamma T_{\Gamma^2}^{\cal R}$,
for $\Gamma'\equiv\Gamma^1\sqcup_\Gamma\Gamma^2$.
This is because of its coarse regularization. 
For a concrete realization one will have to apply EG.

In the EG framework renormalization is as well based on powercounting.
This can be realized, as in the EGF version,
via the so-called scaling degree,
\begin{align}
{\rm sd}(t):=\inf\Big\{\rho 
\,\Big|\,
\lim_{\lambda\to0}\,
{\lambda^{\rho-{\rm dim}}t\big(\tau(\lambda^{-1}\cdot)\big)}
=0,\,\forall\tau\in{\cal D}(\mathbb{R}^{\rm dim})
\Big\},
\end{align}
which counts the inverse power of distributions $t\in{\cal D}'(\mathbb{R}^{\rm dim})$
at the origin of $\mathbb{R}^{\rm dim}$,
i.e., $t(x)=O\big(|x|^{\rm sd}\big)$ as $|x|\to0$.
Then the extension of distributions is characterized as follows.

\begin{Prn}
Let\,
$t^\times\in{\cal S}\big(\mathbb{R}^{\rm dim}\setminus\{0\}\big)$
with scaling degree $\rho:={\rm sd}(t^\times)<\infty$.
Then there exist extensions\,
$t^{\varrho,k}\equiv{\rm ext}^{\varrho,k}(t^\times)\in{\cal S}\big(\mathbb{R}^{\rm dim}\big)$
with the same finite scaling degree\,
${\rm sd}(t^{\varrho,k})=\rho$, 
i.e.~$t^{\varrho,k}(\tau^\times)=t^\times(\tau^\times)$,
$\forall\tau^\times\in{\cal S}\big(\mathbb{R}^{\rm dim}\setminus\{0\}\big)$.
Two cases have to be distinguished.
If the 
distribution $t^\times$ is not singular,
i.e.\! if its singular order\, $\omega$ is negative,
then the 
extended distribution
$t^{\varrho,k}$ is uniquely determined.
Otherwise, if the {\em singular order}\,
$\omega:=\rho-{\rm dim} 
\ge0$,
uniqueness is violated.
Then the extension
depends, firstly,
on an arbitrarily chosen 
tempered
function $\varrho\in{\cal S}\big(\mathbb{R}^{\rm dim}\big)$
which 
satisfies 
$\varrho(0)=1$
and 
$\partial^\alpha\varrho
(0)=0$,
and secondly,
on an arbitrary 
set of constants $k^\alpha\in\mathbb{C}$
which defines the extended distribution on a 
set of
testfunctions, $x\mapsto x^\alpha\varrho(x)$,
i.e.
$t^{\varrho,k}\big(\cdot^\alpha\varrho(\cdot)\big)=k^\alpha$,
$\forall\alpha$, $|\alpha|\le\omega$.
\\[1ex]
\begin{Proof} 
Cf.~Theorems 5.2 and 5.3 in \cite{BF99m}, pp.~23ff.
\end{Proof}
\end{Prn}

\begin{Rem}
More precisely, the extended distribution reads
\begin{align}\label{t=tW+S}
t^{\varrho,k}(\tau)
=t^\times\big(W_{\omega,\varrho}\tau\big)
+\sum_{|\alpha|\le\omega}\frac{k^\alpha}{\alpha!}\,\partial^\alpha\tau(0),
\end{align}
so that 
in the {\em second case},
where an 
associated graph 
would be
called {divergent},
the following projection cannot be the identity, 
$W_{\omega,\varrho}:{\cal S}(\mathbb{R}^{\rm dim})\to{\cal S}_\omega(\mathbb{R}^{\rm dim})$, $\tau\mapsto W_{\omega,\varrho}\tau$, where
$\big(W_{\omega,\varrho}\tau\big)(x)=\tau(x)-\sum_{|\alpha|\le\omega}\frac{\partial^\alpha\tau(0)}{\alpha!}\,x^\alpha\varrho(x)$
and where
${\cal S}_\omega=\big\{\xi\in{\cal S}\,\big|\,\partial^\alpha\xi(0)=0,\,|\alpha|\le\omega\big\}$
denotes test function space
appropriate for the unique extension $t$
in the {\em first case},
i.e.
$t(\tau_\omega^\times)=t^\times(\tau_\omega^\times)$,
$\forall\tau_\omega^\times\in{\cal S}_\omega\big(\mathbb{R}^{\rm dim}\setminus\{0\}\big)$.
\end{Rem}

As the {singular order} 
is supposed to be preserved by the 
extension, 
\eq{t=tW+S} and the following lemma allow to describe 
the difference
of the scalar distributions
$t_{{\rm I}(\Gamma)}^{{\cal R}_i}:=t_{{\rm I}(\Gamma)}^{\varrho,k_i}$
w.r.t.~renormalization schemes ${\cal R}_i$,
$i\le2$,
by a polynomial in derivatives
of the delta distribution, i.e.
\begin{align}\label{t-t=}
t_{{\rm I}(\Gamma)}^{{\cal R}_2}-t_{{\rm I}(\Gamma)}^{{\cal R}_1}=
\sum_{|\alpha|\le\omega_{\Gamma}}{k}_{\Gamma}^{\alpha,{\rm I}(\Gamma)}
\partial^{\alpha}
\delta^{\Pi{\rm V}(\Gamma)},
\end{align}
with coefficients $k^\alpha=k_2^\alpha-k_1^\alpha$
formed by the difference
of the renormalization constants
in ${\cal R}_2$ and ${\cal R}_1$,
provided that $\varrho$ has been fixed.

\begin{Lem}
Let $t\in{\cal S}'(\mathbb{R}^{\dim})$ and ${\rm supp}\,t=\{0\}$.
Then $\exists!\,\big\{k^\alpha\in\mathbb{C}\,\big|\,|\alpha|\le\omega\equiv[{\rm sd}(t)]\big\}$ s.t. $t(x)=\sum_{|\alpha|\le\omega}k^\alpha\prod_{i\le{\rm dim}}\partial^{\alpha_i}\delta(x_i)$.\\[1ex]
\begin{Proof}
Cf.~a textbook on distributions,
e.g.~\S8.4 in \cite{Vl71E}.
\end{Proof}
\end{Lem}

\subsubsection{On the physical meaning\label{sss:PhysMean}}

This subsection
will explain the diagram in Fig.~\ref{Fig:NewVertices},
especially the `laws by physics' arrow, in more detail.
It is devided in two paragraphs,
devoted to the classification of renormalizable theories in pQFT
and to the question of the renormalization ambiguity.

\paragraph{Classification of theories}
By applying the scaling degree's additive properties
which,
based on the scaling degree
$2\rho_{p'}:={\rm sd}\big(\Delta^{+\gamma}_{p'}\big)\le
2|p'|+{\rm sd}\big(\Delta^{+\gamma}_{p}\big)$
of the propagator
$\Delta^{+\gamma}_{p'}\equiv(-1)^{|p'|}\partial_{p'}\partial_{\tilde p'}\Delta^{+\gamma}_p$,
lead to
\begin{align}
\rho_{v}=\sum_{i\le|P(v)|}\rho_{P_i(v)}
\quad\text{and}\quad
\rho_\Gamma=\sum_{k\le|{\rm I}(\Gamma)|}\rho_{{\rm I}_k(\Gamma)},
\end{align}
one determines the singular order $\omega_{\Gamma}=\rho_{\Gamma}-{\rm dim}$
of the graph $\Gamma$, i.e.
\begin{align}
\omega_{\Gamma}=\sum_{k\le|{\rm V}(\Gamma)|}\omega_{{\rm V}_k(\Gamma)}-\omega_{\Omega(\Gamma)}
\quad\text{where}\quad
\omega_{v}:=\rho_{v}-({\rm1+d}).
\end{align}
Powercounting therefore
reproduces the well-known classes, called
\begin{align}
\begin{cases}
\text{non-renormalizable}
\\
\text{renormalizable}
\\
\text{super-renormalizable}
\end{cases}
\text{if}\quad
\begin{cases}
(\exists{v}\in{\rm U})\;\omega_{v}>0,
\\
(\forall{v}\in{\rm U})\;\omega_{v}\le0,\,
(\exists{v}\in{\rm U})\;\omega_{v}=0,
\\
(\forall{v}\in{\rm U})\;\omega_{v}<0.
\end{cases}
\end{align}
or as usual, in numbers of
divergent graphs ${\rm G}_{\rm div}$ and vertices ${\rm V}_{\rm div}$,
\begin{align}
\text{if}
\begin{cases}
\hspace{.4em}|{\rm V}_{\rm div}|
\begin{cases}
=\infty,
\\
<\infty,&\;\;\text{where}
\end{cases}
&{\rm V}_{\rm div}=\{v\in{\rm W}\,|\,(\exists\Gamma\in{\rm G}_{\rm div})\,v=\Omega\Gamma\},\\
\;\big|{\rm G}_{\rm div}\big|\hspace{.67em}
<\infty,
&{\rm G}_{\rm div}=\{\Gamma\in{\rm G}_{\rm W}\,|\,\omega_\Gamma\ge0\}.
\end{cases}
\end{align}
This will be illustrated for a (toy model)
pQFT with $|{\rm U}|=1$ initial vertex which couples $m$ scalar fields $\phi$.

\begin{Exa}\label{Exa:phi^m}
Let ${H}_{\cal I}\propto\;:\!\phi^m\!:$.
Since $\Delta^+_{\phi}(x)=O\big(|x|^{-2}\big)$ as $x\to0$,
$\rho_{\phi}=1$ and $\rho_{\phi^m}=m$.
Then, for a graph $\Gamma$ consisting (exclusively) of $\phi^m$-vertices,
\begin{align}\label{omegaGamma}
\omega_{\Gamma}=|{\rm V}(\Gamma)|\big(m-({\rm1+d})\big)-|{\rm E}(\Gamma)|+({\rm1+d}).
\end{align}
Depending on the considered spacetime dimension 1+d, i.e.\!
\begin{align}
\begin{cases}<m,\\=m,\\>m,\end{cases}
\text{so that}\quad
\begin{cases}{\rm V}_{\rm div}={\rm W},\\
{\rm V}_{\rm div}=\big\{v\in{\rm W}\,\big|\,|{\rm P}(v)|\le m\big\},\\
{\rm G}_{\rm div}\subseteq\big\{\Gamma\in{\rm G}_{\rm W}\,\big|\,
|{\rm V}(\Gamma)|\le\frac{{\rm1+d}}{{\rm1+d}-m}\big\},\end{cases}
\end{align}
\end{Exa}
one reproduces the above classes of renormalizable theories.

\paragraph{Renormalization freedom}

Besides translation invariance, which has been implemented explicitely,
$k_{\Gamma}$ will have to satisfy other symmetries.
For instance, gtS is ensured by
the condition
$k^{\alpha,{\rm I}(\Gamma)}_{\Gamma}=\sigma_{\Gamma}\big({\rm I}(\pi\circ\Gamma)\big)\,k_{\pi(\Gamma)}^{\pi(\alpha),{\rm I}(\pi\circ\Gamma)}$,
$\forall\alpha,\,|\alpha|\le\omega_{\Gamma}$.
The requirements for Lorentz invariance have been investigated in
\cite{BP99l}.
Moreover, the physicists' strategy of selecting new vertices produced by renormalization
is to implement further (physical) symmetries, e.g.~C,P,T invariance, pseudo-unitarity, and gauge invariance;
Scharf \cite{Sc01Q} calls this
{\em the principle of perturbative gauge invariance}.

\begin{Exa} (Scalar QED)
Let ${H}_{\cal I}=e 
:\!\phi^\ast(\partial^\mu\phi)A_\mu\!:-\;e :\!(\partial^\mu\phi^\ast)\phi A_\mu\!:$. 
Then there are two initial vertices, 
\begin{align*}
\hspace*{2pt}
u_1=\parbox[c]{70pt}{\shadowbox
{\epsfig{figure=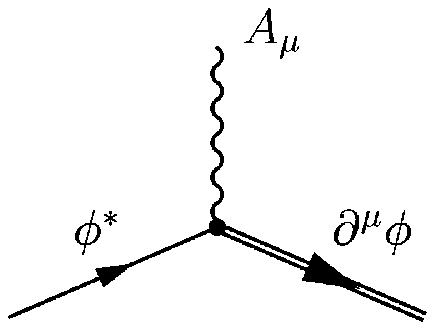,height=.7\linewidth}}}
\hspace{15pt},\hspace{25pt}
u_2=\parbox[c]{70pt}{\shadowbox
{\epsfig{figure=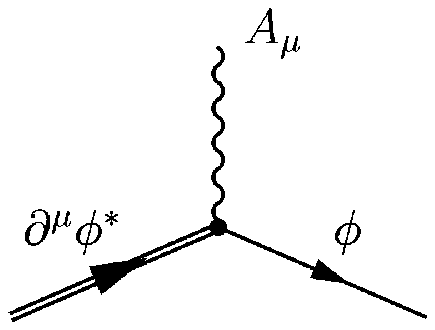,height=.7\linewidth}}}
\hspace*{15pt}.
\end{align*}
Consider ${\rm d}=3$.
Then $\omega_\Gamma=-\omega_{\Omega(\Gamma)}\equiv{\rm1+d}-\rho_{\Omega(\Gamma)}$,
i.e.~a graph $\Gamma$ is divergent only if ${\rm E}(\Gamma)\le4$ and `$\partial\phi$'$\not\in{\rm E}(\Gamma)$.
Therefore,
three candidates ${\rm V}\setminus{\rm U}\subseteq{\rm V}_{\rm div}$
of new physical vertices will be generated by renormalization, 
\begin{align*}
v_1=\parbox[c]{70pt}{\shadowbox
{\epsfig{figure=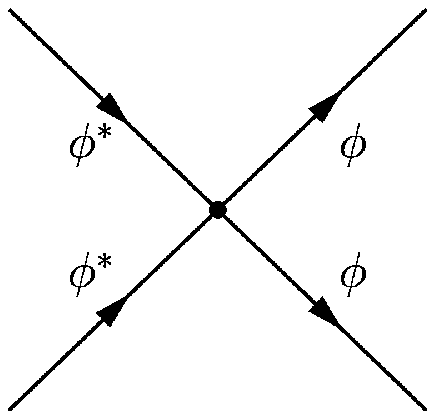,height=.8\linewidth}}}
,\;\;
v_2=\parbox[c]{70pt}{\shadowbox
{\epsfig{figure=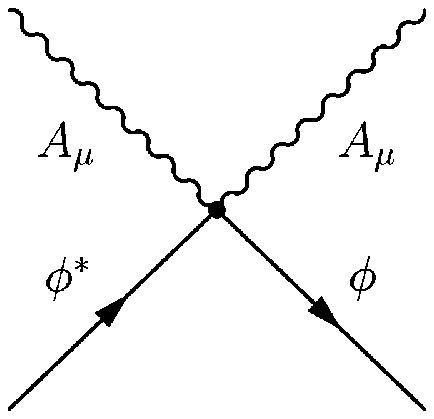,height=.8\linewidth}}}
,\;\;
w_1=\parbox[c]{70pt}{\shadowbox
{\epsfig{figure=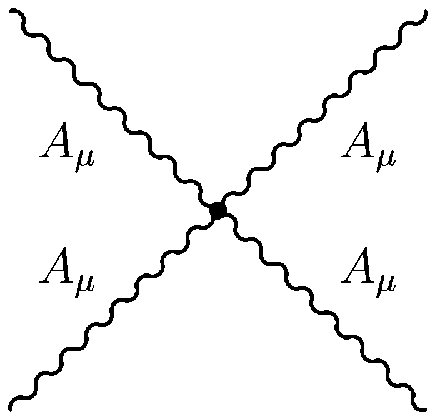,height=.8\linewidth}}}
. 
\end{align*}
However, only $v_1$ and $v_2$ are relevant 
(cf.~\cite{Sc95F}, pp.~335ff.).
In accordance with pseudo-unitarity,
$T_{(v_1)}(x)=
:\!\phi^{\ast\,2}(x)\,\phi^{2}(x)\!:$
where
$a_{v_1}=\lambda\in i\,\mathbb{R}$,
and in accordance with gauge and C-invariance,
$T_{(v_2)}(x)=$
$\frac{1}{i\hbar}\;
g^{\mu\nu}:\!\phi^{\ast}(x)\phi(x)A_\mu(x)A_\nu(x)\!:$
where
$a_{v_2}=\frac{e^2}{i\hbar}$.
In the conventional
approach, one would then
rewrite the interation Hamiltonian
by summing over $\rm V$,
i.e. $H_{\cal I}=i\hbar\sum_{v\in{\rm V}}a_{v}T_{v}$.
\end{Exa}

\subsubsection{EG counterparts of Kreimer's HA\label{sss:Kreimer}}

The goal is to model the change of the renormalization scheme via a HA similar to Kreimer's \cite{Kr98o}.
Formally,
this is in contrast with the literature \cite{GL00c,Pi00t}
where it has been claimed having extracted
Kreimer's HA of renormalization in the EG approach
(which should be impossible).
Similar to Kreimer's approach in \cite{Kr00c}, one will somehow deform
graded commutative versions
of (the three) causally regularizing HAs
in order to construct the aimed HA.
However, the resulting HAs 
are still coassociative but not quasi as Kreimer's.

The graded commutative versions of the causally regularizing HAs
are given as quotient algebras,
$\underline{\cal R}^{\ast}_{{\rm G}_\Gamma}=\widehat{\cal R}^{\ast}_{{\rm G}_\Gamma}\big/\widehat\circ_\Gamma^\ast[\cdot,\cdot]_\sigma^\otimes$,
divided by the ideal which is generated by the graded commutator,
i.e.~$\big(\underline{\cal R}^{\ast}_{{\rm G}_\Gamma},\mathbb{C},+,\underline\circ^\ast_\Gamma,\eta,\Delta,\varepsilon,{\rm S}_{\circ^\ast_\Gamma}\big)$
where the product reads
\begin{align}
\underline T_{\Gamma^1}\;\underline\circ^\ast_\Gamma\;\underline T_{\Gamma^2}=\frac{1}{2}\big[\widehat T_{\Gamma^1}, \widehat T_{\Gamma^2}\big]_{-\sigma}^{\widehat\circ_\Gamma^\ast},
\quad
\forall T_{\Gamma^1}, T_{\Gamma^2}\in{\cal R}_{{\rm G}_\Gamma}^{\sf T}.
\end{align}
Again, one will abbreviate the new representatives,
$T_{\Gamma'}=\underline T_{(\Gamma')}\in\underline{\cal R}_{{\rm G}_\Gamma}^\ast$.
Notice, due to graded commutativity,
$\underline{\cal R}^{\sf T}_{{\rm G}_\Gamma}=\widehat{\cal R}^{\sf T}_{{\rm G}_\Gamma}$ and $\underline\circ_\Gamma^{\sf T}=\widehat\circ_\Gamma^{\sf T}$.

As another prerequisite one introduces the algebra of tuples of strongly
decreasing graphs,
i.e.~$({\cal G}_{\Gamma}^{\supset},\mathbb{C},+,\cdot\,,\emptyset)$,
which is defined as the quotient
of the tensor algebra of graphs divided by some ideal of non-decreasing
graphs,
\begin{align}
{\cal G}_{\Gamma}^{\supset}=\bigoplus_{n\ge0}{\rm G}_{\Gamma}^{\otimes n}\Big/\big(&(...,\Gamma^2)\otimes(\Gamma^1,...),\,
\sqcap\Gamma^2\not\supseteq\sqcap\Gamma^1
;\\ \label{calG-projProp}
&(...,\Gamma^2,\Gamma^1,...)-(...,\Gamma^1,...),\,
\sqcap\Gamma^2=\sqcap\Gamma^1
\big).
\end{align}
The EG counterpart of Kreimer's HA,
providing the transition
$({\cal R}\to{\cal R}')=:{\cal K}$
of renormalization schemes,
will be a HA $\underline{\cal K}^\ast_{{\rm G}_\Gamma}$
that defines an extension of
$\underline{\cal R}^\ast_{{\rm G}_\Gamma}$.
This HA can be constructed in complete analogy
to the latter,
starting with the counterpart of ${\cal R}_{{\rm G}_\Gamma}^{\sf T}$,
i.e.~${\cal K}_{{\rm G}_\Gamma}^{\sf T}:=
\bigoplus_{n\ge0}\mathbb{C}
\big\{k_{\overline\Gamma'}T^{\cal R}_{\Gamma'}\,\big|\,
\overline\Gamma'\in{\cal G}^\supset_\Gamma$,
$\overline\Gamma'_{|\overline\Gamma'|}=\Gamma'$,
$|{\rm V}(\Gamma')|=n\big\}$
where, analogous to \eq{Tbullet:},
\begin{align}
k_{\overline\Gamma'}T^{\cal R}_{\Gamma'}:=(i\hbar)^{\#\Gamma'-|\Gamma'|}{\lsf{\text{$\sf\Sigma$}}}_{{\rm b}(\Omega\Gamma')}\,
{k_{\overline\Gamma'}t_{{\rm I}(\Gamma')}^{\cal R}}\,
N(\Gamma')
\end{align}
which, chosen in accordance with \eq{t-t=},
is determined
by its scalar part,
\begin{align}
\mathbb{T}\ni
k_{\overline\Gamma'}t_{{\rm I}(\Gamma')}^{\cal R}:=
\begin{cases}
t_{{\rm I}(\Gamma')}^{\cal R}&\text{if}\quad\overline\Gamma'=\emptyset,\\
\sum_{|\alpha|\le\omega_{\Gamma'}}k_{\overline\Gamma'\cdot(\Gamma')}^{\alpha,{{\rm I}(\Gamma')}}\,\partial^\alpha\delta^{\Pi{\rm V}(\Gamma')}
&\text{otherwise}.
\end{cases}
\end{align}
The complex
constants
$k_{\overline\Gamma'\cdot(\Gamma')}^{\alpha,{{\rm I}(\Gamma')}}$
with multi-index $\alpha$ and
particle indices
${\rm I}(\Gamma')$,
depending on the product of
the graph $\Gamma'$
with an appropriate tuple
of decreasing graphs $\overline\Gamma'$,
fix the transition
${\cal K}$.
As an empty sum is zero,
$k_{\overline\Gamma'}T_{\Gamma'}^{\cal R}$ vanishes for non-divergent graphs $\Gamma'$, i.e., if $\omega_{\Gamma'}<0$.

The additional structure
in ${\cal K}_{{\rm G}_\Gamma}$
(i.e.~the HA analogous to ${\cal R}_{{\rm G}_\Gamma}$)
can be introduced
as an action of an algebra,
$k_\bullet:{\cal G}_\Gamma^\supset\to{\rm End}({\cal K}_{{\rm G}_\Gamma})$
\big($k_{\overline\Gamma'}T^{\cal R}_{\Gamma'}=0$ if $\sqcap\overline\Gamma'_{|\overline\Gamma'|}\not\supseteq\sqcap\Gamma'$\big),
i.e., $k_{\overline\Gamma^2}\circ k_{\overline\Gamma^1}
=k_{\overline\Gamma^1\cdot\,\overline\Gamma^2}$,
$\forall\overline\Gamma^1,\overline\Gamma^2\in{\cal G}_{\Gamma}^\supset$,
so that, $\forall\overline\Gamma'\in{\cal G}_\Gamma^\supset$,
$k_{\overline\Gamma'}$ defines a HA homomorphism,
i.e.,
$k_{\overline\Gamma'}\circ\bigcirc_\Gamma=\bigcirc_\Gamma\circ(k_{\overline\Gamma'}\otimes k_{\overline\Gamma'})$,
$\Delta\circ k_{\overline\Gamma'}=(k_{\overline\Gamma'}\otimes k_{\overline\Gamma'})\circ\Delta$,
and
${\rm S}_{\circ_\Gamma}\circ k_{\overline\Gamma'}=k_{\overline\Gamma'}\circ{\rm S}_{\circ_\Gamma}$.
Along this line,
starting with Sect.~\ref{sss:GaVD},
one repeats
all the
algebraic constructions
for ${\cal K}$. 
As far as renormalization is involved, the scheme is still ${\cal R}$.
One only has to
require that transition and renormalization exchange, i.e. $k_{\overline\Gamma'}\circ{\rm ren}^{\cal R}_{\Gamma'}:={\rm ren}^{\cal R}_{\Gamma'}\circ k_{\overline\Gamma'}$, whenever $\overline\Gamma'_{|\overline\Gamma'|}\supset\sqcap\Gamma'$.

\begin{Thm}
$\big(\underline{\cal K}^{\ast}_{{\rm G}_\Gamma},\mathbb{C},+,\underline\circ^\ast_\Gamma,\eta,\Delta,\varepsilon,{\rm S}_{\circ^\ast_\Gamma}\big)$
forms a HA which extends
$\underline{\cal R}^{\ast}_{{\rm G}_\Gamma}$, and
$k_\bullet:{\cal G}_\Gamma^\supset\to{\rm End}({\cal K}_{{\rm G}_\Gamma}^\ast)$
defines an action of an algebra 
where, $\forall\overline\Gamma'\in{\cal G}_\Gamma^\supset$,
$k_{\overline\Gamma'}$ is a HA homomorphism.
More precisely,
$(\forall T'\in\underline{\cal K}_{{\rm G}_\Gamma}^{\ast})$
$(\exists!\,T\in\underline{\cal R}^\ast_{{\rm G}_\Gamma})$
$(\exists!\,\overline\Gamma'\in{\cal G}_\Gamma^\supset)$
s.t.
$k_{\overline\Gamma'}T=T'$,
i.e. $k_\bullet:{\cal G}_\Gamma^\supset\to{\rm Hom}(\underline{\cal R}_{{\rm G}_\Gamma}^\ast,\underline{\cal K}_{{\rm G}_\Gamma}^\ast)$
where
$T'\in\underline{\cal R}^\ast_{{\rm G}_\Gamma}\subset\underline{\cal K}_{{\rm G}_\Gamma}^\ast$
requires
$T=T'\equiv k_\emptyset T$.
\\[1ex]
\begin{Proof}
This is the result of
the constructions of the previous subsections,
but in the context of $\cal K$ instead of ${\cal R}$.
\end{Proof}
\end{Thm}
The theorem
gets meaning
when claiming the concrete 
embedding of the new scheme
${\cal R}'$, i.e.
$\underline{\cal R}^{\ast}_{{\rm G}_\Gamma}
\overset{\iota}{\cong}
\underline{\cal R}^{\prime\ast}_{{\rm G}_\Gamma}
\subset
\underline{\cal K}^{\ast}_{{\rm G}_\Gamma}$,
\begin{align}
\iota:
T_{\Gamma'}^{\cal R}\mapsto T_{\Gamma'}^{\cal R'}
:=T_{\Gamma'}^{\cal R}+{\rm S}^{k,1}_{\underline\circ^\ast_\Gamma}
T_{\Gamma'}^{\cal R},
\quad\forall\Gamma'\in{\rm G}_\Gamma,
\end{align}
where ${\rm S}^k_{\underline\circ^\ast_\Gamma}T_{\Gamma'}^{\cal R}
\equiv k_\bullet\circ{\rm S}_{\underline\circ^\ast_\Gamma}T_{\Gamma'}^{\cal R}
:={\rm S}_{\underline\circ^\ast_\Gamma}
\big(
k_{(\Gamma')}T_{\Gamma'}^{\cal R}
\big)$,
$n\equiv|{\rm V}(\Gamma')|$,
and
${\rm S}^{k,1}_{\underline\circ^\ast_\Gamma}T_{\Gamma'}^{\cal R}={\rm S}^{k}_{\underline\circ^\ast_\Gamma}T_{\Gamma'}^{\cal R}$,
except for $n=1$, where
${\rm S}^{k,1}_{\underline\circ^\ast_\Gamma}T_{\Gamma'}^{\cal R}=0$.
The correctness of this setting rests upon the requirement that
${\rm supp}\big(T_{\Gamma'}^{\cal R'}-T_{\Gamma'}^{\cal R}\big)
\subseteq{\rm diag}_n^n$,
which is guaranteed by the following fact. 

\begin{Lem}
${\rm supp}\big({\rm S}^{k,1}_{\underline\circ^\ast_\Gamma}
T_{\Gamma'}^{\cal R}\big)\subseteq{\rm diag}_n^n$.
\\[1ex]
\begin{Proof}
This can be shown by expansion of the antipode, i.e.
\begin{align}
{\rm S}_{\underline\circ^\ast_\Gamma}^k T_{\Gamma'}^{\cal R}
={\rm ren}^{\cal R}_{(\Gamma')}\circ k_{(\Gamma')}
\Big(
- T_{\Gamma'}^{\cal R}
-\sum_{\substack{(\Gamma^1,\Gamma^2)\in{\cal P}_2(\Gamma')}}\sigma_\Gamma(\Gamma^1\sqcup_\Gamma\Gamma^2)\;{\rm S}_{\circ^\ast_\Gamma}^k T_{\Gamma^1}^{\cal R}\;\underline\circ^\ast_\Gamma\; T_{\Gamma^2}^{\cal R}
\Big),
\end{align}
and applying an adequate induction argument.
E.g., if $K^i\in\underline{\cal K}^{\ast}_{{\rm G}_\Gamma}$ and ${\rm supp}\,K^i\subseteq{\rm diag}_{n_i}^{n_i}$, $i\le2$, where $n=n_1+n_2$,
then ${\rm supp}(K^1\,\underline\circ^\ast_\Gamma\,K^2)\subseteq{\rm diag}_n^n$.
\end{Proof}
\end{Lem}
For the case, $\ast={\sf T}$,
one can state the transition between the schemes
explicitely.

\begin{Exa}
Let $P\in{\cal P}_r(\Gamma')$, $r\le n$.
Then $R$ is called a $P$-tree iff
$R\in{\cal G}_\Gamma^{\supset\times r}$,
$(\forall j<r)$ $(R_{ij})_{i\le j}\in{\cal P}_j(\Gamma')$
and $\exists!\,i_j\in\{i_{j-1},i_{j-1}+1\}$
s.t.~$(R_{i_j\,j+1},R_{i_j+1\,j+1})\in{\cal P}_2(R_{i_jj})$
where $i_1=1$,
and $(R_{ir})_{i\le r}=P$.
Applying this notation, one obtains
\begin{align}\label{Forest}
T_{\Gamma'}^{\cal R'}&=T_{\Gamma'}^{\cal R}+
\sum_{r<n}\;
\sum_{\substack{P\in{\cal P}_r(\Gamma')}}
\sigma_{\Gamma'}(P)
\sum_{\substack{R\text{ $P$-tree}}}\;
\underset{i\le r\;}{\bigcirc_\Gamma^{\sf T}}\;
k_{R_i}T_{P_i}^{\cal R}.
\end{align}
For the (most) trivial transition (in this context),
i.e.~$k_{\overline\Gamma'}:=k_{(\Gamma)}$,
$\forall\overline\Gamma'\in{\cal G}_\Gamma^\supset$,
this reduces to
\begin{align}
T_{\Gamma'}^{\cal R'}=T_{\Gamma'}^{\cal R}+\sum_{r<n}\; 
\sum_{\substack{P\in{\cal P}_r(\Gamma')}}
\sigma_{\Gamma'}(P)\;
\underset{i\le r\;}{\bigcirc_\Gamma^{\sf T}}\;
k_{P_i}T_{P_i}^{\cal R},
\end{align}
corresponding to Pinter's result, cf.~Eq.~(14) in \cite{Pi00t}, p.~6.
\end{Exa}
\begin{figure}[h]
\centering
\parbox[c]{90pt}{\shadowbox{\epsfig{figure=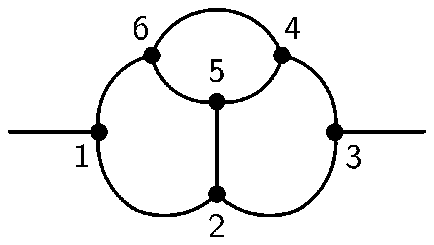,height=.7\linewidth}}}
\hspace{10pt}
\parbox[c]{90pt}{\shadowbox{\epsfig{figure=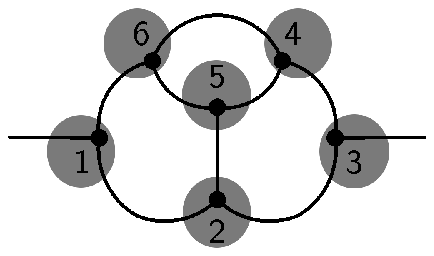,height=.7\linewidth}}}
\hspace{10pt}
\parbox[c]{90pt}{\shadowbox{\epsfig{figure=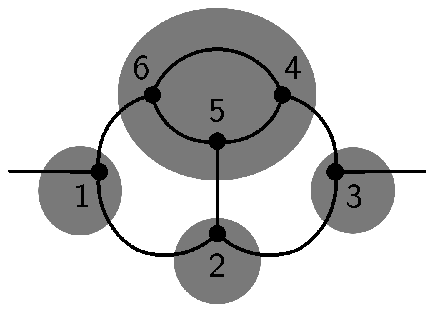,height=.7\linewidth}}}
\\[2.5ex]
\parbox[c]{90pt}{\shadowbox{\epsfig{figure=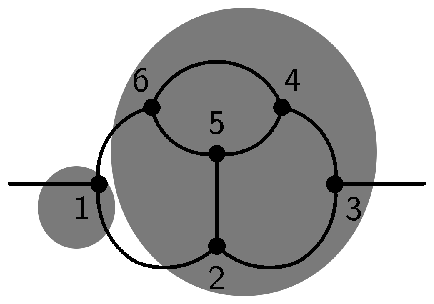,height=.7\linewidth}}}
\hspace{10pt}
\parbox[c]{90pt}{\shadowbox{\epsfig{figure=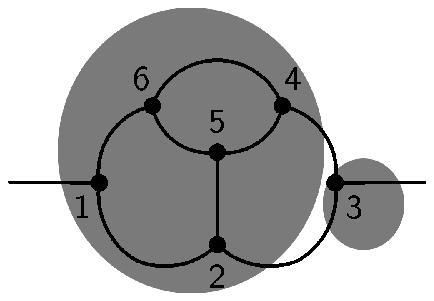,height=.7\linewidth}}}
\hspace{10pt}
\parbox[c]{90pt}{\shadowbox{\epsfig{figure=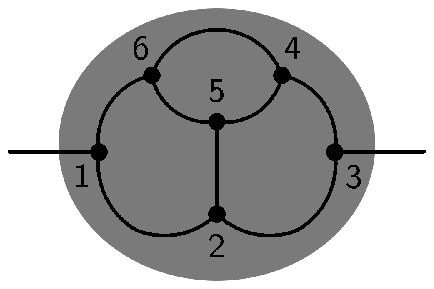,height=.7\linewidth}}}
\\[1ex]
\caption{A sample graph and
partitions $P$ of divergent subgraphs $P_i$.\label{Fig:DivSubGraphs1}}
\end{figure}

Starting the comparison with the conventional approach, the following two examples
will illustrate how overlapping divergences are treated in the EG framework.

\begin{Exa}
Consider ${H}_{\cal I}\propto\phi^3$ for $\rm d=2$
and, especially,
the graph in Fig.~\ref{Fig:DivSubGraphs1}.
Then the singular order of the divergent subgraphs $\Upsilon=P_i$
is given by
$\omega_\Upsilon=3-|{\rm E}(\Upsilon)|$,
i.e.
\begin{align*}
\omega_{\Upsilon}=\begin{cases}1&\text{if}\quad{\rm V}(\Upsilon)\in\{(4,5,6),(1,...,6)\},\\0&\text{otherwise}.\end{cases}
\end{align*}
Notice, 
the two subgraphs with loops,
i.e.~$\big((1,2,5,6),...\big)$ and $\big((2,3,4,5),...\big)$,
which because of their 4 $(>3)$ external lines
are not divergent,
do not contribute to the so-called forest formula \eq{Forest}.
\end{Exa}
Here,
loops obviously do not play the distinguished role they play in the conventional approach
(cf.~Example \ref{Exa:RoleOfLoops}).
That holds true for tree graphs as well.

\begin{figure}[h]
\centering
\parbox[c]{90pt}{\shadowbox{\epsfig{figure=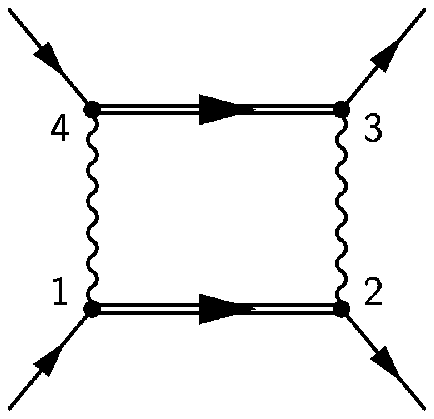,height=.7\linewidth}}}
\hspace{10pt}
\parbox[c]{90pt}{\shadowbox{\epsfig{figure=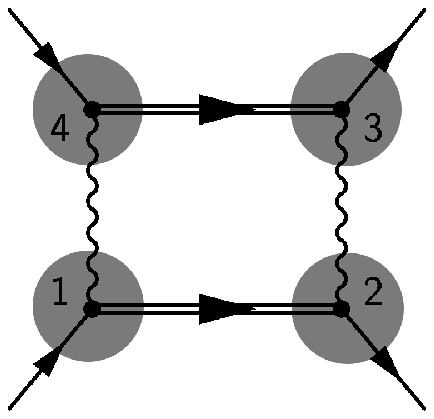,height=.7\linewidth}}}
\hspace{10pt}
\parbox[c]{90pt}{\shadowbox{\epsfig{figure=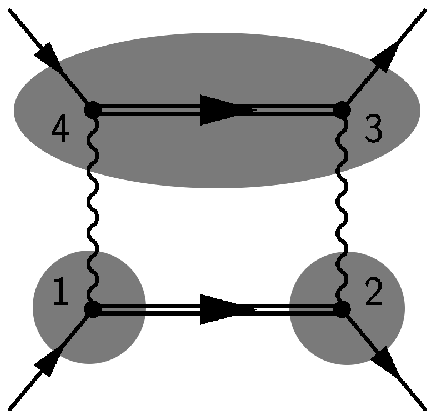,height=.7\linewidth}}}
\\[2.5ex]
\parbox[c]{90pt}{\shadowbox{\epsfig{figure=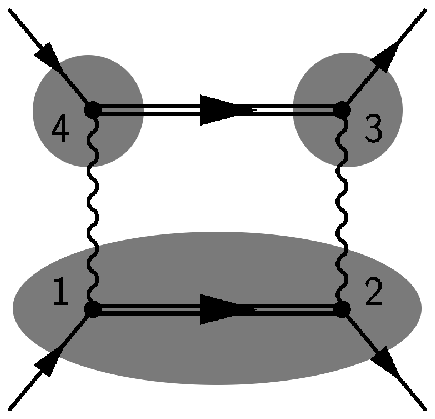,height=.7\linewidth}}}
\hspace{10pt}
\parbox[c]{90pt}{\shadowbox{\epsfig{figure=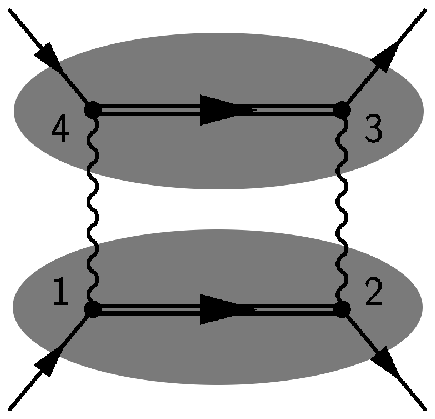,height=.7\linewidth}}}
\hspace{10pt}
\parbox[c]{90pt}{\shadowbox{\epsfig{figure=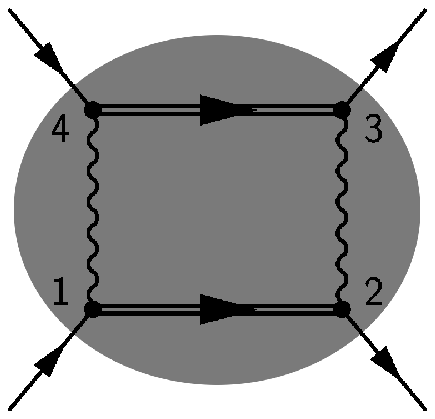,height=.7\linewidth}}}
\\[1ex]
\caption{A sample graph and partitions $P$ of divergent subgraphs $P_i$.\label{Fig:DivSubGraphs2}}
\end{figure}

\begin{Exa}
Consider scalar QED for $\rm d=3$
and, especially, the graph in Fig.~\ref{Fig:DivSubGraphs2}.
Then the singular order of the divergent subgraphs $\Upsilon=P_i$
is given by $\omega_\Upsilon=4-|{\rm E}(\Upsilon)|=0$.
{Notice}, also tree graphs, i.e.~$\big((1,2),...\big)$ and $\big((3,4),...\big)$, do represent divergent subgraphs.
\end{Exa}

Even if the structure of graphs presented here varies from Kreimer's
(which is only pre-Lie),
one can identify the following connection.
\begin{Rem}
Let $R$ be Kreimer's map which is supposed to
leave the short distance singularities unaltered, e.g.,
if ${\rm reg}(\varepsilon)=\sum_{n=-p}^\infty c_n\varepsilon^n$,
$p>0$,
denotes a regularization for $\varepsilon\to0$
then
$R:{\rm reg}(\varepsilon)\mapsto{\rm reg}(\varepsilon)-\sum_{n\ge0}c_n'\varepsilon^n$.
Then one observes
the following
correspondence between the presented and Kreimer's HA,
\begin{align}
k_\bullet\leftrightarrow R
\quad\big(\text{i.e. }
{\rm S}_{\underline\circ^\ast_\Gamma}^k\leftrightarrow S_R\big)
\quad\text{and}\qquad
{\rm ren}^{\cal R}\leftrightarrow{\rm id}-R,
\end{align}
reflecting
the algebraic structure
and, resp.,
the function of renormalization.

In order to ensure that the deformed antipode $S_R$ defines a homomorphism
(which, for ${\rm S}_{\underline\circ^\ast_\Gamma}$, is fulfilled for free)
the map $R$ (and hence, ${\rm id}-R$)
has to satisfy 
a multiplicativity constraint,
the so-called Rota-Baxter relation (cf.~\cite{KE04i}),
\begin{align}
p(XY)+p(X)p(Y)=p\big(Xp(Y)\big)+p\big(p(X)Y\big).
\end{align}
\end{Rem}
Moreover, one can motivate the quasiness of Kreimer's deformed HA.

\begin{Rem}
In the spirit of Kreimer's presentation, e.g.~in \cite{Kr00c},
one may interprete the functor
${\rm Hom}(\underline{\cal R}_{{\rm G}_\Gamma}^\ast,\underline{\cal K}_{{\rm G}_\Gamma}^\ast)$
as a deformation of HAs induced by deforming
\begin{align}
\begin{cases}\text{either the antipode},&{\rm S}_{\underline\circ^\ast_\Gamma}\mapsto
{\rm S}^k_{\underline\circ^\ast_\Gamma}
\equiv k_\bullet\circ{\rm S}_{\circ^\ast_\Gamma},\\
\text{or the coproduct},&
\Delta\mapsto\Delta^k\equiv(k_\bullet\otimes k_\bullet)\circ\Delta,
\end{cases}
\end{align}
while keeping the coproduct/antipode, resp., fixed.
Now it is obvious why the deformed HA in Kreimer's original approach is only quasi;
as one easily checks,
$({\rm id}\otimes\Delta^k)\circ\Delta^kT_{\Gamma'}^{\cal R}=(\Delta^k\circ{\rm id})\circ\Delta^kT_{\Gamma'}^{\cal R}$ is true
only if the transition $k_\bullet$ is trivial.
\end{Rem}
A more detailed presentation of the content of this subsection will be given elsewhere \cite{La04o}.

%% file: p-o

\section{Conclusions and outlook}

About ten years ago HAs have entered the framework of pQFT. At that time Wick and Feynman contractions (i.e.~$\amalg$ and $\Pi$) have been identified, by Rota and Stein \cite{RS94p,RS94v}, as examples for Laplace pairs in the context of combinatorial HAs. Since then, the formal perturbative expansion could have been considered as given by a Laplace HA. However, only recently this has been done, cf.~\cite{Fa00o,BF03l}. The knowledge about it did not get that popular among physicists as Kreimer's Hopf algebraic description of renormalization \cite{Kr98o} in the late nineties.

In this paper, both these HA approaches to pQFT (i.e.~the forth/seventh and the eighth entry in the table of Fig.~1) get represented in the EG framework. The causally regularizing HAs (enties 5--7) can therefore be regarded as a bridge between them. Furthermore, the HAs that model normalization and localization of
free fields (entries 2--3) appear as the stucture which Cliffordization twists into the Laplace HA representing the perturbative expansion.

Supplementing the presented scenario of HAs, which apparently underlies the EG approach to pQFT, some topics for further research in the field
turn up immediately (enumerated from the physical to the mathematical one):
\begin{itemize}
\item
In the spirit of Kreimer's more recent work \cite{Kr02n,Kr02s}, one might also think about implementing Scharf's principle \cite{Sc01Q} of perturbative gauge invariance into the HA context.
\item
Laplace pairs can be interpreted as objects in the theory of quantum groups (e.g., \cite{BO02q,BS02q,BF03l}), namely as so-called R-forms (cf.~Chap.~VIII in \cite{Ka95Q}). Investigations in this direction still seem to be fruitful.
\item
Following an idea of Kontsevich \cite{Ko02r}, one might try to understand renormalization as a dual procedure to the compactification of the configuration space \cite{FM94a}, see also \cite{BK04t}.
\end{itemize}

%% file: p-app

\begin{appendix}

\section{Appendix}

\subsection{Calculations}

\begin{Lem}\label{Lem:Delta^(N)gtS}
$\Delta^{({\sf N})}$, i.e.~$\Delta$ and $\Delta^{\sf N}$, are gtS.\\[1ex]
\begin{Proof}
The two claims can be proven at once, where the possible empty alternative is symbolized by brackets.
Let $X\subseteq{\rm P}^n$, $n\in\mathbb{N}$,
and let $\pi\in S_n$ be a permutation on $X$.
Then
\begin{align*}
\Delta^{({\sf N})}\big(\varphi^{\bigcirc\pi(X)}\big)&=\sum_{P_\pm\in{\cal P}^0_2(\pi(X))}\sigma_{\pi(X)}(P_\pm)\cdots\\
&=\sum_{P_\pm\in{\cal P}^0_2(X)}\sigma_{\pi(X)}(X)\;\sigma_{X}(P_\pm)\cdots
=\sigma(\pi)\;\Delta^{({\sf N})}\big(\varphi^{\bigcirc X}\big),
\end{align*}
where the following identity for sets of partitions,
i.e.~${\cal P}^0_2={\cal P}^0_2\circ\pi$, $\forall\pi\in S_n$,
and following the group representation rule for the grading sign,
$\sigma_Y(Z)=\sigma_Y(X)\,\sigma_X(Z)$, $\forall X,Y,Z\in{\rm P}^n$,
have been applied.
\end{Proof}
\end{Lem}

\begin{Lem}\label{DeltaN-coass}
$\Delta$ and $\Delta^{\sf N}$
are coassociative where, for the first coproduct,
$\varepsilon$ also satisfies the counit property.\\[1ex]
\begin{Proof}
The three calculations below verify the claims.
Let $X\in{\rm P}^n$, $n\in\mathbb{N}$.
Firstly,
\begin{align*}
\big({\rm id}\otimes\Delta\big)\circ\Delta\big(\varphi^{\bigcirc X}\big)
&=\sum_{P\in{\cal P}^0_2(X)}
\sigma_X(P)\;
\varphi^{\bigcirc P_1}\otimes\Delta\big(\varphi^{\bigcirc P_2}\big)\\
&=\sum_{P\in{\cal P}^0_2(X)}\sigma_X(P)\;
\varphi^{\bigcirc P_1}\otimes
\sum_{Q\in{\cal P}^0_2(P_2)}\sigma_{P_2}(Q)\;
\varphi^{\bigcirc Q_1}\otimes\varphi^{\bigcirc Q_2}\\
&=\sum_{(P_1,Q_1,Q_2)\in{\cal P}^0_3(X)}\sigma_{X}(P_1,Q_1,Q_2)\;
\varphi^{\bigcirc P_1}\otimes\varphi^{\bigcirc Q_1}\otimes\varphi^{\bigcirc Q_2}\\
&\;\,\vdots\\
&=\big(\Delta\otimes{\rm id}\big)\circ\Delta\big(\varphi^{\bigcirc X}\big),
\end{align*}
secondly,
\begin{align*}
\big({\rm id}\otimes\varepsilon\big)\circ\Delta\big(\varphi^{\bigcirc X}\big)
&=\sum_{P\in{\cal P}^0_2(X)}
\sigma_X(P)\;
\varphi^{\bigcirc P_1}\otimes\varepsilon\big(\varphi^{\bigcirc P_2}\big)\\
&=\sigma_X(X\otimes\emptyset)\;
\varphi^{\bigcirc X}\otimes1=\varphi^{\bigcirc X}\\
&\;\,\vdots\\
&=\big(\varepsilon\otimes{\rm id}\big)\circ\Delta\big(\varphi^{\bigcirc X}\big),
\end{align*}
which have been rather standard, and thirdly,
\begin{align*}
\big({\rm id}&\otimes\Delta^{\sf N}\big)\circ\Delta^{\sf N}\big(\varphi^{\bigcirc X}\big)\\
&=\sum_{P\in{\cal P}^0_2(X)}
\sigma_X(P)\;
{\rm pr}_+\circ\varphi^{\bigcirc P_1}\otimes{\rm pr}_-\circ\Delta^{\sf N}\big(\varphi^{\bigcirc P_2}\big)\\
&=\sum_{P\in{\cal P}^0_2(X)}\sigma_X(P)\;
\varphi_+^{\bigcirc P_1}\otimes
\sum_{Q\in{\cal P}^0_2(P_2)}\sigma_{P_2}(Q)\;
{\rm pr}_-\circ\varphi_+^{\bigcirc Q_1}\otimes{\rm pr}_-\circ\varphi_-^{\bigcirc Q_2}\\
&=\sum_{P\in{\cal P}^0_2(X)}\sigma_X(P)\;
\varphi_+^{\bigcirc P_1}\otimes
\sigma_{P_2}(\emptyset\otimes P_2)\;
1\otimes\varphi_-^{\bigcirc P_2}\\
&=\sum_{P\in{\cal P}^0_2(X)}\sigma_{X}(P)\;
\varphi_+^{\bigcirc P_1}\otimes1\otimes\varphi_-^{\bigcirc P_2}\\
&\;\,\vdots\\
&=\big(\Delta^{\sf N}\otimes{\rm id}\big)\circ\Delta^{\sf N}\big(\varphi^{\bigcirc X}\big),
\end{align*}
where the projection properties, ${\rm pr}_\pm^2={\rm pr}_\pm$ and ${\rm pr}_\mp\circ{\rm pr}_\pm=0$, have been applied.
\end{Proof}
\end{Lem}

\begin{Lem}\label{DeltaN,epsilon-homom}
$\Delta^{({\sf N})}$ and $\varepsilon$ define homomorphisms w.r.t.~the products $\sssf{\Box}\in\{\circ,\,\sssf{N}\,\}$.\\[1ex]
\begin{Proof}
Let $X^1,X^2\subset{\rm P}^n$, $n\in\mathbb{N}$.
Then
\begin{align}
\varepsilon\big(\varphi^{\Box X^1}\;\sssf{\Box}\;\varphi^{\Box X^2}\big)
\equiv\varepsilon\big(\varphi^{\Box X^1}\big)\,\varepsilon\big(\varphi^{\Box X^2}\big)
=\begin{cases}
1&\text{if}\quad X^1=X^2=\emptyset,\\
0&\text{otherwise},
\end{cases}
\end{align}
and
\begin{align*}
&\Delta^{({\sf N})}\Big(\varphi^{\Box X^1}\;\sssf{\Box}\;\varphi^{\Box X^2}\Big)
=\sum_{P\in{\cal P}^0_2(X^1\otimes X^2)}
\sigma_{X^1\otimes X^2}(P)\;\varphi_{(+)}^{\Box P_1}\otimes\varphi_{(-)}^{\Box P_2}
\\
&=\sum_{\substack{P\in{\cal P}^0_2(X^1\otimes X^2)\\P_i=P^1_i\otimes P^2_i,\,i\le2\\P^j_i\in{\cal P}^0_2(X^j),\,j\le2}}
\sigma_{X^1}(P^1)\,\sigma_{X^2}(P^2)\;
\underbrace{
\varphi_{(+)}^{\Box P^1_1}\circ\varphi_{(+)}^{\Box P^2_1}\otimes\varphi_{(-)}^{\Box P^1_2}\circ\varphi_{(-)}^{\Box P^2_2}}_{=
\big(\varphi_{(+)}^{\Box P^1_1}\otimes\varphi_{(-)}^{\Box P^1_2}\big)\Box\big(\varphi_{(+)}^{\Box P^2_1}\otimes\varphi_{(-)}^{\Box P^2_2}\big)
}\\
&=\Delta^{({\sf N})}\Big(\varphi^{\Box X^1}\Big)\;\sssf{\Box}\;\Delta^{({\sf N})}\Big(\varphi^{\Box X^2}\Big),
\end{align*}
where 
the composition laws for the grading sign (based on the group representation) have been applied.
\end{Proof}
\end{Lem}

\subsection{Illustrations}

\begin{Exa}\label{Exa:Graphs}
Illustrated for two graphs of QED, cf.~Example \ref{Exa:QED}, e.g.~for\\[-2.5ex]
\begin{align*}
\Gamma^1&\equiv(X^1,\Lambda^1)=\parbox[c]{50pt}{
\shadowbox{\epsfig{figure=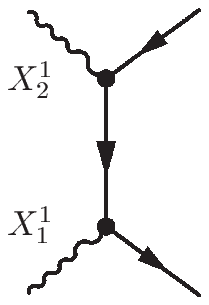,height=\linewidth}}},
&\Gamma^2&\equiv(X^2,\Lambda^2)=\parbox[c]{50pt}{
\shadowbox{\epsfig{figure=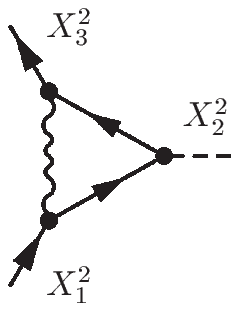,height=\linewidth}}}
\intertext{with vertices $(\overline\psi,\psi,A)$, except of $X^2_2=(\overline\psi,\psi,A^{\rm cl})$, and matrices of lines}
\Lambda^1&=\begin{pmatrix}
\emptyset&\{(1,2)\}\\
\{(2,1)\}&\emptyset
\end{pmatrix},
&\Lambda^2&=\begin{pmatrix}
\emptyset&\{(2,1)\}&\{(3,3)\}\\
\{(1,2)\}&\emptyset&\{(2,1)\}\\
\{(3,3)\}&\{(1,2)\}&\emptyset
\end{pmatrix},
\end{align*}
the concatenation with the matrix
\begin{align*}
\Lambda^{12}=\begin{pmatrix}
\{(2,1)\}&\emptyset&\emptyset\\
\emptyset&\emptyset&\{(1,2)\}\end{pmatrix},
\end{align*}
results in the graph\\[-4ex]
\begin{align*}
\Gamma^1\,\sqcup_{\Lambda^{12}}\,\Gamma^2&\equiv
\Bigg(X,\begin{pmatrix}
\Lambda^1&\Lambda^{12}\\
\widetilde\Lambda^{12}&\Lambda^2
\end{pmatrix}
\Bigg)
=\parbox[c]{20pt}{
\shadowbox{\epsfig{figure=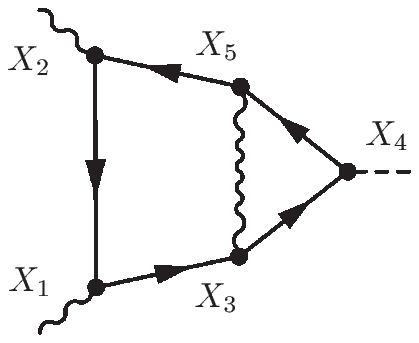,height=3\linewidth}}}\hspace{6.7em}.\\[-1.5ex]
\end{align*}
\end{Exa}

\begin{Exa}\label{Exa:Dyson}
Consider the following quantum model,
where the time evolution is determined by the Schr{\"o}dinger equation,
$i\hbar\,\partial_t\varphi(t,\vec x)=H(t)\varphi(t,\vec x)$,
and a interaction Hamiltonian,
$H(t)=\int_{\mathbb{R}^{\rm d}}dx\,{H}(t,\vec x)$.
The $S$-matrix 
${S}=\lim_{t_\pm\to\pm\infty}U(t_+,t_-)$
is the limit of an unitary operator $U$
which describes the time evolution of an incoming free field.
By repeatedly applying
the integrated Schr{\"o}dinger equation,
$\varphi(t,\vec x)=\varphi(s,\vec x)-\frac{i}{\hbar}\int_s^tdt_1\,H(t_1)\varphi(t_1,\vec x)$,
one obtains the so-called {\em Dyson series}
\begin{align*}
U(t,s)=1+\sum_{n=1}^\infty\frac{(-i)^n}{\hbar^n}\int_s^{t}dt_1\dots\int_s^{t_{n-1}}dt_n\,H(t_1)\cdots H(t_n).
\end{align*}
By inserting the {time-ordering} operation, i.e.
\begin{align}\label{T-ordering}
{\sf T}\big(h(t_1,\dots,t_n)\big)=\sum_{\pi\in {\cal S}_n}
\theta_{(t_{\pi_1},\dots, t_{\pi_n})}\,
\sigma(\pi)\, h(t_{\pi_1},\dots,t_{\pi_n}),
\end{align}
where
\begin{align}
\theta_{(t_1,\dots, t_n)}
:=\theta(t_1-t_2)\cdots\theta(t_{n-1}-t_n)
=\begin{cases}1&\text{if}\quad t_1>\dots>t_n,\\
0&\text{otherwise}\end{cases}
\end{align}
denotes a product of step
functions,
the limit can be performed, i.e.
\begin{align*}
S&=1+\sum_{n=1}^\infty\frac{1}{n!(i\hbar)^n}\int_{\mathbb{R}^{1+{\rm d}}} dt_1\,dx_1
\dots\int_{\mathbb{R}^{1+{\rm d}}} dt_n\,dx_n{\rm T}\big({H}(t_1,\vec x_1)\cdots{H}(t_n,\vec x_n)\big).
\end{align*}
\end{Exa}

\end{appendix}

%% file: p.bbl
\begin{thebibliography}{10}

\bibitem{BK04t}
C.~Bergbauer and D.~Kreimer.
\newblock {The Hopf algebra of rooted trees in Epstein-Glaser renormalization}.
\newblock {\em \rm arXiv:hep-th/0403207}, 22 Mar 2004.

\bibitem{BP99l}
K.~Besser, G.~Pinter, and D.~Prange.
\newblock Lorentz invariant renormalization in causal perturbation theory.
\newblock {\em \rm arXiv:hep-th/9903266v2}, 23 Apr 1999.

\bibitem{Bo00g}
F.~Boas.
\newblock Gauge theories in local causal perturbation theory. {Ph.D.} thesis,
  {Hamburg University}, 1999.
\newblock {\em \rm arXiv:hep-th/0001014}, 4 Jan 2000.

\bibitem{BP57u}
N.~N. Bogoliubow and O.~S. Parasiuk.
\newblock {\"U}ber die {M}ultiplikation der {K}ausalfunktionen in der
  {Q}uantentheorie der {F}elder.
\newblock {\em Acta Math.}, 97:227--266, 1957.

\bibitem{BS55p}
N.~N. Bogoliubow and D.~W. Schirkow.
\newblock Probleme der {Q}uantentheorie der {F}elder.
\newblock {\em Fortschritte der Physik}, 3:439--495, 1955.

\bibitem{BF03l}
C.~Brouder, B.~Fauser, A.~Frabetti, and R.~Oeckl.
\newblock Let's twist again.
\newblock {\em \rm arXiv:hep-th/0311253v1}, 26 Nov 2003.

\bibitem{BF04n}
C.~Brouder, A.~Frabetti, and C.~Krattenthaler.
\newblock Non-commutative {H}opf algebra of formal diffeomorphisms.
\newblock {\em \rm arXiv:math.QA/0406117}, 7 Jun 2004.

\bibitem{BO02q}
C.~Brouder and R.~Oeckl.
\newblock Quantum groups and quantum field theory: {I. The free scalar field}.
\newblock {\em \rm arXiv:hep-th/0208118}, 16 Aug 2002.

\bibitem{BS02q}
C.~Brouder and W.~Schmitt.
\newblock Quantum groups and quantum field theory: {III. Renormalization}.
\newblock {\em \rm arXiv:hep-th/02010097}, 10 Oct 2002.

\bibitem{BF97i}
R.~Brunetti and K.~Fredenhagen.
\newblock Interacting quantum fields in curved space: renormalizability of
  $\varphi^4$.
\newblock {\em \rm arXiv:gr-qc/9701048}, 21 Jan 1997.

\bibitem{BF99m}
R.~Brunetti and K.~Fredenhagen.
\newblock Microlocal analysis and interacting quantum field theories:
  renormalization on physical backgrounds.
\newblock {\em Commun.Math.Phys.}, 208:623--661, 2000.
\newblock \rm arXiv:math-ph/9903028, 12 Mar 1999.

\bibitem{BF96T}
R.~Brunetti, K.~Fredenhagen, and M.~K{\"o}hler.
\newblock The microlocal spectrum condition and the {W}ick polynomials of free
  fields.
\newblock {\em Commun. Math. Phys.}, 180:312, 1996.

\bibitem{CD82A}
Y.~Choquet-Bruhat, C.~DeWitt-Morette, and M.~Dillard-Bleick.
\newblock {\em Analysis, Manifolds and Physics}.
\newblock North-Holland, 1982.

\bibitem{DF98a}
M.~D\"utsch and K.~Fredenhagen.
\newblock A local (perturbative) construction of observables in gauge theories:
  the example of {QED}.
\newblock {\em Commun.Math.Phys.}, 203:71--105, 1999.
\newblock \rm arXiv:hep-th/9807078, 18 Nov 1998.

\bibitem{DH95C}
M.~D\"utsch, T.~Hurth, and G.~Scharf.
\newblock Causal construction of {Yang-Mills} theories {IV}.
\newblock {\em Nuovo Cim.}, 108A:737, 1995.

\bibitem{KE04i}
K.~Ebrahimi-Fard, L.~Guo, and D.~Kreimer.
\newblock Integrable renormalization {II}: the general case.
\newblock {\em \rm arXiv:hep-th/0403118v1}, 10 Mar 2004.

\bibitem{EG73t}
H.~Epstein and V.~Glaser.
\newblock The role of locality in perturbation theory.
\newblock {\em Ann. Inst. Henri Poincar{\'e}}, 19(3):211--295, 1973.

\bibitem{Fa00o}
B.~Fauser.
\newblock On the {H}opf algebraic origin of {W}ick normal-ordering.
\newblock {\em \rm arXiv:hep-th/0007032}, 5 Jul 2000.

\bibitem{FM94a}
W.~Fulton and R.~MacPherson.
\newblock A compactification of configuration spaces.
\newblock {\em Ann. Math.}, 139:183--225, 1994.

\bibitem{GL00c}
J.~M. Gracia-Bondia and S.~Lazzarini.
\newblock {Connes-Kreimer-Epstein-Glaser} renormalization.
\newblock {\em \rm arXiv:hep-th/0006106}, 15 Jun 2000.

\bibitem{Gr00q}
N.~Grillo.
\newblock Quantization of the gravitation field, characterization of the
  physical subspace and unitarity in causal quantum gravity.
\newblock {\em \rm arXiv:hep-th/9911118 v2}, 22 Jun 2000.

\bibitem{He66p}
K.~Hepp.
\newblock Proof of the {Bogoliubov-Parashiuk} theorem on renormalization.
\newblock {\em Commun. Math. Phys.}, 2:301--326, 1966.

\bibitem{Ka95Q}
C.~Kassel.
\newblock {\em Quantum groups}.
\newblock Springer-Verlag, 1995.

\bibitem{KS97Q}
A.~Klimyk and C.~Schm{\"u}dgen.
\newblock {\em Quantum groups and their respresentations}.
\newblock Springer-Verlag, 1997.

\bibitem{Ko02r}
M.~Kontsevich.
\newblock Talk at {IHES} conference.
\newblock {\em ``Renormalization, theory and perspectives''}, Oct. 14--18 2002.

\bibitem{Kr95o}
F.~Krahe.
\newblock On the algebra of ghost fields.
\newblock {\em \rm preprint DIAS-STP-95-02}.
\newblock \rm arXiv:hep-th/9502097, 15 Feb 1995.

\bibitem{Kr95a}
F.~Krahe.
\newblock A causal approach to massive {Yang-Mills} theories.
\newblock {\em Acta Phys. Polonica B}, 66:2453, 1996.
\newblock \rm preprint DIAS-STP-95-01. arXiv:hep-th/9508038, 9 Aug 1995.

\bibitem{Kr98o}
D.~Kreimer.
\newblock On the {H}opf algebra structure of perturbative quantum field theory.
\newblock {\em Adv. Theor. Math. Phys.}, 2:303, 1998.
\newblock \rm arXiv:q-alg/9707029 v4, 12 Feb 1998.

\bibitem{Kr99o}
D.~Kreimer.
\newblock On overlapping divergences.
\newblock {\em Commun. Math. Phys.}, 204:669, 1999.
\newblock \rm arXiv:hep-th/9810022 v4, 12 Feb 1999.

\bibitem{Kr00c}
D.~Kreimer.
\newblock Combinatorics of (perturbative) quantum field thoery.
\newblock {\em \rm arXiv:hep-th/0010059}, 9 Oct 2000.

\bibitem{Kr02n}
D.~Kreimer.
\newblock New mathematical structures in renormalizable quantum field theories.
\newblock {\em \rm arXiv:hep-th/0211136 v1}, 14 Nov 2002.

\bibitem{Kr02s}
D.~Kreimer.
\newblock Structures in {F}eynman graphs - {H}opf algebras and symmetries.
\newblock {\em \rm arXiv:hep-th/0202110 v1}, 18 Feb 2002.

\bibitem{La03c}
A.~Lange.
\newblock {Causal perturbative Quantum Field Theory in the Epstein Glaser
  approach: Graphs and Hopf algebras}. {Ph.D.~thesis}, {Universit{\"a}t
  Leipzig}, 2003.

\bibitem{La04o}
A.~Lange.
\newblock On an analog of {Kreimer's Hopf algebra}.
\newblock {\em In preparation}.

\bibitem{Ob92M}
M.~Oberguggenberger.
\newblock {\em {Multiplication of distributions and applications to partial
  differential equations}}.
\newblock Number 259 in {Pitman Research Notes in Mathematics Series}. {Longman
  Scientific \& Technical}, 1992.

\bibitem{Pi00t}
G.~Pinter.
\newblock The {H}opf algebra structure of {C}onnes and {K}reimer in
  {E}pstein-{G}laser renormalization.
\newblock {\em \rm arXiv:hep-th/0012057}, 7 Dec 2000.

\bibitem{RS94p}
G.-C. Rota and J.~A. Stein.
\newblock Plethystic {Hopf} algebras.
\newblock {\em Proc. Natl. Acad. Sci. USA}, 91:13057--13061, 1994.

\bibitem{RS94v}
G.-C. Rota and J.~A. Stein.
\newblock Plethystic {Hopf} algebras and vector symmetric functions.
\newblock {\em Proc. Natl. Acad. Sci. USA}, 91:13061--13066, 1994.

\bibitem{Sc95F}
G.~Scharf.
\newblock {\em Finite quantum electrodynamics: the causal approach}.
\newblock Springer-Verlag, 1995.

\bibitem{Sc01Q}
G.~Scharf.
\newblock {\em Quantum gauge theories: a true ghost story}.
\newblock Wiley-Interscience, 2001.

\bibitem{Sc54s}
L.~Schwartz.
\newblock {Sur l'impossibilit{\'e} de la multiplication des distributions}.
\newblock {\em {Comptes Rendus Acad. Sci. Paris}}, 239:847--848, 1954.

\bibitem{Vl71E}
V.~S. Vladimirov.
\newblock {\em Equations of mathematical physics}.
\newblock M. Dekker, 1971.

\bibitem{VD88T}
V.~S. Vladimirov, Y.~N. Drozzinov, and O.~I. Zavialov.
\newblock {\em Tauberian theorems for generalized functions}.
\newblock Mathematics and its applications, Soviet series. Kluwer Academic
  Publishers, 1988.

\bibitem{Zi69c}
W.~Zimmermann.
\newblock Convergence of {Bogoliubov's} method of renormalization in momentum
  space.
\newblock {\em Commun. Math. Phys.}, 15:208--234, 1969.

\bibitem{Zi96Q}
J.~Zinn-Justin.
\newblock {\em Quantum field theory and critical phenomena}.
\newblock Clarendon Press, 1996.

\end{thebibliography}
